\documentclass[onecolumn]{aastex701}
\usepackage{amsmath}

\newcommand{\msun}{\text{M}_\odot}

\graphicspath{{figures/}}
\newcommand{\berkeley}{\affiliation{Department of Astronomy, University of California, Berkeley, CA 94720, USA}}
\newcommand{\warsaw}{\affiliation{{Astronomical Observatory, University of Warsaw,
Al.~Ujazdowskie~4, 00-478~Warszawa, Poland}}}
\newcommand{\warwick}{\affiliation{Department of Physics, University of Warwick, Gibbet Hill
Road, Coventry, CV4~7AL,~UK}}
\newcommand{\villanova}{\affiliation{Villanova University, Department of Astrophysics and Planetary
Sciences, 800 Lancaster Ave., Villanova, PA 19085, USA}}
\defcitealias{synthpopI}{J. Kl{\"u}ter \& M. J. Huston et al.}

\shorttitle{Black Hole Microlensing}
\shortauthors{Huston et al.}

\begin{document}

\title{A Search For Stellar-mass Black Holes Via Astrometric Microlensing II: 2012-2015 Keck Candidates}

\author[0000-0003-4591-3201]{Mace J. Huston}
\berkeley
\email[show]{mhuston@berkeley.edu}

\author[0000-0001-9611-0009]{Jessica R. Lu}
\berkeley
\email[show]{jlu.astro@berkeley.edu}

\author[0000-0002-6406-1924]{Casey Y. Lam}
\altaffiliation{NHFP Einstein Fellow}
\affiliation{Observatories of the Carnegie Institution for Science, Pasadena, CA 91101, USA}
\email{clam@carnegiescience.edu}
 
\author[0000-0003-0547-8444]{J. Nijaid Arredondo}
\affiliation{Department of Physics,
University of Illinois at Urbana-Champaign, Urbana, Illinois 61801, USA}
\email{josena2@illinois.edu}

\author[0000-0002-0287-3783]{Natasha S. Abrams}
\berkeley
\email{nsabrams@berkeley.edu}

\author[0009-0005-9846-8561]{Shep Brooke}
\berkeley
\email{shcesysa@berkeley.edu}

\author[0009-0007-2112-8619]{Sage H. Remulla}
\berkeley 
\email{shremull@berkeley.edu}

\author[0000-0002-6786-8774]{Eran Ofek}
\affiliation{Weizmann Institute of Science, Rehovot 76100, Israel}
\email{eran.ofek@weizmann.ac.il}

\author[0000-0002-7226-0659]{Michael S. Medford}
\berkeley
\email{michaelmedford@gmail.com}

\author[0000-0002-9915-8195]{Fatima Abdurrahman}
\affiliation{Independent Researcher}
\email{TheDrFatima@gmail.com}

\author[0000-0002-2350-4610]{Shrihan Agarwal}
\affiliation{Department of Astronomy, University of Chicago, Chicago, IL 60615, US}
\berkeley
\affiliation{Lawrence Berkeley National Laboratory, 1 Cyclotron Road, Berkeley, CA 94720, USA}
\email{shrihan@uchicago.edu}

\author[0000-0003-0652-1862]{Edward Broadberry}
\affiliation{Department of Astronomy, University of Maryland, College Park, MD 20742, USA}
\email{edbroad@umd.edu}

\author[0000-0001-6384-7450]{Matthew Freeman}
\berkeley
\email{matthew.s.r.freeman@gmail.com}

\author[0000-0003-2874-1196]{Matthew W. Hosek Jr.}
\affiliation{UCLA Department of Physics and Astronomy, Los Angeles, CA 90095, USA}
\email{mwhosek@astro.ucla.edu}

\author[0000-0001-5341-0765]{Siyao Jia}
\berkeley
\email{siyaojia07@gmail.com}

\author[0000-0002-5029-3257]{Sean K. Terry}
\affiliation{Department of Astronomy, University of Maryland, College Park, MD 20742, USA}
\affiliation{Code 667, NASA Goddard Space Flight Center, Greenbelt, MD 20771, USA}
\email{skterry@umd.edu}

\collaboration{17}{(Leading Authors)}

\author[0000-0001-5207-5619]{Andrzej Udalski}
\warsaw
\email{udalski@astrouw.edu.pl}

\author[0000-0001-7016-1692]{Przemek Mr{\'o}z}
\warsaw
\email{pmroz@astrouw.edu.pl}

\author[0000-0002-9245-6368]{Rados{\l}aw Poleski}
\warsaw
\email{rpoleski@astrouw.edu.pl }

\author[0000-0002-2335-1730]{Jan Skowron}
\warsaw
\email{jskowron@astrouw.edu.pl}

\author[0000-0002-0548-8995]{Micha{\l} K. Szyma{\'n}ski}
\warsaw
\email{} 

\author[0000-0002-7777-0842]{Igor Soszy{\'n}ski}
\warsaw
\email{soszynsk@astrouw.edu.pl}

\author[0000-0002-2339-5899]{Pawe{\l} Pietrukowicz}
\warsaw
\email{pietruk@astrouw.edu.pl} 

\author[0000-0003-4084-880X]{Szymon Koz{\l}owski}
\warsaw
\email{simkoz@astrouw.edu.pl}   

\author[0000-0001-6364-408X]{Krzysztof Ulaczyk}
\warwick
\email{} 

\author[0000-0002-9326-9329]{Krzysztof A. Rybicki}
\warsaw
\email{rybicki.kris@gmail.com}

\author[0000-0002-6212-7221]{Patryk  Iwanek}
\warsaw
\email{piwanek@astrouw.edu.pl} 

\author[0000-0002-3051-274X]{Marcin Wrona}
\warsaw \villanova
\email{mwrona@astrouw.edu.pl} 

\author[0000-0002-1650-1518]{Mariusz Gromadzki}
\warsaw
\email{marg@astrouw.edu.pl} 

\author[0000-0002-8911-6581]{Mateusz J. Mr{\'o}z}
\warsaw
\email{mmroz@astrouw.edu.pl}

\author[0009-0006-7826-2416]{Micha{\l} Ma{\l}kowski}
\warsaw
\email{mmalkowski@astrouw.edu.pl}

\collaboration{16}{(The OGLE Collaboration)}

\author[0009-0000-8865-4778]{Fumio Abe}
\affiliation{Institute for Space-Earth Environmental Research, Nagoya University, Nagoya 464-8601, Japan}
\email{abe@isee.nagoya-u.ac.jp}

\author[0000-0001-8043-8413]{David P.~Bennett}
\affiliation{Code 667, NASA Goddard Space Flight Center, Greenbelt, MD 20771, USA}
\affiliation{Department of Astronomy, University of Maryland, College Park, MD 20742, USA}
\email{bennett.moa@gmail.com}

\author{Aparna Bhattacharya} 
\affiliation{Code 667, NASA Goddard Space Flight Center, Greenbelt, MD 20771, USA}
\affiliation{Department of Astronomy, University of Maryland, College Park, MD 20742, USA}
\email{abhatta5@umd.edu}

\author[0000-0002-8131-8891]{Ian A. Bond}
\affiliation{ School of Mathematical and Computational Sciences, Massey University, Auckland 0745, New Zealand}
\email{I.A.Bond@massey.ac.nz}

\author{Kotaro Daimon}
\affiliation{Department of Earth and Space Science, Graduate School of Science, The University of Osaka, Toyonaka, Osaka 560-0043, Japan}
\email{} 

\author[0009-0007-1381-3384]{Ryusei Hamada}
\affiliation{Department of Earth and Space Science, Graduate School of Science, The University of Osaka, Toyonaka, Osaka 560-0043, Japan}
\email{} 

\author[0000-0003-4776-8618]{Yuki Hirao}
\affiliation{Institute of Astronomy, Graduate School of Science, The University of Tokyo, 2-21-1 Osawa, Mitaka, Tokyo 181-0015, Japan}
\email{hirao@iral.ess.sci.osaka-u.ac.jp}

\author{Shuma Makida}
\affiliation{Department of Earth and Space Science, Graduate School of Science, The University of Osaka, Toyonaka, Osaka 560-0043, Japan}
\email{} 

\author[0000-0003-2267-1246]{Stela Ishitani Silva}
\affiliation{Department of Physics, The Catholic University of America, Washington, DC 20064, USA}
\affiliation{Code 667, NASA Goddard Space Flight Center, Greenbelt, MD 20771, USA}
\email{stela.ishitani@gmail.com}

\author[0000-0001-9818-1513]{Shota Miyazaki}
\affiliation{Institute of Space and Astronautical Science, Japan Aerospace Exploration Agency, 3-1-1 Yoshinodai, Chuo, Sagamihara, Kanagawa 252-5210, Japan}
\email{miyazaki@ir.isas.jaxa.jp}

\author[0000-0003-1978-2092]{Yasushi Muraki}
\affiliation{Institute for Space-Earth Environmental Research, Nagoya University, Nagoya 464-8601, Japan}
\email{ysa_muraki@topaz.plala.or.jp}

\author{Tutumi Nagai}
\affiliation{Department of Earth and Space Science, Graduate School of Science, The University of Osaka, Toyonaka, Osaka 560-0043, Japan}
\email{} 

\author[0009-0005-3414-455X]{Kansuke Nunota}
\affiliation{Department of Earth and Space Science, Graduate School of Science, The University of Osaka, Toyonaka, Osaka 560-0043, Japan}
\email{nunota@iral.ess.sci.osaka-u.ac.jp}

\author{Ryo Ogawa}
\affiliation{Department of Earth and Space Science, Graduate School of Science, The University of Osaka, Toyonaka, Osaka 560-0043, Japan}
\email{} 

\author{Ryunosuke Oishi}
\affiliation{Department of Earth and Space Science, Graduate School of Science, The University of Osaka, Toyonaka, Osaka 560-0043, Japan}
\email{} 

\author{Hideaki Ose}
\affiliation{Department of Earth and Space Science, Graduate School of Science, The University of Osaka, Toyonaka, Osaka 560-0043, Japan}
\email{} 

\author[0000-0001-8472-2219]{Greg Olmschenk}
\affiliation{Code 667, NASA Goddard Space Flight Center, Greenbelt, MD 20771, USA}
\email{greg@olmschenk.com}

\author[0000-0003-2388-4534]{Cl\'ement Ranc}
\affiliation{Sorbonne Universit\'e, CNRS, UMR 7095, Institut d'Astrophysique de Paris, 98 bis bd Arago, 75014 Paris, France}
\email{ranc@iap.fr}

\author[0000-0001-5069-319X]{Nicholas J. Rattenbury}
\affiliation{Department of Physics, University of Auckland, Private Bag 92019, Auckland, New Zealand}
\email{n.rattenbury@auckland.ac.nz }

\author[0000-0002-1228-4122]{Yuki K. Satoh}
\affiliation{College of Science and Engineering, Kanto Gakuin University, Yokohama, Kanagawa 236-8501, Japan}
\email{sato.yuki@nit.ac.jp}

\author[0000-0002-4035-5012]{Takahiro Sumi}
\affiliation{Department of Earth and Space Science, Graduate School of Science, The University of Osaka, Toyonaka, Osaka 560-0043, Japan}
\email{sumi@ess.sci.osaka-u.ac.jp}

\author[0000-0002-5843-9433]{Daisuke Suzuki}
\affiliation{Department of Earth and Space Science, Graduate School of Science, The University of Osaka, Toyonaka, Osaka 560-0043, Japan}
\email{dsuzuki@ess.sci.osaka-u.ac.jp }

\author{Takuto Tamaoki}
\affiliation{Department of Earth and Space Science, Graduate School of Science, The University of Osaka, Toyonaka, Osaka 560-0043, Japan}
\email{} 


\author{Chihiro Ueda}
\affiliation{Department of Earth and Space Science, Graduate School of Science, The University of Osaka, Toyonaka, Osaka 560-0043, Japan}
\email{} 

\author{Paul J. Tristram}
\affiliation{University of Canterbury Mt.¥ John Observatory, P.O. Box 56, Lake Tekapo 8770, New Zealand}
\email{tristram.p@gmail.com} 

\author[0000-0002-9881-4760]{Aikaterini Vandorou}
\affiliation{Code 667, NASA Goddard Space Flight Center, Greenbelt, MD 20771, USA}
\affiliation{Department of Astronomy, University of Maryland, College Park, MD 20742, USA}
\email{katievan@umd.edu}

\author[0000-0001-7692-0581]{Hibiki Yama}
\affiliation{Department of Earth and Space Science, Graduate School of Science, The University of Osaka, Toyonaka, Osaka 560-0043, Japan}
\email{yama@kusastro.kyoto-u.ac.jp}

\collaboration{50}{(The MOA Collaboration)}

\begin{abstract}
The Milky Way is expected to host $\sim$10$^8$ stellar-mass
black holes with an uncertain binary fraction.
The only proven method to detect isolated stellar mass black holes is
gravitational microlensing. 
Here we report the results of a
microlensing search for black holes with photometry and astrometry.
By combining 10 years of seeing-limited photometry from OGLE and MOA with diffraction-limited
photometry and astrometry from adaptive optics
imagers at the W.~M.~Keck Observatory, we constrain lens masses for OGLE-2012-BLG-0169, OGLE-2014-BLG-0613/MOA-2015-BLG-041, OGLE-2015-BLG-0029/MOA-2015-BLG-170, and OGLE-2015-BLG-0211.
Of the four long-duration microlensing events monitored, we ruled out black hole lenses
in 3 events, which likely have stellar or white dwarf lenses. OGLE-2015-BLG-0211 remains a black hole candidate with a poorly constrained lens mass with a 1$\sigma$ upper mass limit of 3.2$M_\odot$ and a 3$\sigma$ upper mass limit of 21.6$M_\odot$.
This event suffered from poor observing conditions and significant astrometric reference frame uncertainties, but its analysis may benefit from additional astrometric data in the upcoming Gaia Data Release 4.
Of the six long-timescale ($t_E>100$ days) microlensing events from this work and previous studies, one black hole has been confirmed with a second not ruled out.
We briefly examine Galactic model simulations and find that our result agrees with current expectations.
Ultimately, we need a larger sample of isolated black holes to constrain their formation processes.
This will be possible in the
coming years with Rubin and Roman, as well as improved
astrometry from JWST and
large, ground-based telescopes equipped with 
adaptive optics.
\end{abstract}

\keywords{}

\section{Introduction} \label{sec:intro}

Stellar mass black holes (BHs) are produced when a massive star dies.
However, the exact relation between a star's initial mass and the remnant type and mass after death is not well understood \citep{Heger:2003}.
The initial-final mass relation (IFMR) mapping the star's initial mass to the remnant's type and mass cannot be directly measured.
Instead, we must rely on statistical constraints through measurements of the stellar initial mass function (IMF) and the black hole mass function (BHMF).

Between $10^8 - 10^9$ stellar mass BHs are predicted to exist within the Milky Way \citep{Agol:2002, Elbert:2018}. 
About 60 candidate BHs in accreting binary systems have been discovered in the Milky Way, although only about a third of these candidates have been dynamically confirmed \citep{CorralSantana:2016, Casares:2014}. 
So far, 3 BHs in non-interacting binary systems have been discovered in the Milky Way field via Gaia astrometry and confirmed via radial velocities \citep{El-Badry2023a,El-Badry2023b,Tanikawa2023,Chakrabarti2023,GaiaCollaboration2024}. A few additional detached BHs in binary systems have been identified in Milky Way globular clusters \citep{Giesers2018,Giesers2019,Whitaker2026}.
Beyond the Milky Way, the gravitational wave (GW) instruments LIGO and Virgo detect BH-BH binary mergers.
There were 153 confirmed measurements of BH-BH merger GW events in the first four observing runs through O4 of Advanced LIGO and Advanced Virgo \citep{TheLIGOScientificCollaboration2025d}.
However, the host galaxy cannot be identified for the vast majority of GW BHs as the localization area is too large \citep{Abbott:2018}; thus the IMFs in these galaxies are unknown. 
These detection methods produce a selection bias where nearly all BHs detected to date are in binary systems. This makes it difficult to infer the IFMR as the binary fractions, formation, and destruction channels are very uncertain.

Isolated BHs are difficult to detect with no luminous companion or energetic eruptions and yet are critical for understanding the IFMR. While the binary fraction of black holes in the Milky Way is not yet well constrained, it is expected that the majority ($>80$\%; see e.g. \citealp{Wiktorowicz2019}) of Galactic BHs have no companions.
The only demonstrated technique to find these isolated stellar-mass BHs is gravitational microlensing. 

When a foreground massive object (such as a black hole) passes in front of a background ``source'' star, the source's light rays are deflected into multiple images.
Lensing events produced by stellar-mass lenses and sources in our galaxy are typically unresolved by current telescopes (with image separations on the order of milliarcseconds or smaller), and are thus known as ``microlensing'' events \citep{Paczynski:1986a}.
However, microlensing produces a transient event that is detectable with current instruments.
Photometrically, the source will be magnified and its apparent brightness will increase for a time; astrometrically, its position will appear to shift \citep{Hog:1995, Miyamoto:1995, Walker:1995}.

In a microlensing event, the images produced are focused around a ring surrounding the lens, which we quantify with the angular Einstein ring radius $\theta_{\rm E} = \sqrt{\kappa M_l \pi_{\rm rel}}$. Here,
$\kappa = 4G/c^2 {\rm AU} \simeq 8.144\  {\rm mas}\ M_\odot^{-1}$ is a constant, $M_l$ is the lens mass, $\pi_{\rm rel} = \pi_{\rm l} - \pi_{\rm s}$ is the relative lens-source parallax, the lens and source parallaxes are $\pi_{\rm s} = {\rm au}/D_{\rm s}$ and  $\pi_{\rm l} = {\rm au}/D_{\rm s}$, and $D_{\rm s}$ and $D_{\rm l}$ are the distances to the lens and source, respectively. The event occurs over a timescale defined by the Einstein ring radius crossing time $t_E = \theta_E/\mu_{rel}$, where $\mu_{rel}$ is the lens-source relative proper motion. Another key parameter is the microlensing parallax, which may be measured via a change in the light curve due to Earth's annual motion for long events or with simultaneous observation from Earth and space. It is quantified as $\pi_E = \pi_{\rm rel}/\theta_E$. See \citet{Lu2026} for more details on the microlensing parameters and mathematics used in this work.

A photometric signal alone is insufficient to measure the mass of the lens and prove its black hole nature, as the only routine observable is the timescale, which is a degenerate function of mass, distance, and proper motion. 
The astrometric signal yields a direct measurement of the Einstein radius
which, when combined with photometry, can be used to measure the lens mass.
The astrometric microlensing shift is expected to be $\sim$1 milli-arcsecond (mas) for a 10$M_\odot$ BH located at 4 kpc passing in front of a star at 8 kpc. This signal is just within the capabilities of current space-based instruments and ground-based adaptive optics systems. 
Ultimately, astrometric and photometric lensing has the promise of delivering a sample of BH detections that can be used to constrain their IFMR, kick velocity distribution, and multiplicity.

Each year, over a thousand microlensing events are detected photometrically with wide-field, optical imaging surveys pointed at the Galactic bulge such as the Optical Gravitational Lensing Experiment (OGLE; \citealp{Udalski:2015b}).
Population synthesis simulations suggest that events with Einstein crossing times longer than 100 days consist of $\sim$20-30\% BHs \citep{Abrams2025}.
The astrometric microlensing signal lasts much longer than the photometric magnification, as the former scales with the source-lens separation $u$ (in Einstein radii) as $u^{-1}$, while the latter scales as $u^{-4}$ \citep{Dominik:2000}. 
BH candidates can therefore be selected from OGLE and followed up with high-precision astrometry, as has been proven feasible by \citet{Lu:2016}.

Only one BH has been confirmed with this method thus far. OGLE-2011-BLG-0462 (OB110462) was first alerted in 2011 by the OGLE Early Warning System and identified as a high-probability BH candidate based on its long timescale ($\sim$150 days). Astrometric follow-up with the Hubble Space Telescope confirmed the lens to be a black hole \citep{Lam2022,Sahu2022,Mroz2022}, with a mass of $6.0^{+1.2}_{-1.0} M_\odot$ measured by \citet{Lam2023} and $7.15\pm0.83 M_\odot$ measured by \citet{Sahu2025}, which are consistent to 1$\sigma$.

We present the results of our search for black holes using astrometric
microlensing measurements for a sample of 4 OGLE-selected events.
The rest of this paper is organized in the following manner.
In \S\ref{sec:obs}, we present the data used in our analysis, including 
laser-guided adaptive optics
observations from the W. M. Keck Observatory obtained over 4+ years.
The astrometric analysis of each event is presented in \S\ref{sec:ast_analysis}.
We describe our microlens model fitting methods in
\S\ref{sec:modeling}.
We report our limits on astrometric microlensing signals and one
weak black hole candidate in \S\ref{sec:results}.
Finally, we discuss implications for future black hole microlensing experiments in \S\ref{sec:discussion} and population constraints in \S\ref{sec:disc_res_sample}.
We conclude our findings in \S\ref{sec:conclusions}.


\begin{deluxetable*}{llllllrr}[htb!]
\tablecaption{Black Hole Candidates}
\label{tab:targets}
\tablehead{
  \colhead{Target} &
  \colhead{Nickname} &
  \colhead{R.A. (J2000)} &
  \colhead{Dec. (J2000)} & 
  \colhead{l (deg.)} & 
  \colhead{b (deg.)} & 
  \colhead{I$_{base}$} & 
  \colhead{Kp$_{base}$}
}
\startdata
OGLE-2012-BLG-0169 & OB120169 & 17:49:51.38 & -35:22:28.0 & -5.057 & -4.090 & 19.38 & 17.92 \\
OGLE-2014-BLG-0613 & OB140613 & 17:53:57.68 & -28:34:21.6 & 1.251 & -1.382 & 18.22 & 14.28 \\
OGLE-2015-BLG-0029 & OB150029 & 17:59:46.60 & -28:38:41.8 & 1.828 & -2.523 & 15.13 & 12.36 \\
OGLE-2015-BLG-0211 & OB150211 & 17:29:26.18 & -30:58:54.3 & -3.606 & 1.862 & 17.29 & 11.27 \\
\enddata
\tablecomments{Position on the sky and baseline I magnitude as reported by OGLE, along with the baseline Kp magnitude measured by NIRC2.}
\end{deluxetable*}

\section{Observations and Image Reduction} \label{sec:obs}

We selected four long-duration photometric microlensing events for
follow-up astrometric monitoring: 
OGLE-2012-BLG-0169 (OB120169),
OGLE-2014-BLG-0613 (OB140613), 
OGLE-2015-BLG-0029 (OB150029), and
OGLE-2015-BLG-0211 (OB150211).
Their basic properties are listed in Table \ref{tab:targets}. 
OB120169 was examined by \citet{Lu:2016}, but it is re-presented here with additional photometry and astrometry data obtained after initial publication,
tightening the mass constraint. 

We
identified the longest duration events with reasonable magnification
($\gtrsim$0.5 mag)   
and baseline magnitudes $I\lesssim19$ in the OGLE EWS.
We ultimately selected new targets OB140613, OB150029, and OB150211, as they had 
long Einstein crossing times ($t_{\rm E} > 120 {\rm \, days}$), 
which suggests the probability of finding a BH lens is $\sim$40\% 
for each target \citep{Lam:2020}. 
Since the astrometric monitoring
needed to start near the events' photometric peak, the selections could not be done using
the full light curves but were instead performed based on modeling
the rising part of the light curves prior to their peaks.
Although low microlensing parallax ($\pi_E\lesssim0.1$) suggests a high-mass lens and thus may be used in the selection black hole candidates \citep{Lam:2020}, it is often not well-constrained until the full light curve has been observed.

\subsection{OGLE Photometry \label{sec:OGLE photometry}}

Photometry was obtained from the 
Optical Gravitational Lensing Experiment survey
\citep[OGLE,][]{Udalski:2015b}. 
OGLE is a continuous, long term
survey carried out with the 1.3-m Warsaw telescope at the Las Campanas
Observatory in Chile. The survey is currently in its fourth phase
(OGLE-IV), with the telescope equipped with a 32-CCD mosaic camera,
and focuses on monitoring stars toward the Galactic bulge for
microlensing,
as detailed by \citet{Udalski:2015b}. Currently, the OGLE survey
discovers, in real time, around 2000 microlensing events per year with its
Early Warning System \citep[EWS,][]{Udalski:2003}\footnote{\url{http://ogle.astrouw.edu.pl/ogle4/ews/ews.html}}.
The $I$-band light curves used in this study come from an independent
off-line reduction, optimized for these events and using an improved
lens position, which used the OGLE photometric
pipeline and Difference Image Analysis (DIA) package
\citep{Wozniak:2000}.

\subsection{MOA Photometry}
OB140613 and OB150029 were also alerted by the Microlensing Observations in Astrophysics \citep[MOA,][]{Bond:2001b,Sumi2003} survey as MOA-2015-BLG-041 and MOA-2015-BLG-170, respectively. The second phase of this long-term survey toward the Galactic bulge is conducted on the 1.8-m MOA-II telescope at the Mount John University Observatory in New Zealand. For these two events, we acquired the R- and V-band data from an independent reduction. 
In the Section \ref{sec:modeling}, we perform the fit with instrumental MOA R-band magnitudes. The V-band data were excluded from the fits due to poor quality (given the strong reddening toward the bulge) and sparse cadence. In this work, we disregard the baseline magnitude in MOA-R, and use the OGLE-I magnitude for optical lens brightness limit analysis.

\subsection{Keck Astrometry and Photometry}
\label{sec:astrometryobs}

\begin{figure*}[htb!]
  \centering
   \includegraphics[width=0.49\textwidth]{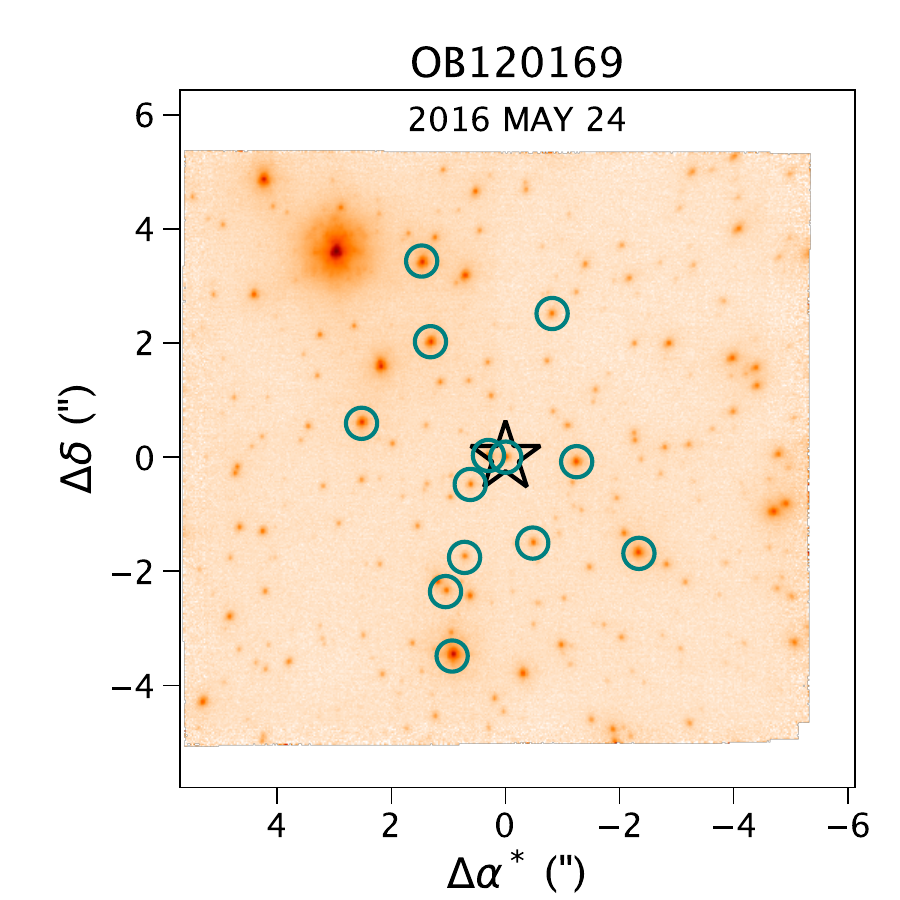} 
   \includegraphics[width=0.49\textwidth]{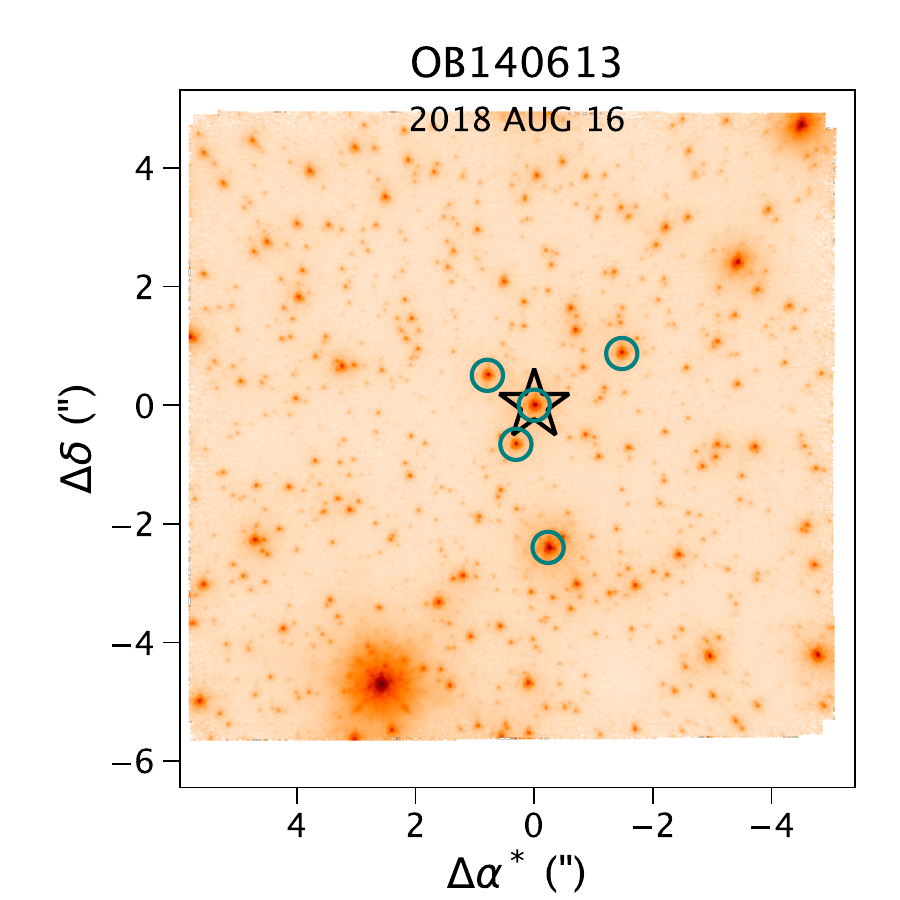} 
   \includegraphics[width=0.49\textwidth]{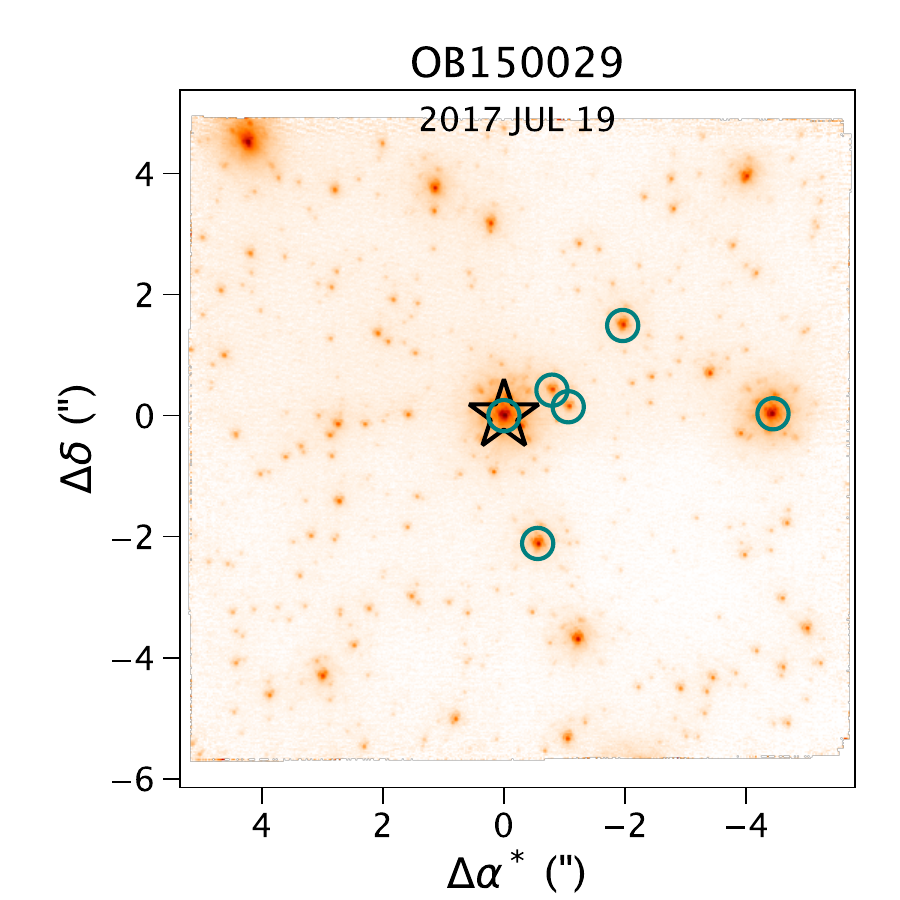} 
   \includegraphics[width=0.49\textwidth]{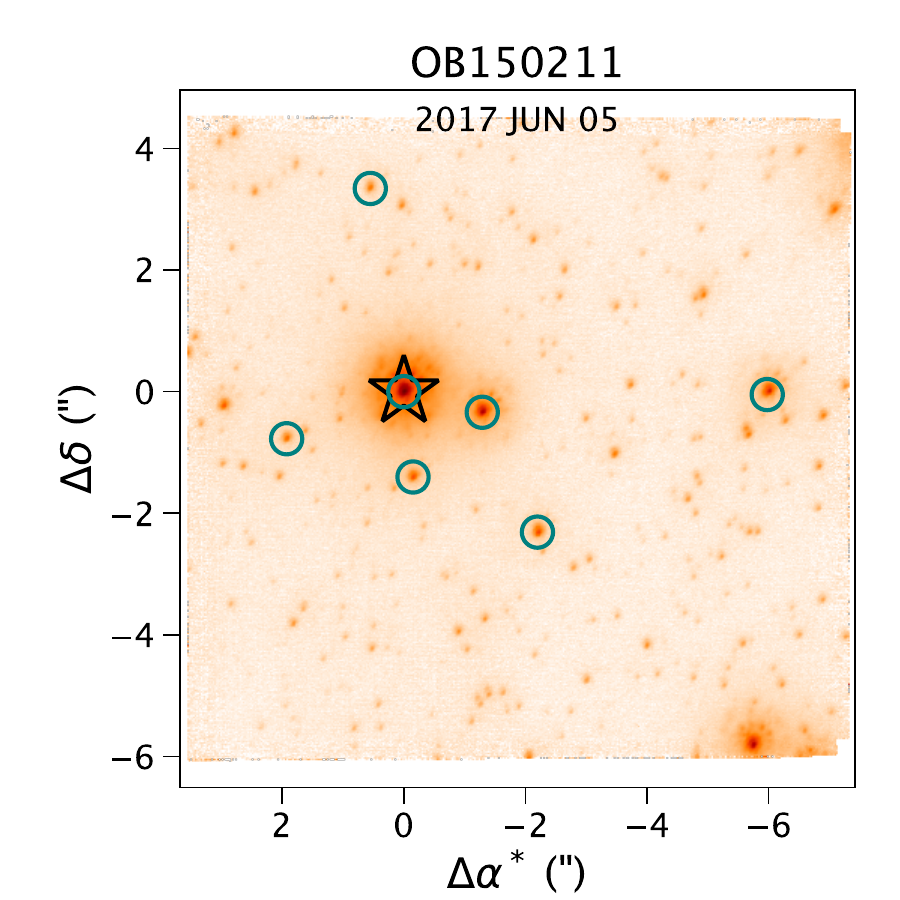}
\caption{NIRC2 Kp images ($10'' \times 10''$) of each of the four targets. 
The microlensing target is marked with a {\em black star} and the PSF stars are marked with {\em cyan circles}. 
\label{fig:targets}
}
\end{figure*}

Astrometric measurements for the targets were obtained with the
W.~M.~Keck Observatory, Keck I and II 10 m telescopes using the laser guide
star adaptive optics (AO) system \citep{vanDam:2006, Wizinowich:2006},
which provides near-diffraction limited images at near-infrared
wavelengths. 
Each target was observed at or shortly after its photometric peak, with continued monitoring over a 4+ year period
in order to both detect the maximum of the
astrometric signal and
the baseline proper motions well after the event. 

Table \ref{tab:ast_obs} lists the individual observation epochs.
Images (Figure \ref{fig:targets}) were mainly obtained using the NIRC2 (PI: K. Matthews) instrument using
the narrow camera with a plate scale of 9.971 $\pm$ 0.004 mas pix$^{-1}$, a $10''$ field of view \citep{Service:2016} and the Kp filter 
($1.95-2.30 \mu$m). 
Data taken after 2019 April 17 were obtained with the newer OSIRIS
imager instrument \citep{Larkin:2006,Arriaga:2018} with a plate scale of
9.942 $\pm$ 0.003 mas pix$^{-1}$, a $20''$ field of view \citep{Freeman2023}, and the Kp filter. 
Each epoch of observation consisted of 10$-$80 exposures, each with
10$-$60 s integration times, based on the image quality,
atmospheric conditions, and saturation avoidance. 
The images were dithered over a 0\farcs7 box, and
at least three images were obtained at each dither position in order
to increase the AO performance stability. 
There were significant variations in seeing (0\farcs4$-$1\farcs5) and the resulting image resolution (50-115 mas)
over the observations and sometimes within an epoch of observations, as shown in Table \ref{tab:ast_obs}.
Therefore, images with a FWHM$>$1.25$\times$FWHM$_{min}$ were
thrown out, where FWHM$_{min}$ was the lowest FWHM delivered in that epoch, and are not included in Table \ref{tab:ast_obs}.

The astrometric images were analyzed in a manner similar to
\citet{Lu:2016} which we summarize here, with differences noted. 
Raw images were processed using the Keck Adaptive optics Imaging (KAI) reduction pipeline
\citep{lu_2021_6677744}, which calibrates for flat-field
variations, dark current variations, sky subtraction, cosmic-ray
removal, instrumental distortion, and achromatic differential
atmospheric refraction (DAR). 
All cleaned exposures were co-added and weighted by Strehl, 
using the IRAF routine \verb/Drizzle/ \citep{Fruchter:2002} 
to produce both a final combined image and three subset images (with a randomly selected $1/3$ of the data) used for error analysis.
The updated KAI version used in this work includes a non-linearity correction for NIRC2 imaging to correct for saturation and updated distortion solutions for the OSIRIS imager from \citet{Freeman2023}. 

\begin{deluxetable*}{ccccc@{$\;\pm\;$}cc@{$\;\pm\;$}ccc}
\tabletypesize{\footnotesize}
\tablecaption{Astrometric Observations \label{tab:ast_obs}}
\tablehead{
  \colhead{Object} & 
  \colhead{Date} & 
  \colhead{$t_{\mathrm{int}}$} & 
  \colhead{$N_{\mathrm{exp}}$} & 
  \twocolhead{FWHM} &
  \twocolhead{Strehl} &
  \colhead{$\sigma_{\mathrm{ast}}$} & 
  \colhead{$\sigma_{\mathrm{phot}}$}
  \\ 
  \colhead{ } & 
  \colhead{ } & 
  \colhead{(s)} & 
  \colhead{ } & 
  \twocolhead{(mas)} &
  \colhead{ } & 
  \colhead{ } & 
  \colhead{(mas)} & 
  \colhead{(mag)}
}
\startdata
OB120169 & 2012-06-23 & 60 & 9 & 62 & 3 & 0.35 & 0.12 & 0.19 & 0.11 \\
 & 2012-07-10 & 60 & 22 & 65 & 3 & 0.29 & 0.03 & 0.13 & 0.03 \\
 & 2013-04-30 & 60 & 32 & 60 & 3 & 0.39 & 0.06 & 0.13 & 0.11 \\
 & 2013-07-15 & 60 & 12 & 74 & 6 & 0.36 & 0.07 & 0.17 & 0.08 \\
 & 2015-05-05 & 60 & 17 & 69 & 4 & 0.34 & 0.08 & 0.12 & 0.09 \\
 & 2015-06-07 & 60 & 30 & 55 & 3 & 0.52 & 0.11 & 0.15 & 0.13 \\
 & 2016-05-24 & 60 & 27 & 64 & 4 & 0.36 & 0.09 & 0.13 & 0.08 \\
 & 2016-07-14 & 10 & 54 & 54 & 3 & 0.62 & 0.11 & 0.16 & 0.13 \\
OB140613 & 2015-06-07 & 40 & 64 & 54 & 3 & 0.41 & 0.04 & 0.07 & 0.11 \\
 & 2015-06-29 & 40 & 18 & 54 & 3 & 0.36 & 0.05 & 0.11 & 0.07 \\
 & 2016-04-17 & 40 & 22 & 87 & 4 & 0.16 & 0.02 & 0.18 & 0.06 \\
 & 2016-05-24 & 40 & 60 & 63 & 2 & 0.29 & 0.03 & 0.08 & 0.18 \\
 & 2016-08-02 & 30 & 61 & 51 & 2 & 0.42 & 0.06 & 0.06 & 0.13 \\
 & 2017-06-05 & 20 & 67 & 58 & 3 & 0.37 & 0.03 & 0.10 & 0.13 \\
 & 2017-07-14 & 30 & 38 & 56 & 3 & 0.36 & 0.05 & 0.07 & 0.12 \\
 & 2018-05-11 & 40 & 47 & 61 & 3 & 0.31 & 0.06 & 0.09 & 0.22 \\
 & 2018-08-16 & 30 & 37 & 51 & 2 & 0.42 & 0.05 & 0.08 & 0.50 \\
 & 2019-04-17 & 30 & 21 & 61 & 3 & 0.33 & 0.04 & 0.11 & 0.07 \\
OB150029 & 2015-06-07 & 30 & 50 & 50 & 2 & 0.44 & 0.04 & 0.07 & 0.11 \\
 & 2015-07-23 & 30 & 39 & 50 & 2 & 0.42 & 0.04 & 0.06 & 0.09 \\
 & 2016-05-24 & 30 & 29 & 57 & 5 & 0.31 & 0.08 & 0.08 & 0.11 \\
 & 2016-07-14 & 30 & 43 & 51 & 3 & 0.36 & 0.05 & 0.07 & 0.14 \\
 & 2017-05-21 & 30 & 20 & 65 & 4 & 0.28 & 0.06 & 0.11 & 0.15 \\
 & 2017-07-14 & 30 & 21 & 63 & 3 & 0.23 & 0.03 & 0.11 & 0.25 \\
 & 2017-07-19 & 10 & 78 & 53 & 3 & 0.36 & 0.05 & 0.08 & 0.19 \\
 & 2018-08-21 & 30 & 42 & 62 & 3 & 0.26 & 0.03 & 0.10 & 0.27 \\
 & 2019-04-17 & 30 & 42 & 60 & 4 & 0.30 & 0.05 & 0.04 & 0.15 \\
OB150211 & 2015-05-05 & 30 & 13 & 80 & 5 & 0.19 & 0.03 & 0.56 & 0.18 \\
 & 2015-06-07 & 40 & 35 & 67 & 3 & 0.30 & 0.03 & 0.27 & 0.12 \\
 & 2015-06-29 & 40 & 30 & 67 & 4 & 0.27 & 0.03 & 0.34 & 0.08 \\
 & 2015-07-24 & 40 & 27 & 63 & 4 & 0.30 & 0.04 & 0.29 & 0.13 \\
 & 2016-05-02 & 40 & 28 & 101 & 6 & 0.13 & 0.02 & 0.99 & 0.10 \\
 & 2016-07-14 & 30 & 40 & 67 & 4 & 0.26 & 0.03 & 0.25 & 0.08 \\
 & 2016-08-02 & 30 & 35 & 82 & 5 & 0.18 & 0.03 & 0.39 & 0.19 \\
 & 2017-06-05 & 30 & 26 & 73 & 4 & 0.23 & 0.02 & 0.30 & 0.06 \\
 & 2017-06-08 & 10 & 18 & 69 & 4 & 0.25 & 0.03 & 0.44 & 0.18 \\
 & 2017-07-19 & 10 & 20 & 71 & 4 & 0.16 & 0.02 & 0.54 & 0.10 \\
 & 2018-05-11 & 30 & 10 & 106 & 8 & 0.11 & 0.02 & 0.97 & 0.17 \\
 & 2018-08-02 & 30 & 20 & 111 & 7 & 0.10 & 0.02 & 0.97 & 0.14 \\
 & 2018-08-16 & 30 & 13 & 75 & 5 & 0.20 & 0.03 & 0.66 & 0.27 \\
 & 2019-06-28 & 30 & 18 & 83 & 6 & 0.13 & 0.02 & 0.44 & 0.05 \\
 & 2020-05-25 & 18 & 13 & 84 & 9 & 0.15 & 0.03 & 0.59 & 0.21
\enddata

\tablecomments{$t_{int}$: integration time per exposure.
$N_{exp}$: number of exposures used in the combined image.
$\sigma_{ast}$/$\sigma_{phot}$: median astrometric/photometric errors for reference stars in the final combined image.
All images were obtained with the NIRC2 imager except those taken after 2019 Apr 20,
which were obtained with the OSIRIS imager.
}
\end{deluxetable*}

\subsection{Other Data Sources}

We searched for additional data in the 2MASS, Gaia, KMTNet, PanSTARRS, Spitzer, UKIDSS GPS, and VISTA surveys. 
No additional data from these sources were found for OB120169.
Photometric measurements for OB140613, OB150029, and OB150211 were found in several of these sources; however, they had poor temporal coverage and/or large photometric uncertainties. These three targets are in the Gaia DR3 catalog \citep{GaiaCollaboration2023}, with sparse photometry and no published epoch astrometry; Gaia DR4 (anticipated 2026 Dec) will provide epoch astrometry, and reanalysis should be performed when it is available.
Most notably, OB140613, OB150029, and OB150211 had photometric measurements obtained from \textit{Spitzer} as part of a microlensing campaign \citep{Yee:2015a, CalchiNovati:2015}.
However, there are likely systematic errors in the Spitzer photometry, along with underestimated error bars (\citealp{Koshimoto:2019}; E. Bachelet, private communication). Additionally, the Spitzer imaging of these occurred over periods of a few weeks, providing limited temporal coverage and no baseline measurements for these long-timescale events.
For these reasons, no additional sources of data beyond the OGLE, MOA (where available), and Keck measurements were included when fitting the events.

\section{Astrometric Data Analysis}
\label{sec:ast_analysis}

Astrometry and photometry were extracted from the combined images using the Anisoplanatic and Instrumental Reconstruction of Off-axis PSFs for AO (AIROPA) package for point-spread function (PSF) fitting \citep{Witzel:2016,Ciurlo2022,Turri2022,Terry2023}.
AIROPA is based on Starfinder \citep{Diolaiti:2000} and uses a small sample of ``PSF stars'' selected by the user to derive a mean PSF over the entire image. 
While AIROPA has the capability of fitting for a variable PSF over the field of view, our analysis assumed a constant PSF because we lack information on the atmospheric turbulence profile for some epochs.
Stars are identified in an image as sources that are fit by the PSF
with a minimum correlation of 0.8 between the star image and PSF fit and that are 5$\sigma$ above the background noise of the image.
After stars are identified, PSF extraction is repeated, using the same PSF stars; but now with all other known stars subtracted to produce a cleaner PSF. 
This process is repeated three times and the final starlist is also cleaned of fake sources whose peak-pixel flux exceeds the flux contribution of any brighter neighboring star.  

We found that the choice of PSF stars is critically
important and must be optimized based on maximizing the number of
bright, isolated stars and minimizing the distance between the star
and our microlensing target. Figure \ref{fig:targets} shows a
NIRC2 image around each target and the PSF stars used for source
extraction. Typically, the astrometric precision near the center 
and the residuals after PSF fitting were improved by selecting a
smaller number of stars at small radii rather than a larger number of 
stars further out. Using the 4--6 PSF stars, AIROPA was run on the 
combined image and a starlist was created containing raw pixel
positions and fluxes for all the stars. Starlists were photometrically
calibrated using VVV \citep[VISTA Variables in the Via Lactea, Data Release 4.2;][]{Minniti2023} Ks-band data. 

AIROPA was also run on the subset images in order to empirically
determine astrometric and photometric uncertainties. For the
astrometric error, we adopted the error on the mean of the positions
in the three subset images. For the photometric error, we adopted the
RMS error of the fluxes in the three subset images. 
We find that these choices produce the most accurate errors for
linearly moving, non-variable stars over multiple-year timescales. 
The resulting astrometric precisions from each epoch
of data are shown in Table \ref{tab:ast_obs}.
Astrometric uncertainties are dominated by systematic errors at the bright end, likely due to imperfections in our PSF estimation, and
limited signal-to-noise and crowding at the faint end. 

\subsection{Cross-Epoch Matching and Transformations}\label{sec:flystar}

The individual starlists from each epoch must be cross-matched and
transformed onto a common coordinate system to track their motion.
This task can be challenging given that there are no stationary stars
in the field of view. 
Ideally, we would use high-precision Gaia astrometry to establish a
precise and accurate astrometric reference frame.
However, the NIRC2 field of view is small ($\sim 10''$) and bulge
fields are often extincted such that the density of Gaia sources is
too low and the resulting precision of the reference frame is
compromised. Thus, Gaia DR3 is used only as a starting point for the
astrometric transformations. 

We use the FlyStar\footnote{\url{https://github.com/MovingUniverseLab/flystar}} astrometric alignment package to perform this analysis. FlyStar iteratively fits 2-d transformations between the reference frame and each starlist, fits a stellar motion model including proper motion and parallax for each star and re-fits the transformations, repeating this process several times. The process is similar to that of \citet{Lu:2016} and \citet{Lam2022b}, but we upgraded FlyStar to include parallax in the motion models instead of strictly linear proper motion after parallax-like residuals appeared in several stars a preliminary alignment result.

For each target, we first query the Gaia DR3 database \citep{GaiaCollaboration2023} for all sources
inside the NIRC2 field of view. 
We find 9, 6, 6, and 3 matches with Gaia sources
for OB120169, OB140613, OB150029, and OB150211, respectively, including the
target when it is present in Gaia; these Gaia sources are typically faint with G$>$17 mag. 
This is our initial ``reference star'' set, which are the subset of catalog stars used in
the reference frame transformation.
Each epoch of NIRC2 data is cross-matched and transformed,
iteratively, into the Gaia reference frame using the Gaia DR3 source
positions and proper motions and allowing for an affine (full 1st
order) bi-variate transformation of the NIRC2 positions.
After this first pass, the NIRC2 starlists are then cross-matched to
each other
and positions, proper
motions, and parallax are extracted to establish a
new reference frame with 3-10$\times$ more reference stars available. 

Second and third passes of iterative cross-matching and transformations are performed, allowing for 2nd-order transformations, tightening the match criteria, performing outlier rejection, and beginning to exclude the reference stars with poor measurements.
A fourth pass also using 2nd-order transformations is completed after reducing the set of stars
used in the transformation to just those that are bright, close to the target, and with small positional and proper motion errors. We limit reference stars to those with 0.5 mas positional error and 0.5 mas/yr proper motion errors for all targets.
The exact choice of magnitude and distance criteria varies across the targets depending on the densities and luminosities of the stars in the field of view.
For OB120169, we select stars with Kp$<$20 mag and a distance of $<$4 arcsec from the target; 18 mag and 7 arcsec for OB140613;  20.0 mag and 2.5 arcsec for OB150029; and 18.5 mag and 5 arcsec for OB150211.
Ultimately, the filtered list of reference stars used in the transformations
are 47, 36, 47, and 35 for OB120169, OB140613,
OB150029, and OB150211, respectively. 

At the end of the fourth pass, we estimate the positional transformation uncertainties via a full-sample bootstrap with replacement. The uncertainties for each star are added in quadrature to their measurement errors. We then re-fit their motion models with the adjusted errors.
To evaluate the transformation fit quality, we examine the residuals for bright stars near the target (the same magnitude and distance limits used for reference star selection) and compute the median reduced chi-squared on their photometry and astrometry. 
If the result is not $\tilde{\chi}^2\lesssim$1.0, this suggests that the uncertainties produced from the extraction are underestimated. 
We return to the beginning of pass 4 and inflate the input measurement errors with additive and/or multiplicative error terms on all stars in the field as needed. Additive error effectively applies an error floor, where the brightest (most well-measured) stars are limited to a certain level of precision and thus down-weighted relative to the un-modified error reference frame calculation. Multiplicative error scales errors up proportionally to their original errors, which maintains the relative weights of each star in the reference calculation. The optimal selection depends on the magnitude distributions of the reference stars and where the target star lies within that, so we test the application of both to each target. 

In the astrometry, we applied an additive error of 0.02 mas for OB120169 and none for the other targets. For photometry, we applied an additive error of 0.01 mag for OB120169 and none for OB140612 and OB150211. For OB150029 photometry, we applied an additive error of 0.03 mag and multiplicative error of 1.3.
When the  median reduced chi-squared reaches approximately 1.0, this indicates that the astrometric transformations and uncertainties are accurate.

After the FlyStar alignment process, the reference frame has strayed from the Gaia frame established at the start of iteration 1. 
Limiting our sample to stars within the radial distance cut used for reference star selection, Gaia provides 2, 1, 2, and 1 reference star matches with parallax measurements and 3, 1, 2, and 2 with proper motions for OB120169, OB140613, OB150029, and OB150211, respectively. 
For the targets with at least one Gaia-matched star besides the target itself, we compute the variance-weighted mean offsets between the Gaia and Flystar parallaxes and proper motions. For OB140613, we have no stars with parallax and proper motion except the target itself. Fortunately, this target is in a significantly denser field than the other targets, and thus we can compare the median parallaxes and proper motions of 72 stars in the radial-cut region to that of a Galactic model catalog generated via SynthPop (\citetalias{synthpopI}, \citeyear{synthpopI}; \citealp{Huston2026}) and trimmed to the reference star K$_{\rm p}$ magnitude limit.

We then apply these reference frame offsets ($\Delta v_x$, $\Delta v_y$, $\Delta \pi$) to the pass 4 epoch astrometry ($x_{0}$, $y_{0}$) to produce the final star table:
\begin{equation}
\begin{split}
    x = x_0 + \Delta v_x (t-t_0) + \Delta\pi P_x(t,\alpha,\delta); \\
    y = y_0 + \Delta v_y (t-t_0) + \Delta\pi P_y(t,\alpha,\delta),
\end{split}
\end{equation}
where $t$ is the time of observation, $t_0$ is the reference time for the motion model fits, and $P_{[x,y]}(t,\alpha,\delta)$ is the parallax vector component in the $x$ or $y$ direction at the time of observation and the target's right ascension and declination (see \citealp{Lu2026}, \S3.2 for a full description of the parallax vector and its calculation).
The uncertainties on the proper motion shift are small ($\leq$0.4 mas/yr), as each field has multiple well-measured proper motions. The parallax reference frame has significant uncertainties of up to 0.4 mas, as most Gaia parallaxes in the field are poorly measured. 
We note that the absolute proper motion does not impact a mass measurement, while parallax does. For example, a positive parallax reference frame offset in the data can cause an astrometric microlensing model fitter to assign the source a parallax that is too high, thus placing the source too close and biasing derived parameters including mass. On the other hand, only the relative lens-source proper motion matters in the mass calculation, not their reference frame. However, any interpretation of final individual lens or source proper motions is affected by the reference frame, so caution must be used when drawing conclusions about kinematic implications including natal kicks.
We record the uncertainty in the Gaia parallax shift where available via propagation of the parallax measurement uncertainties from Gaia and Keck. For OB140613, we estimate the Galactic model-based parallax frame uncertainty via full-sample bootstrap with replacement. These parallax reference frame uncertainties are used in the astrometry fitting process as described in \S\ref{sec:phot_ast_fits}.

\section{Microlens Modeling}
\label{sec:modeling} 

To constrain the mass of the lens, we fit the photometry and astrometry as a function of time using a microlensing model with a point source and a point lens via the Bayesian Analysis of Gravitational Lensing Events \citep[BAGLE;][]{Lu2026} package.

We note that all astrometry is analyzed assuming that
the observed centroid measurements are of the combined lens and source and are
uncontaminated by neighbor light.
This would be the case if the neighbor contributions in the small Keck
AO aperture are negligible.
Ideally, we would include dilution of the astrometric signal due to
luminous contaminants.
However, this requires a more computationally complex
model that fits for the number of neighbor stars and the positions and
fluxes of each neighbor.
For simplicity, we neglect neighbors in the
astrometry.
This is conservative as the consequence of neglecting astrometric
signal dilution is that we may underestimate the lens mass.
As we show in \S\ref{sec:results}, only one of our targets (OB150029) shows
evidence of significant blending in the Keck AO aperture.

\subsection{BAGLE Microlensing Modeler/Fitter}
\label{sec:fit_method}

We fit the data to the microlensing model using Bayesian inference and the nested sampling algorithm \texttt{MultiNest} \citep{Feroz:2009,Feroz:2013} via \texttt{pyMultiNest} \citep{Buchner2014}, which is able to track multi-modal solutions that are common in microlensing events and provide posterior samples. 
For each event we perform two fits, one with seeing-limited photometry only, and another with seeing-limited photometry plus Keck photometry and astrometry. 
Seeing-limited microlensing photometric lightcurves often show evidence for systematic and correlated noise likely from both astrophysical and experimental sources. Thus, we also fit for additive and multiplicative error modifiers for the OGLE and MOA data sets (not Keck). We initially explored Gaussian process models \citep[see e.g.][]{Li:2019,Golovich:2022,Lam2022b} to capture this red noise instead but found that priors must be very finely tuned, and for these four targets, the models failed either by destroying part of the microlensing signal or producing no significant fit quality improvement.

During the Bayesian inference process, each pass of \texttt{MultiNest} was run five times, to ensure that the reported uncertainties on the posteriors were accurate.
At times, the ellipsoidal sampling approximation breaks down, and the uncertainties are significantly underestimated; this is a known issue with \texttt{MultiNest} (see Appendix A.9 of \citealp{Nelson:2020}).
Thus, the logZ (the evidence, which accounts for both the log-likelihood and the prior) of the five runs were compared to verify the values and uncertainties were consistent. 
We ran all fits presented here with 1000 live points, an evidence tolerance of 0.1, and a sampling efficiency of 0.8.

\subsection{PSPL: OGLE(+MOA) Photometry Only}
\label{sec:phot_fits}

We first fit all seeing-limited photometry with a point-source, point-lens (PSPL) model with parallax.
The basic PSPL parallax photometry-only model includes 7 parameters:
time of projected lens-source closest approach ($t_0$), 
lens-source projected distance at closest approach in units of Einstein radii ($u_0$), 
Einstein crossing time ($t_E$), 
microlensing parallax ($\vec{\pi}_E = [\pi_{E,E}$, $\pi_{E,N}]$), 
source flux fraction ($b_{SFF} = f_{\rm source}/(f_{\rm source}+f_{\rm lens} + f_{\rm neighbors})$), 
and baseline magnitude ($m_{base}$). 
More details about the microlensing model can be found in \citet{Lu2026}.
For targets also observed by MOA, there are additional $b_{SFF}$ and $m_{base}$ parameters.
An additional parameter per data set for additive error $\varepsilon_a$ and/or for multiplicative error $\varepsilon_m$ may be included, where the adjusted error $\sigma$ is calculated from the input errors $\sigma_0$ as $\sigma = \varepsilon_m\sqrt{\sigma_0^2 + \varepsilon_a}$.
Note, we use parameters defined in the heliocentric reference frame
(unlike the pseudo-geocentric frame commonly used in exoplanet microlensing; see e.g. \citealp{Gould:2004})
in order to make easier comparisons between photometric and
astrometric fit parameters (see \citealp{Lu2026}).
However, all observations and the predicted model flux and position
are compared in the geocentric frame.

To determine which model to use, the data are first fit with the base model, the additive error model, the multiplicative error model, and the additive+multiplicative error model. 
The model that best fits the data is selected by comparing the Bayesian Information Criterion (BIC) of the different models.
For OGLE data only, the additive+multiplicative error model showed significant improvement over the no-error modification model ($\Delta$BIC$>$10) and moderate improvement ($\Delta$BIC$>$2) over the additive error only model for OB120169. For OB150211, the additive error only model provided a significant improvement over the base model ($\Delta$BIC$>$10), while multiplicative error did not result in further improvement.
The additive+multiplicative error model also produced a significantly minimized BIC ($\Delta$BIC$>$10) for OB140613 and OB150029 with both OGLE and MOA photometry and separate error adjustment terms per instrument. 
The results of the photometry-only fits for each target using their respective models are presented in \S\ref{sec:results}.
The priors used for each target's final fits are presented in Appendix \ref{app:priors}.

There are two things to note from these fits.
First, $M_L$ cannot be derived from the results of the photometry-only fits alone.
Second, there are strong correlations and degeneracies, with that of $t_E$, $\pi_E$, and $b_{SFF}$ being particularly notable \citep{DiStefano:1995,Sumi2011}. 
Most previous microlens model fitting of photometry-only light curves have employed Galactic priors based on a model of the stellar positions, proper motions, and masses along the line of sight; however, these are still extremely uncertain \citep[e.g.][]{Shan2019,Lam:2020}. Even in the case of finite source effects enabling a $\theta_E$ measurement, an assumption must typically be made for the source distance \citep[see e.g.][]{Gould1994}. 
We instead utilize high-resolution photometry and astrometry from Keck to break these degeneracies and correlations. 
An astrometric fit can directly measure $\theta_E$, $\pi_S$, and $\pi_E$ to break the mass-distance degeneracy. 
High-resolution imaging also reduces blended flux from unrelated neighbors, which can help to mitigate the $t_E$-$\pi_E$-$b_{SFF}$ degeneracy.
Our method, however, still relies on a Galactic model to establish reasonable priors for some microlensing parameters including parallaxes and angular Einstein ring radii.


\subsection{PSPL: Photometry $+$ Astrometry Fits}
\label{sec:phot_ast_fits}

Next, we jointly fit the OGLE(+MOA) photometry with Keck photometry and astrometry.
In addition to the $7(+2)$ base parameters involved in the photometry-only and the optional error modifier terms, the astrometry fit adds $7$ additional parameters: the Einstein radius, $\theta_E$, the source position at time $t_0$ in an arbitrary reference frame, $\vec{x}_{S,0} = (x_{S,0,E}, x_{S,0,N})$, the source proper motion $\vec{\mu}_S = (\mu_{S,E}, \mu_{S,N})$, the source parallax $\pi_S$,
and the parallax reference frame offset $\pi_{\rm ref}$ (see \S\ref{sec:flystar}). 
All other quantities (such as lens mass $M_L$) can be derived from these fit values. 

The full OGLE photometry + Keck photometry and astrometry fits with additive and multiplicative errors have 17
free parameters. If MOA photometry is present, there are 21 parameters.
For OB150211, we exclude multiplicative error, resulting in 16 parameters for the joint fit.
Unlike \citet{Lu:2016}, we do not use the
previous photometry-only fit posteriors as priors in this fit and instead 
jointly fit all data sets simultaneously.
The results of the joint photometry and astrometry fits are presented in \S\ref{sec:results}.
The priors used in our fits are provided in Appendix \ref{app:priors}.

\section{Results}
\label{sec:results}

\label{res:pspl}

The results for the point-source, point-lens (PSPL) photometric and
astrometric fits are presented for each target below. We first present OB150211, where a BH lens is not ruled out. OB140613 shows a possible astrometric microlensing signal but a lens mass too low to be a BH. OB120169 and OB150029 show little or no significant astrometric deviation, ruling out a black hole lens. OB140613 and OB150029 are probable stellar lenses, while OB120169 may likely be either a star or a white dwarf.

\subsection{OB150211}
\label{sec:res_ob150211}

\paragraph{Multi-Modal Solutions}
For OB150211, the OGLE-only and OGLE+Keck joint fits show a two-mode solution with
$u_0 > 0$ and $u_0 < 0$ (Tables \ref{tab:OB150211_pspl_phot_fit} \& \ref{tab:OB150211_pspl_fit}).
The parameters in common across the OGLE-only vs. OGLE+Keck fits show
discrepancies, indicating some tension between the photometry and
astrometry.

\begin{deluxetable*}{lrrrcrrrc}[htb!]
\tablecaption{OB150211 PSPL Fits with only OGLE Photometry 
  \label{tab:OB150211_pspl_phot_fit}}
\tabletypesize{\scriptsize}
\tablehead{
  \colhead{} & 
  \multicolumn{4}{c}{$u_0 < 0$} &
  \multicolumn{4}{c}{$u_0 > 0$} \\ 
  \colhead{Parameter} & 
  \colhead{$\mathcal{L}_{max}$} & 
  \colhead{MAP} &
  \colhead{Median} &
  \colhead{68\% CI} &
  \colhead{$\mathcal{L}_{max}$} & 
  \colhead{MAP} &
  \colhead{Median} &
  \colhead{68\% CI}
}
\startdata
log$\mathcal{L}$ & 1981.59 & 1978.15 & 1961.50 & & 1981.18 & 1977.41 & 1980.25 &  \\ 
$\tilde{\chi}^2$ & 0.98 & 0.95 & 1.03 & & 1.00 & 0.96 & 0.98 &  \\ 
log$\mathcal{Z}$ & & & 1946.2 & & & & 1945.7 &  \\ 
$N_{dof}$ & & & 719 & & & & 719 &  \\ 
\hline 
$t_0$ (MJD) & 57208.35 & 57206.52 & 57208.65 & [-1.85, 1.26]  & 57212.50 & 57214.70 & 57212.64 & [-1.74, 2.57]  \\ 
$u_0$ & -0.62 & -0.60 & -0.56 & [-0.05, 0.07]  & 0.42 & 0.41 & 0.44 & [-0.08, 0.09]  \\ 
$t_E$ (days) & 111.0 & 111.0 & 116.2 & [-5.0, 7.7]  & 129.7 & 127.7 & 126.7 & [-10.5, 12.0]  \\ 
$\pi_{E,E}$ & 0.054 & 0.025 & 0.042 & [-0.021, 0.017]  & 0.008 & 0.012 & 0.012 & [-0.030, 0.028]  \\ 
$\pi_{E,N}$ & -0.032 & -0.044 & -0.003 & [-0.027, 0.037]  & -0.031 & -0.055 & -0.027 & [-0.036, 0.026]  \\ 
$b_{SFF,I}$ & 1.158 & 1.090 & 0.999 & [-0.179, 0.133]  & 0.694 & 0.707 & 0.741 & [-0.173, 0.219]  \\ 
$I_{base}$ (mag) & 17.294 & 17.293 & 17.294 & [-0.001, 0.001]  & 17.294 & 17.294 & 17.294 & [-0.001, 0.001]  \\ 
$\varepsilon_{a,I}$ (mmag) & 9.5 & 10.0 & 9.5 & [-0.6, 0.6]  & 9.2 & 9.8 & 9.5 & [-0.7, 0.7]  \\ 
\tableline
$I_{src}$ (mag) & 17.135 & 17.199 & 17.294 & [-0.135, 0.215]  & 17.691 & 17.670 & 17.620 & [-0.281, 0.288] 
\enddata
\tablecomments{The points of
  maximum likelihood ($\mathcal{L}_{max}$), Maximum A Posteriori (MAP), and the median value are given. The 68\% confidence interval values
  are presented as [$x_{16\%} - x_{Median}$, $x_{84\%} - x_{Median}$]
  for some parameter x. 
  Some parameters have $\mathcal{L}_{max}$ and/or MAP values that fall outside the
  68\% CI region. Note that the median values are marginalized over each individual parameter rather than corresponding to a single n-dimensional model sampled by the fitter.
}
\end{deluxetable*}

\begin{figure*}[h]
\centering
\includegraphics[width=\textwidth]{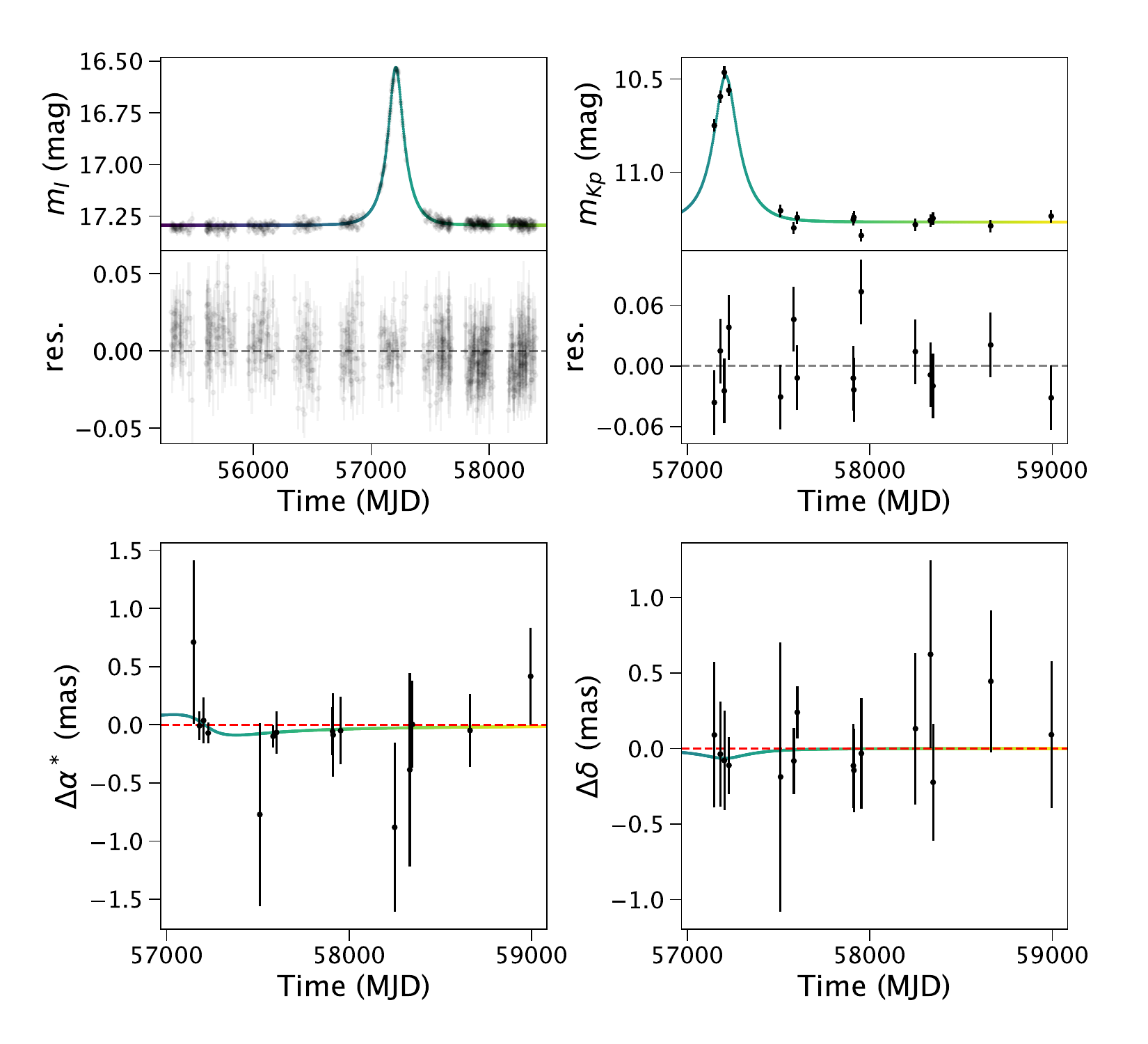}
\caption{The photometric + astrometric best-fit model for OB150211.
The color of the model lines correspond to time, as indicated by the
lower axes of the magnitude and position plots.
\emph{Top left:} the OGLE I-band magnitude of the target (black ticks)
overlaid with the model, with residuals on the bottom.
\emph{Top right:} the Keck Kp magnitude of the target (black ticks)
overlaid with the model, with residuals on the bottom.
\emph{Bottom row:} the difference in RA \emph{(left)} and DEC
\emph{(right)} with the measured motion
(black ticks) and lensed
model (colored line) where the unlensed motion (proper motion and parallax) has been removed (red line).}
\label{fig:ob150211_results}
\end{figure*}

The negative $u_0$ solution is preferred in the OGLE + Keck
fits based on the median solution $\log\mathcal{L}$ and the Bayes factor (BF) of 1.7
relative to the positive $u_0$ solution, which is weak evidence. 
While the negative $u_0$
solution is preferred, the positive $u_0$ solution,
with a similar mass, is still likely.
We adopt the preferred mode for figures showing a point solution
and the maximum likelihood values are used as the best-fit (Figure
\ref{fig:ob150211_results}); however, we sample from the full global
posterior, including all modes, for the discussion and all other figures. 

\begin{figure*}
\centering
\includegraphics[width=0.9\textwidth]{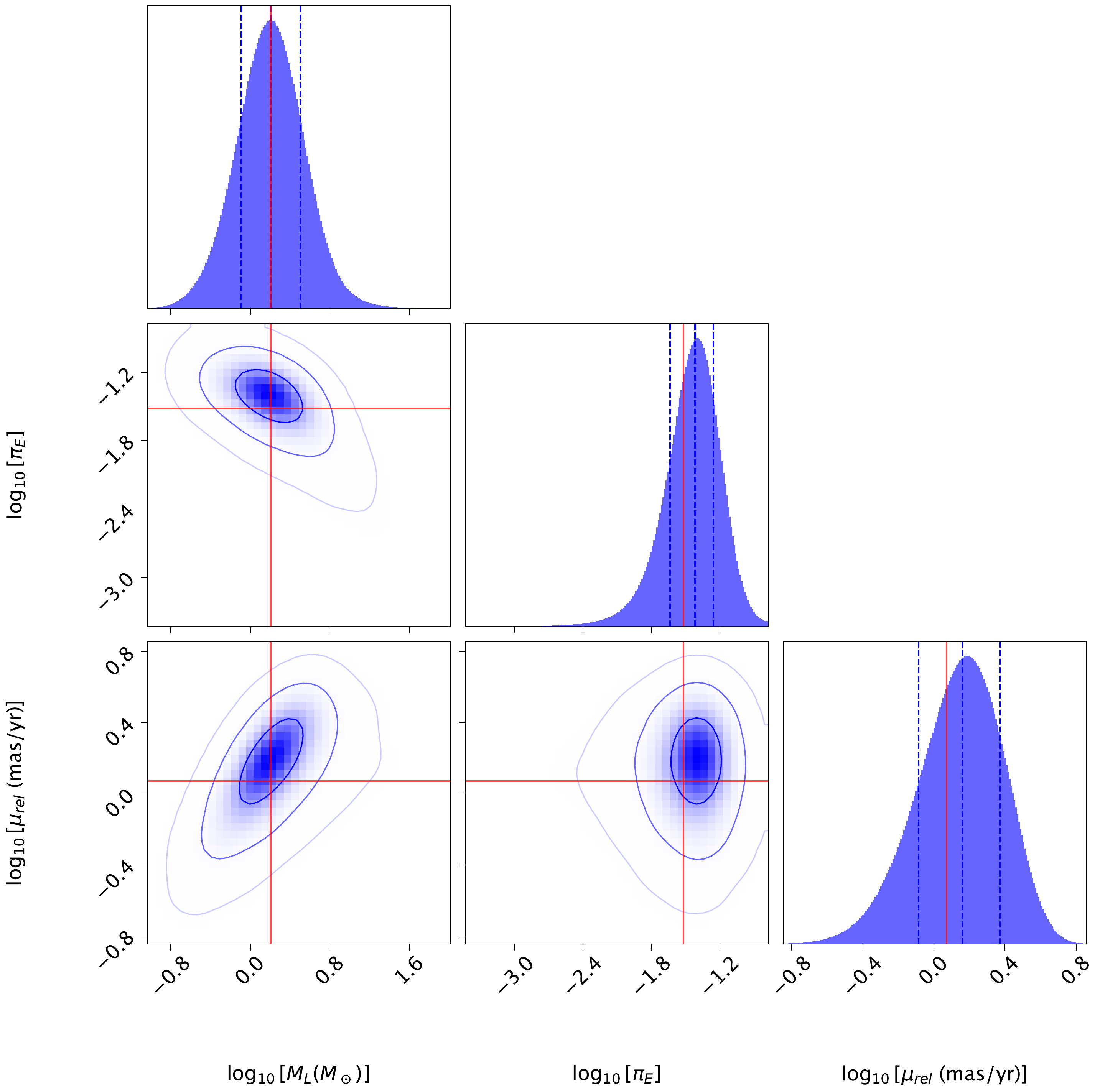}
\caption{
Global posterior probability distributions for OB150211. 
The median of the
1D marginalized posterior probability is shown in red.
}
\label{fig:ob150211_corner}
\end{figure*}

\begin{figure}
  \centering
  \includegraphics[width=0.48\textwidth]{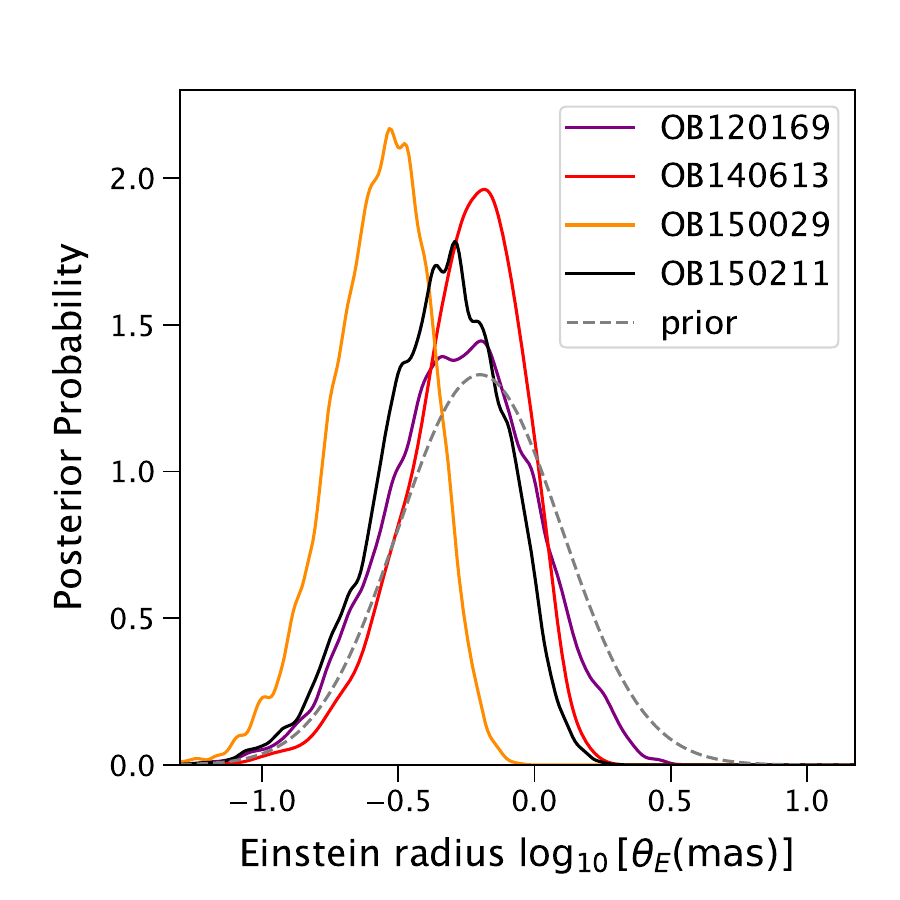}
  \includegraphics[width=0.48\textwidth]{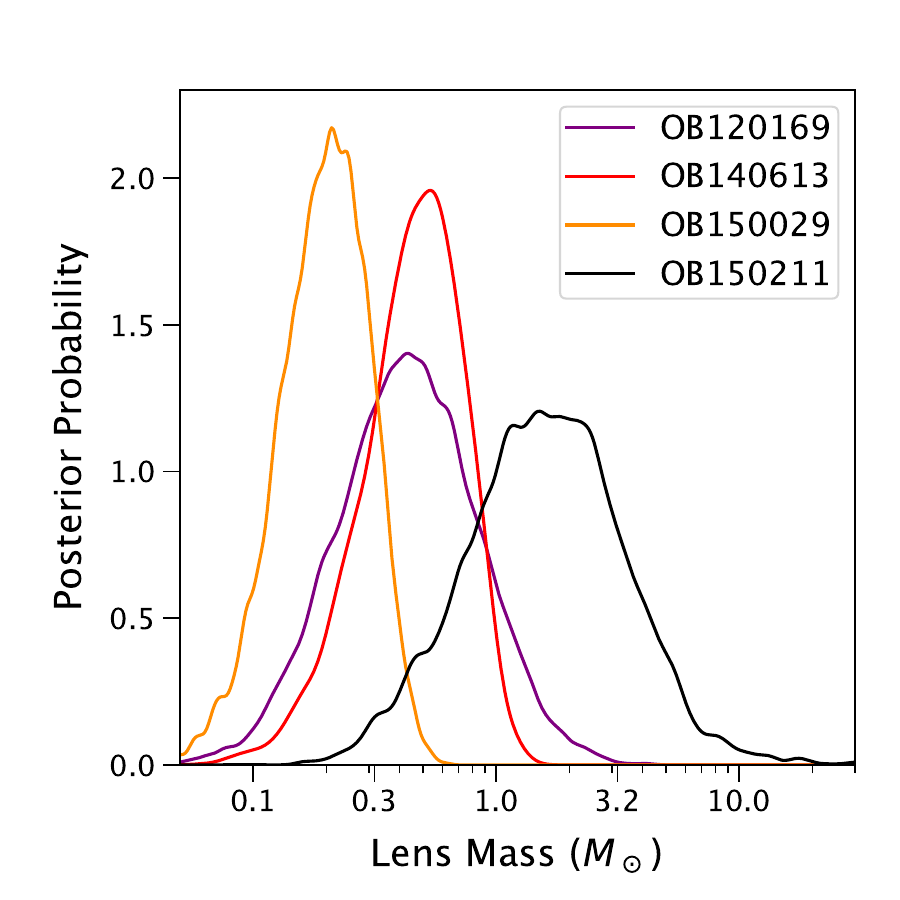}
  \caption{
    \label{fig:masses}
    Einstein radius (left) and lens mass (right) posterior probability distributions for all four
    targets from the joint photometry-astrometry fits. The $\theta_E$ panel also shows the prior used for all joint fits.
    }
\end{figure}

The quality of the best fit is good when we include the additive error term on seeing-limited photometry, yielding
$\tilde{\chi}^2 = 0.98$.
The maximum likelihood, MAP, and median point estimates differ 
from each other significantly, indicating that the results are likely
still in the prior-dominated regime. Figure \ref{fig:masses} shows the $\theta_E$ prior and posterior and the lens mass posterior for all four events.
Figure \ref{fig:ob150211_corner} shows the global photometry + astrometry fit posteriors for $\log M_L$ and two parameters with strong correlations, $\pi_{\rm E}$ and $\mu_{\rm rel}$.

\begin{deluxetable*}{lrrrcrrrcrrrc}
\tablecaption{OB150211 PSPL Fits with OGLE + Keck Photometry and Astrometry  \label{tab:OB150211_pspl_fit}}
\tabletypesize{\scriptsize}
\tablehead{
  \colhead{} & 
  \multicolumn{4}{c}{$u_0 < 0$} &  
  \multicolumn{4}{c}{$u_0 > 0$} \\ 
  \colhead{Parameter} & 
  \colhead{$\mathcal{L}_{max}$} & 
  \colhead{MAP} &
  \colhead{Median} &
  \colhead{68\% CI} &
  \colhead{$\mathcal{L}_{max}$} & 
  \colhead{MAP} &
  \colhead{Median} &
  \colhead{68\% CI} &
}
\startdata
log$\mathcal{L}$ & 2219.45 & 2213.89 & 2205.13 & & 2219.09 & 2213.15 & 2215.14 &  \\ 
$\tilde{\chi}^2$ & 1.00 & 1.01 & 1.00 & & 0.98 & 1.02 & 0.97 &  \\ 
log$\mathcal{Z}$ & & & 2159.3 & & & & 2158.7 &  \\ 
$N_{dof}$ & & & 755 & & & & 755 &  \\ 
$\pi-$volume & & & 0.50 & & & & 0.50 &  \\ 
\hline 
$t_0$ (MJD) & 57209.83 & 57210.97 & 57208.92 & [-1.82, 1.20]  & 57211.41 & 57211.58 & 57211.80 & [-1.12, 1.82]  \\ 
$u_0$ & -0.54 & -0.48 & -0.51 & [-0.03, 0.04]  & 0.51 & 0.51 & 0.48 & [-0.05, 0.04]  \\ 
$t_E$ (days) & 117.9 & 126.3 & 121.4 & [-3.8, 5.4]  & 120.9 & 120.8 & 123.2 & [-5.1, 7.1]  \\ 
$\pi_{E,E}$ & 0.041 & 0.038 & 0.032 & [-0.015, 0.015]  & 0.031 & 0.036 & 0.026 & [-0.016, 0.016]  \\ 
$\pi_{E,N}$ & -0.012 & -0.000 & -0.000 & [-0.020, 0.031]  & 0.001 & 0.009 & -0.007 & [-0.032, 0.016]  \\ 
$b_{SFF,I}$ & 0.963 & 0.814 & 0.881 & [-0.100, 0.068]  & 0.891 & 0.887 & 0.829 & [-0.120, 0.092]  \\ 
$I_{base}$ (mag) & 17.294 & 17.293 & 17.294 & [-0.001, 0.001]  & 17.294 & 17.293 & 17.294 & [-0.001, 0.001]  \\ 
$\varepsilon_{a,I}$ (mmag) & 9.0 & 9.1 & 9.5 & [-0.6, 0.7]  & 9.3 & 9.0 & 9.5 & [-0.6, 0.6]  \\ 
$b_{SFF,Kp}$ & 0.98 & 0.89 & 0.91 & [-0.10, 0.06]  & 0.92 & 0.94 & 0.86 & [-0.13, 0.09]  \\ 
$Kp_{base}$ (mag) & 11.26 & 11.28 & 11.27 & [-0.01, 0.01]  & 11.27 & 11.29 & 11.27 & [-0.01, 0.01]  \\ 
$\theta_E$ (mas) & 0.48 & 1.30 & 0.50 & [-0.21, 0.29]  & 0.27 & 1.52 & 0.46 & [-0.20, 0.31]  \\ 
$\pi_S$ (mas) & 0.091 & 0.099 & 0.111 & [-0.021, 0.021]  & 0.111 & 0.067 & 0.112 & [-0.021, 0.021]  \\ 
$\mu_{S,\alpha*}$ (mas/yr) & -3.54 & -3.13 & -3.47 & [-0.08, 0.14]  & -3.46 & -3.34 & -3.42 & [-0.10, 0.17]  \\ 
$\mu_{S,\delta}$ (mas/yr) & -7.20 & -7.39 & -7.23 & [-0.12, 0.15]  & -7.27 & -7.29 & -7.27 & [-0.16, 0.10]  \\ 
$x_{S0,\alpha*}$ (mas) & 27.44 & 27.29 & 27.47 & [-0.08, 0.09]  & 27.34 & 27.45 & 27.38 & [-0.08, 0.08]  \\ 
$x_{S0,\delta}$ (mas) & -106.68 & -106.32 & -106.60 & [-0.14, 0.14]  & -106.64 & -106.52 & -106.67 & [-0.12, 0.12]  \\ 
\tableline
$M_L$ ($\msun$) & 1.4 & 4.3 & 1.6 & [-0.8, 1.5]  & 1.1 & 5.0 & 1.6 & [-0.8, 1.7]  \\ 
$\pi_L$ (mas) & 0.112 & 0.147 & 0.133 & [-0.024, 0.027]  & 0.119 & 0.124 & 0.131 & [-0.025, 0.025]  \\ 
$\pi_{rel}$ (mas) & 0.021 & 0.049 & 0.019 & [-0.010, 0.016]  & 0.009 & 0.057 & 0.016 & [-0.009, 0.014]  \\ 
$\mu_{L,\alpha*}$ (mas/yr) & -4.97 & -6.91 & -4.65 & [-0.81, 0.56]  & -4.29 & -7.82 & -4.46 & [-0.90, 0.69]  \\ 
$\mu_{L,\delta}$ (mas/yr) & -6.78 & -7.34 & -7.23 & [-0.89, 0.71]  & -7.31 & -8.37 & -6.98 & [-0.68, 0.63]  \\ 
$\mu_{rel,\alpha*}$ (mas/yr) & 1.43 & 3.77 & 1.21 & [-0.60, 0.85]  & 0.83 & 4.48 & 1.07 & [-0.76, 0.96]  \\ 
$\mu_{rel,\delta}$ (mas/yr) & -0.42 & -0.05 & -0.02 & [-0.78, 1.03]  & 0.04 & 1.08 & -0.28 & [-0.76, 0.74]  \\ 
$I_{src}$ (mag) & 17.334 & 17.517 & 17.431 & [-0.080, 0.131]  & 17.419 & 17.424 & 17.497 & [-0.114, 0.170]  \\ 
$Kp_{src}$ (mag) & 11.28 & 11.41 & 11.37 & [-0.07, 0.13]  & 11.35 & 11.36 & 11.43 & [-0.11, 0.17] 
\enddata
\tablecomments{Note, both the point of
  maximum likelihood and the MAP value are given and the 68\% confidence interval values
  are presented as [$x_{16\%} - x_{Median}$, $x_{84\%} - x_{Median}$]
  for some parameter x. 
  Some parameters, such as $t_0$, $\pi_S$, $b_{SFF,Kp}$,
  and $Kp_{src}$, have MAP values that fall outside the
  68\% CI region.}
\end{deluxetable*}

\paragraph{Possible Black Hole Solution}
Independently from this work, OB150211 was examined by \cite{Rybicki2024} in their analysis of photometry from OGLE and Spitzer along with priors from a Galactic model to estimate the probability of a sample of events being dark remnants. They found a probability of 28\% for a flat mass prior and 5\% for a \citet{Kroupa2001} initial mass function prior and concluded that the lens likely resides in the Galactic bulge. 

From our astrometric monitoring, OB150211 has a global median best-fit Einstein radius of 
$\theta_E = 0.49$ mas with a 68\% credible range of [0.28 - 0.79] mas.
The resulting mass posterior in Figure
\ref{fig:masses}
shows that the median lens mass
is $M_L = 1.60$ M$_\sun$ [0.82 - 3.17 M$_\sun$, 68\% CI]. The 99.73\% CI is [0.19, 21.6] M$_\sun$. 
Note, the MAP solution gives a higher lens mass of 4.3
M$_\sun$.
Given the large allowed range of masses from the fit, a black hole lens is allowed but not favored.
The lens mass is poorly constrained, with only a brown dwarf or free floating planet lens being ruled out.

\begin{figure*}
\centering
\includegraphics[width=0.9\textwidth]{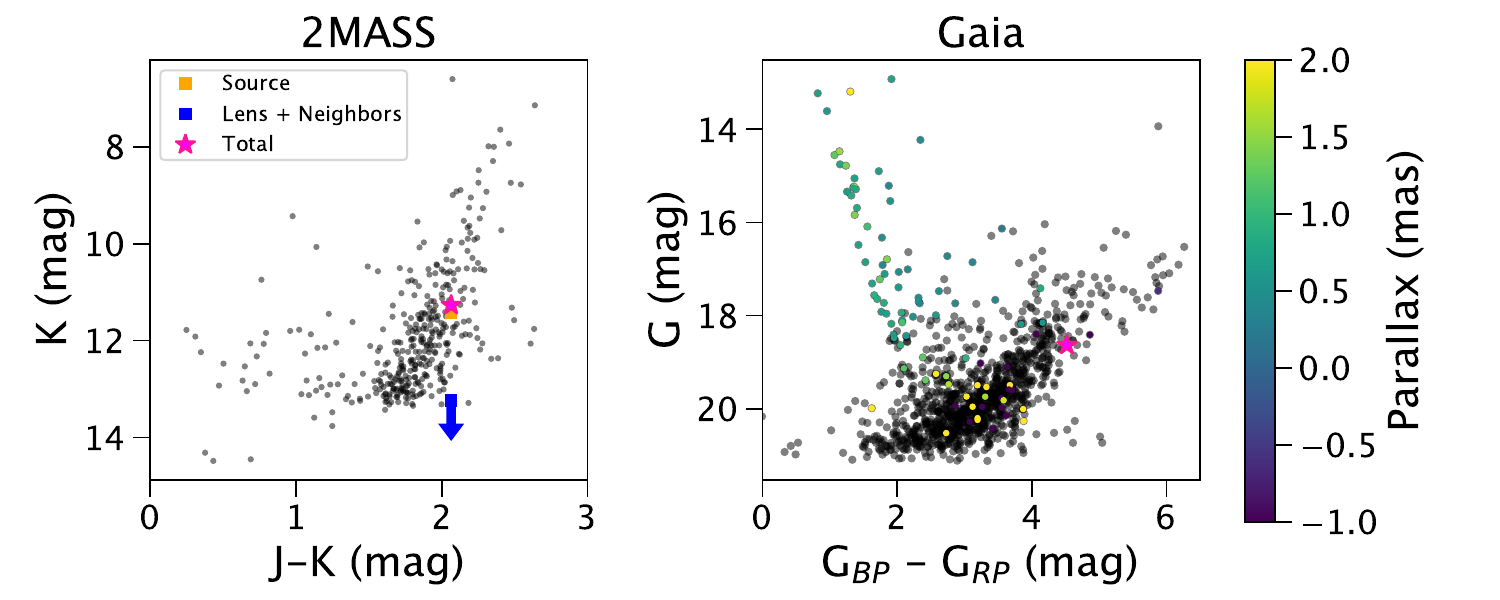}
\caption{Color-magnitude diagrams for OB150211 from 
2MASS ({\em left}) and Gaia ({\em right}). The matched source from
2MASS and Gaia are shown as a {\em magenta star}. Using the best-fit 
$b_{SFF,Kp}$, the source flux ({\em orange square}) is separated from the lens + neighbors
flux ({\em blue square}), assuming that source flux fraction is the
same in the J- and K-bands. The target's Gaia DR3 parallax is negative and consistent with zero.
}
\label{fig:ob150211_cmd}
\end{figure*}

Further support for the dark lens scenario 
comes from the high source flux fraction.
The best-fit solution indicates that most of the light in
both the OGLE and Keck aperture is coming from the source as the
source flux fraction (blending) is $b_{SFF,OGLE} = 86$\% [75 - 94\%, 68\%
CI] and $b_{SFF,Kp} = 89$\% [77 - 97\%, 68\% CI].
Thus, within the uncertainties, the lens luminosity is consistent with
0.
The source is a red-giant, likely in the far bulge, and any luminous
lens is likely on the main sequence as shown in Figure
\ref{fig:ob150211_cmd}.
Alternatively, any extra non-source
luminosity could also come from a lens or source companion.
A luminous source or lens companion could dilute
the astrometric signal, but depending on system architecture, it could also alter the signal in observable ways \citep[see e.g.][]{Bhadra2026}. However, the photometric
and astrometric fits do not show signs of binary-like residuals (Figure
\ref{fig:ob150211_results}).  

One final support for the isolated compact object or black hole nature of OB150211
comes from the combination of $t_E$ and $\pi_E$.
OB150211 has an Einstein crossing time of $t_E = 122$ [118 - 128, 68\% CI]
days, and the microlensing
parallax is $\pi_E = 0.038$ [0.023 - 0.056, 68\% CI],
which is consistent with black holes found in
population synthesis runs from PopSyCLE as shown in Figure
\ref{fig:tE_piE}.

\paragraph{Probability that OB150211 is a Black Hole} 
Following the method from \citet{Lam2022}, we evaluate the probability that the lens is a black hole (BH),
neutron star (NS), dark white dwarf (WD), luminous lens (star; LL), or 
brown dwarf (BD) by combining population simulations with 
the observed constraints on the lens mass $M_L$,
distance $d_L$, and magnitude $mag_L$ as determined from the baseline
magnitudes ($Kp_{base}$, $I_{base}$) and source flux fractions
($b_{SFF,Kp}$, $b_{SFF,I}$).
For a given $d_L$ and $M_L$ drawn from the posterior sample, we 
predict the expected ${Kp}_L$ and ${I}_L$ magnitudes using a
Galaxia model of
the Galaxy to randomly select from the age distribution at this
distance \citep{Sharma:2011,Lam:2020}, 
stellar evolution and atmosphere models
from SPISEA \citep{Hosek:2020}, and
a 3D extinction map that distributes extinction from a 2D map \citep{Surot2020} along the line of sight according to an exponential disk model.
We assume solar metallicity as the metallicity has a minimal effect
on the magnitudes within our other uncertainties.
If the predicted magnitude for a posterior sample point is brighter than the
lens flux upper limit from the same sample point, then a luminous lens is not
allowed, and it must be dark. 
If the predicted magnitude is fainter, a
luminous lens is allowed for that sample. 
This method is similar to that used by \citet{Wyrzykowski:2016} for Gaia photometry-only dark lens candidate analysis, but 
since we have astrometry, we can sample mass and distance posteriors directly.
Thus, the only Galactic model assumptions we apply are the age distribution
and extinction value at a given position.

\begin{figure}
\centering
\includegraphics[width=0.47\textwidth]{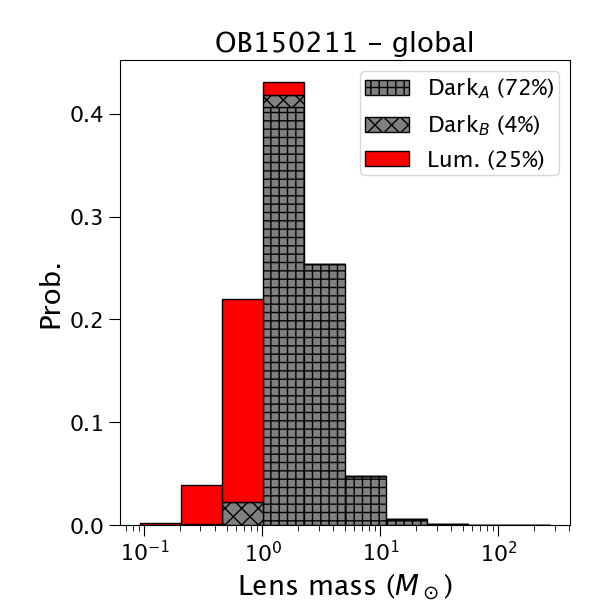}
\caption{ \label{fig:ob150211_dark_prob}
Probability that the lens is dark or luminous vs. lens mass. Those
lenses of a given mass and distance with a predicted magnitude
brighter than the magnitude allowed by the model fit are considered
dark (``Dark$_B$''), as well as objects with masses and ages that indicate they would have evolved to their final compact object state (``Dark$_A$''). The global (all modes) probability distribution is shown.
}
\end{figure}

After fully sampling the fit posteriors, we build up a probability distribution for the dark
or luminous lens hypothesis as a function of mass (Figure \ref{fig:ob150211_dark_prob}). 
Note that MISTv1.2-based \citep{Dotter2016,Choi2016} SPISEA models exclude 
initial masses below 0.1$M_\odot$, so any lower mass object is considered dark and labeled a ``brown dwarf,'' although some low mass stars exist in this regime.
The models include luminous white
dwarfs when they are young, but the vast majority of white dwarfs are
also considered dark. 
Adopting a dark lens type transition from brown dwarfs to white dwarfs at
0.1 $M_\odot$, white dwarfs to neutron stars at $1.2 M_\odot$ and
neutron stars to BHs at $2.2 M_\odot$, the relative probabilities of
LL:BD:WD:NS:BH are 26:0:9:33:32.

\paragraph{Resolving the Lens and Source} 
The nature of OB150211 may be confirmed in the future as the 
lens and source separate over time. 
The relative proper motion is
1.450 mas yr$^{-1}$ [0.821 - 2.351 mas yr$^{-1}$, 68\% CI].
The source and lens can be resolved with diffraction-limited, adaptive
optics imaging from Keck at 2 $\mu m$ in $\sim$2049, at 1 $\mu m$ with
adaptive optics improvements by 2032, and with thirty-meter class telescopes like the Extremely Large Telescope (ELT) and Giant Magellan Telescope (GMT) at first science light the early 2030s. 

It may be possible to improve the fit accuracy and rule out additional 
parameter space with
additional archival data sources such as Gaia DR4 and with improvements
to point-source extraction techniques applied to the adaptive optics
data. 
Additionally, improved Galactic model simulations (see e.g. \S\ref{sec:disc_res_sample} and \citealp{Huston2026})
may provide more realistic priors for key parameters like parallax and Einstein radius.
Finally, this source has not been detected at X-ray wavelengths in serendipitous/slew 
observations by
Chandra, XMM-Newton, or eROSITA. 
Detection of X-rays from ISM accretion via targeted observation might
provide additional support for OB150211's black hole nature.

\subsection{OB140613}
\label{sec:res_ob140613}

\begin{figure*}
\centering
\includegraphics[width=\textwidth]{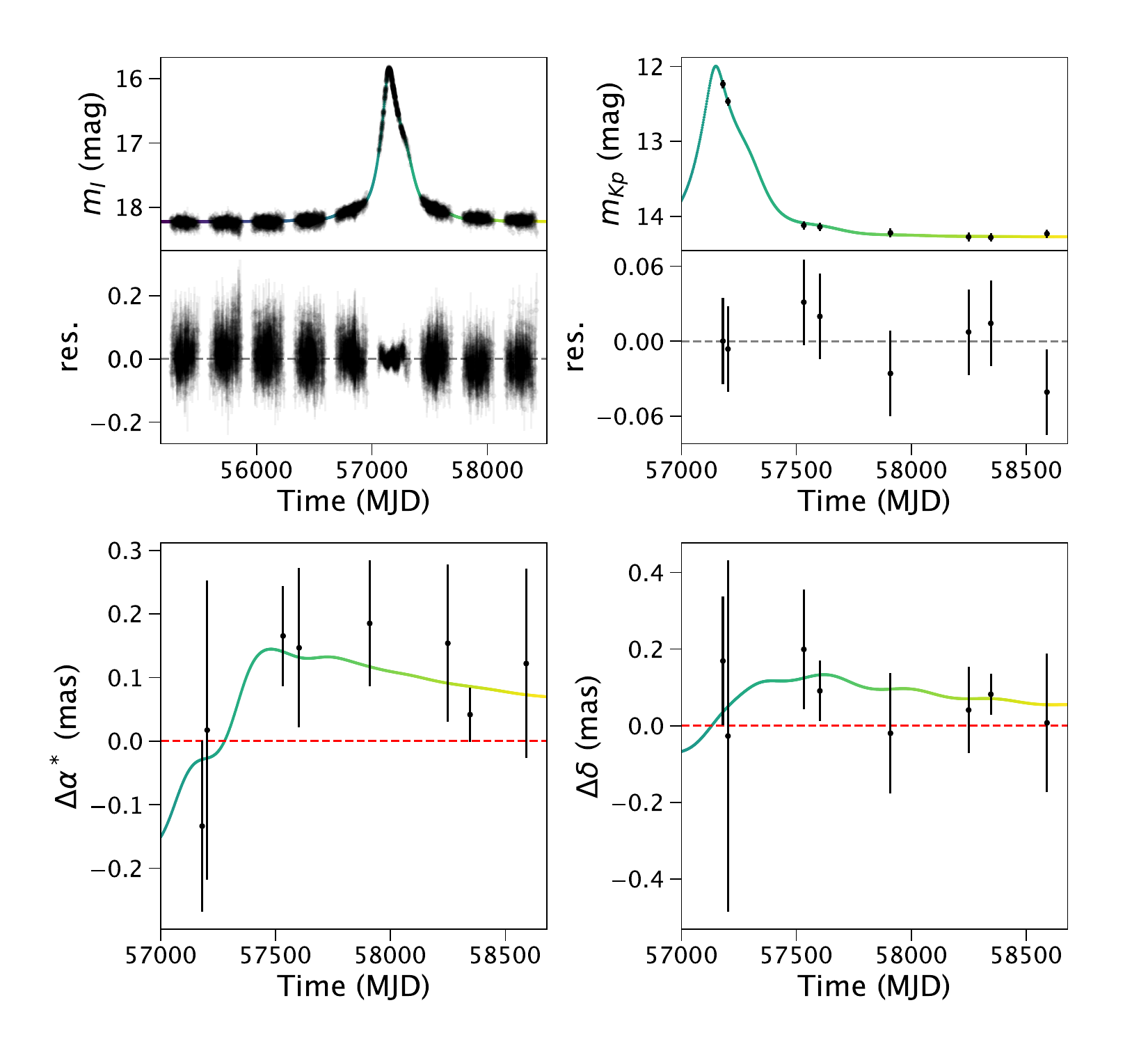}
\caption{The photometric + astrometric best-fit model for OB140613, following the conventions of Fig. \ref{fig:ob150211_results}.
\emph{Top left:} Keck image on August 16, 2018.
\emph{Top right:} the I-band magnitude of the target.
\emph{Bottom row:} the difference in RA \emph{(left)} and DEC \emph{(right)} from the top left's origin.}
\label{fig:ob140613_results}
\end{figure*}

OB140613 is our longest duration event with an Einstein crossing time
of 
$t_E = 294.7$ days [292.4 - 297.0 days, 68\% CI] (Figure \ref{fig:ob140613_results}).
Best-fit parameters for both the photometry-only fits and the photometry + astrometry
fits are shown in Table \ref{tab:OB140613_pspl_fit}.
For both fits, MultiNest reports a two-mode solution with
$u_0 \simeq 0.14$. One solution has positive $\pi_{\rm E,N}$ and a high source flux fraction, while the other has negative $\pi_{\rm E,N}$ and a lower source flux fraction.
The less blended, positive $\pi_{E,N}$ solution is very strongly preferred with a Bayes factor of $\sim10^{110}$, and thus we only present this mode in detail. 
Here, the source flux fractions are $b_{SFF,I} = 0.927$ [0.913 - 0.940, 68\% CI] and $b_{SFF,Kp} = 0.845$ [0.814 - 0.876, 68\% CI].
This preferred solution has a shorter $t_E$ at 295 days than the secondary solution at 344 days.
The parameters in common across the fits all agree well, indicating no significant tension between photometry and astrometry.

\begin{deluxetable*}{lrrrcrrrc}[htb!]
\tablecaption{OB140613 PSPL fits with seeing-limited photometry only and joint photometry-astrometry analysis 
  \label{tab:OB140613_pspl_fit}}
\tabletypesize{\scriptsize}
\tablehead{
  \colhead{} & 
  \multicolumn{4}{c}{OGLE + MOA} &
  \multicolumn{4}{c}{OGLE + MOA + Keck} \\ 
  \colhead{Parameter} & 
  \colhead{$\mathcal{L}_{max}$} & 
  \colhead{MAP} &
  \colhead{Median} &
  \colhead{68\% CI} &
  \colhead{$\mathcal{L}_{max}$} & 
  \colhead{MAP} &
  \colhead{Median} &
  \colhead{68\% CI}
}
\startdata
log$\mathcal{L}$ & 74049.35 & 74044.13 & 74050.25 & & 74194.92 & 74184.98 & 74195.09 &  \\ 
$\tilde{\chi}^2$ & 1.00 & 1.00 & 1.00 & & 1.00 & 1.00 & 1.00 &  \\ 
log$\mathcal{Z}$ & & & 73973.3 & & & & 74088.3 &  \\ 
$N_{dof}$ & & & 62367 & & & & 62382 &  \\ 
\hline 
$t_0$ (MJD) & 57192.13 & 57191.82 & 57192.19 & [-0.29, 0.30]  & 57192.19 & 57192.65 & 57192.19 & [-0.29, 0.29]  \\ 
$u_0$ & 0.14 & 0.14 & 0.14 & [-0.00, 0.00]  & 0.14 & 0.14 & 0.14 & [-0.00, 0.00]  \\ 
$t_E$ (days) & 295.4 & 297.9 & 294.7 & [-2.3, 2.4]  & 294.7 & 292.5 & 294.7 & [-2.3, 2.3]  \\ 
$\pi_{E,E}$ & -0.122 & -0.120 & -0.122 & [-0.001, 0.001]  & -0.122 & -0.123 & -0.122 & [-0.001, 0.001]  \\ 
$\pi_{E,N}$ & 0.089 & 0.087 & 0.089 & [-0.001, 0.002]  & 0.089 & 0.092 & 0.089 & [-0.002, 0.001]  \\ 
$b_{SFF,I}$ & 0.919 & 0.908 & 0.927 & [-0.014, 0.013]  & 0.927 & 0.942 & 0.927 & [-0.014, 0.013]  \\ 
$I_{base}$ (mag) & 18.220 & 18.220 & 18.220 & [-0.001, 0.001]  & 18.220 & 18.220 & 18.220 & [-0.001, 0.001]  \\ 
$\varepsilon_{a,I}$ (mmag) & 7.3 & 7.5 & 7.2 & [-0.4, 0.4]  & 7.3 & 7.9 & 7.3 & [-0.4, 0.4]  \\ 
$\varepsilon_{m,I}$ & 1.62 & 1.62 & 1.62 & [-0.02, 0.02]  & 1.62 & 1.59 & 1.62 & [-0.02, 0.02]  \\ 
$b_{SFF,R}$ & 0.48 & 0.48 & 0.49 & [-0.01, 0.01]  & 0.486 & 0.495 & 0.486 & [-0.007, 0.007]  \\ 
$R_{base}$ (mag) & -10.21 & -10.21 & -10.21 & [-0.00, 0.00]  & -10.213 & -10.212 & -10.213 & [-0.000, 0.000]  \\ 
$\varepsilon_{a,R}$ (mmag) & 19.5 & 20.5 & 19.5 & [-0.5, 0.5]  & 19.5 & 19.6 & 19.5 & [-0.5, 0.5]  \\ 
$\varepsilon_{m,R}$ & 1.23 & 1.22 & 1.23 & [-0.01, 0.01]  & 1.23 & 1.23 & 1.23 & [-0.01, 0.01]  \\ 
$b_{SFF,Kp}$ & & & &  & 0.84 & 0.89 & 0.85 & [-0.03, 0.03]  \\ 
$Kp_{base}$ (mag) & & & &  & 14.28 & 14.28 & 14.28 & [-0.02, 0.02]  \\ 
$\theta_E$ (mas) & & & &  & 0.56 & 0.48 & 0.58 & [-0.24, 0.30]  \\ 
$\pi_S$ (mas) & & & &  & 0.107 & 0.130 & 0.105 & [-0.020, 0.021]  \\ 
$\mu_{S,\alpha*}$ (mas/yr) & & & &  & -3.28 & -3.23 & -3.28 & [-0.06, 0.05]  \\ 
$\mu_{S,\delta}$ (mas/yr) & & & &  & -5.85 & -5.83 & -5.85 & [-0.05, 0.06]  \\ 
$x_{S0,\alpha*}$ (mas) & & & &  & -12.47 & -12.55 & -12.47 & [-0.08, 0.09]  \\ 
$x_{S0,\delta}$ (mas) & & & &  & 6.45 & 6.40 & 6.45 & [-0.11, 0.10]  \\ 
 & & & &  & & & &  \\ 
\tableline
$M_L$ ($\msun$) & & & &  & 0.5 & 0.4 & 0.5 & [-0.2, 0.2]  \\ 
$\pi_L$ (mas) & & & &  & 0.192 & 0.203 & 0.195 & [-0.043, 0.049]  \\ 
$\pi_{rel}$ (mas) & & & &  & 0.085 & 0.073 & 0.088 & [-0.036, 0.046]  \\ 
$\mu_{L,\alpha*}$ (mas/yr) & & & &  & -2.72 & -2.75 & -2.70 & [-0.21, 0.26]  \\ 
$\mu_{L,\delta}$ (mas/yr) & & & &  & -6.26 & -6.18 & -6.28 & [-0.18, 0.14]  \\ 
$\mu_{rel,\alpha*}$ (mas/yr) & & & &  & -0.56 & -0.48 & -0.58 & [-0.30, 0.24]  \\ 
$\mu_{rel,\delta}$ (mas/yr) & & & &  & 0.41 & 0.36 & 0.42 & [-0.17, 0.22]  \\ 
$I_{src}$ (mag) & 18.311 & 18.325 & 18.303 & [-0.016, 0.016]  & 18.303 & 18.285 & 18.302 & [-0.016, 0.017]  \\ 
$R_{src}$ (mag) & -9.42 & -9.41 & -9.43 & [-0.02, 0.02]  & -9.429 & -9.448 & -9.430 & [-0.016, 0.017]  \\ 
$Kp_{src}$ (mag) & & & &  & 14.46 & 14.41 & 14.46 & [-0.03, 0.03] 
\enddata
\tablecomments{The points of
  maximum likelihood ($\mathcal{L}_{max}$), Maximum A Posteriori (MAP), and the median value are given. The 68\% confidence interval values
  are presented as [$x_{16\%} - x_{Median}$, $x_{84\%} - x_{Median}$]
  for some parameter x. 
  Some parameters have $\mathcal{L}_{max}$ and/or MAP values that fall outside the
  68\% CI region.
}
\end{deluxetable*}

The Einstein radius is $\theta_E = 0.58$ mas 
[0.34 - 0.89 mas, 68\% CI], which is very similar to the $\theta_E = 0.56$ mas [0.33 - 0.86 mas, 68\% CI] of the secondary solution.
The relative proper motion between the source and lens is
$\mu_{rel} = 0.72$ mas yr$^{-1}$ [0.42 - 1.1 mas yr$^{-1}$] and is consistent with the
observed and simulated distribution of proper motions for disk and
bulge stars along the line of sight.
Although OB140613 is the longest event in our sample,
the microlensing parallax is $\pi_E = 0.151$ [0.150 - 0.153, 68\% CI], which is much larger than typical black
holes found in population synthesis runs from PopSyCLE as shown in
Figure \ref{fig:tE_piE}.

The resulting lens mass is 0.47 M$_\odot$ [0.28 - 0.72 M$_\odot$, 68\%], with
a
3$\sigma$ upper limit of 1.27 M$_\odot$, making it
highly improbable that OB140613 is a black hole.
The preferred solution's low blending in both OGLE-I and Kp is consistent with a low-mass stellar or white dwarf lens.
We apply the method described in \S\ref{sec:res_ob150211} to evaluate the probabilities of different lens types, which results in 75\% star probability, 24\% white dwarf, and 1\% brown dwarf.
The slow proper motion means that the lens and source likely won't be
resolved until ELT and other 30-m class telescopes are available in the early 2030s or
current 10-m class telescopes are equipped with adaptive optics
systems that deliver diffraction-limited images at optical wavelengths.

The PSPL model is a
reasonable, but not excellent fit to the data, with correlated residuals visible in the OGLE photometry (see Figure \ref{fig:ob140613_results}). The trend is not visible in the sparser Keck data or the dense but more uncertain MOA data.
This suggests that the PSPL model may not be correct and that a binary
source or binary lens may be required.
Then the blend flux ($\sim$15\%) suggested by the Keck fit may come from a luminous lens or a companion to either the source or lens. 
We attempted to fit the data with a binary source-point lens model in BAGLE, but the information supplied by the data was insufficient to put any meaningful constraints on a flux ratio or projected separation for the potential binary source system.

\subsection{OB120169}

Based on the joint photometry + astrometry fits (Table \ref{tab:OB120169_pspl_fit}), OB120169 is a long duration event with
$t_E = 176.3$ days [166.3 - 190.6 days, 68\% CI] and a moderate microlensing parallax of $\pi_E =
0.16$ [0.12 - 0.18, 68\% CI].
There is significant tension between the OGLE photometry-only fit and
the OGLE + Keck, photometry + astrometry fit.
The OGLE-only fit yields $b_{SFF,I} = 1.04$ [0.94 - 1.14, 68\% CI], consistent with no blended lens or neighbor flux. However, the Keck high-resolution images (Figure \ref{fig:targets}) show that there are nearby stars that would be blended in OGLE ($\sim0.6''$ typical seeing).
Because it separates these neighbor stars from the source, Keck has a higher $b_{SFF}$ than OGLE. In the joint fit, $b_{SFF,Kp}$ is limited to a maximum value of 1, so $b_{SFF,I}$ is pushed to a lower value than the initial photometry-only fit found.
The Keck photometry is well fit by an unblended PSPL model with $b_{SFF,Kp}$ =
0.91 [0.79 - 0.97, 68\% CI], while OGLE's blend becomes $b_{SFF,I} = 0.57$ [0.50 - 0.63, 68\% CI].
Thus, the final joint fit's $t_E$ is longer than the photometry-only fit value of 139.9 days [132.0 - 150.2 days, 68\% CI].

\begin{figure}
\centering
\includegraphics[width=\textwidth]{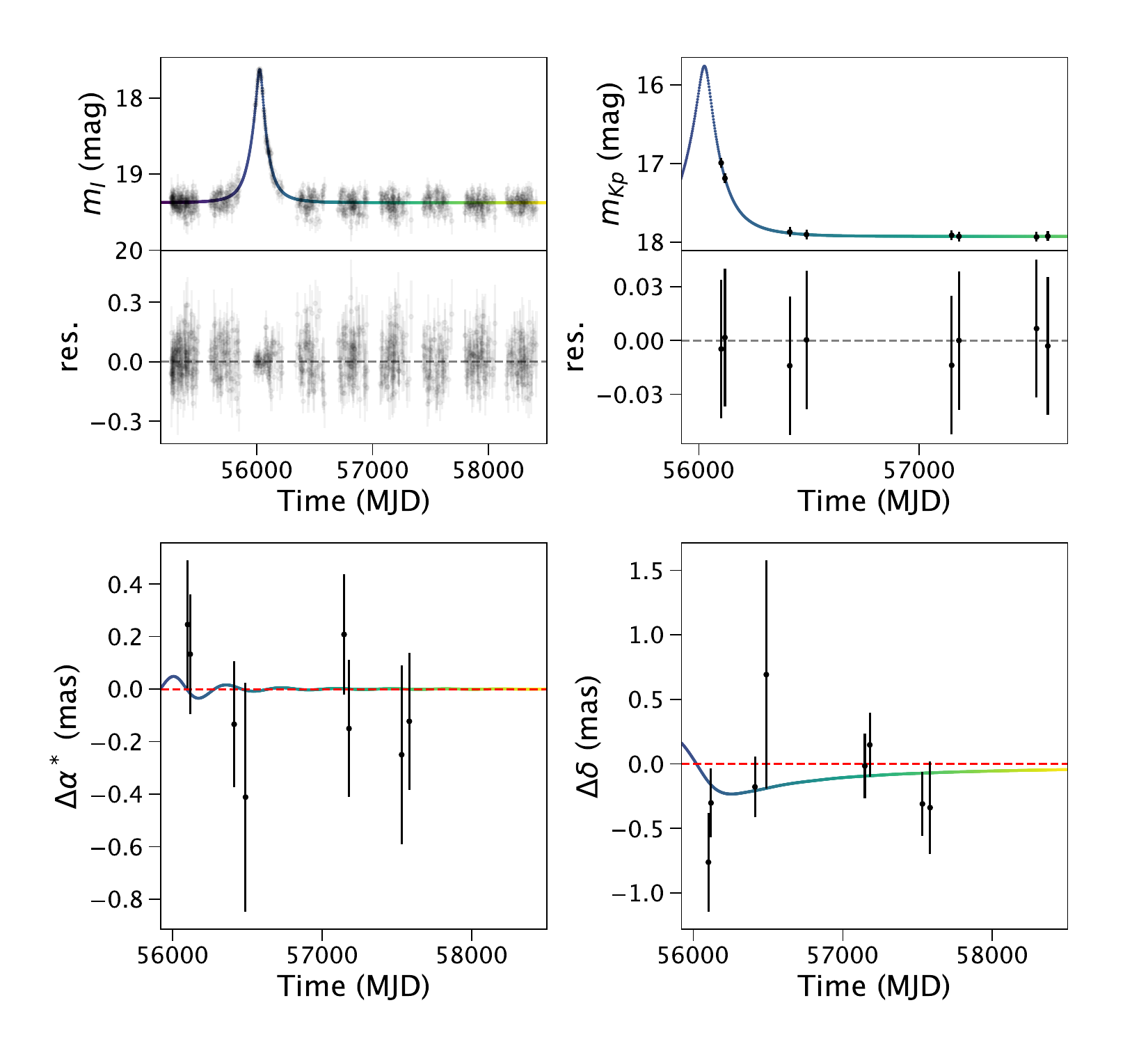}
\caption{The photometric + astrometric best-fit model for OB120169, following the conventions of Fig. \ref{fig:ob150211_results}.
\emph{Top left:} Keck image on May 24, 2016.
\emph{Top right:} the I-band magnitude of the target.
\emph{Bottom row:} the difference in RA \emph{(left)} and DEC \emph{(right)} from the top left's origin.}
\label{fig:ob120169_results}
\end{figure}

Compared to the analysis presented for this target by \citet{Lu:2016}, the
time baseline of the observations has increased by $>2\times$, and we
have improved control over systematic errors. 
However, the Einstein radius is too small to be significantly measured with
$\theta_E = 0.5$ mas [0.3 - 1.0 mas, 68\% CI].
This source was first observed in our pilot survey and, unfortunately,
suffered from poor weather and an insufficient number of observations
when the astrometric signal was likely largest.

The lens mass is 0.43 M$_\odot$ [0.22 - 0.83 M$_\odot$, 68\% CI].
The upper limit (99.73\% confidence) is $M_L < 2.9$ M$_\odot$, and the large mass uncertainty does
not allow us to distinguish between a luminous stellar lens or a
compact remnant lens (Figure \ref{fig:masses}).
Another potential indication of a dark lens is that the
source flux fraction in the Keck AO imaging indicates that the lens
has a very low luminosity.
We apply the method described in \S\ref{sec:res_ob150211} to evaluate the probabilities of different lens types, which results in 40\% star probability, 55\% white dwarf, 2\% neutron star and 3\% brown dwarf.
With a slow relative proper motion between the source and lens of
1.1 mas yr$^{-1}$ [0.6 - 2.0 mas yr$^{-1}$, 68\% CI], and the two may be resolvable in with new instruments and AO upgrades in the next couple of decades.

\begin{deluxetable*}{lrrrcrrrc}
\tabletypesize{\scriptsize}
\tablecaption{OB120169 PSPL fits}
\label{tab:OB120169_pspl_fit}
\tablehead{
  \colhead{} & 
  \multicolumn{4}{c}{OGLE only} &
  \multicolumn{4}{c}{OGLE + Keck} \\ 
  \colhead{Parameter} & 
  \colhead{$\mathcal{L}_{max}$} & 
  \colhead{MAP} & 
  \colhead{Median} & 
  \colhead{68\% CI} &
  \colhead{$\mathcal{L}_{max}$} & 
  \colhead{MAP} & 
  \colhead{Median} & 
  \colhead{68\% CI}
}
\startdata
log$\mathcal{L}$ & 890.95 & 887.53 & 891.10 & & 1019.69 & 1012.89 & 1020.08 & \\ 
$\tilde{\chi}^2$ & 0.99 & 0.98 & 1.01 & & 1.02 & 1.01 & 1.00 &  \\ 
log$\mathcal{Z}$ & & & 857.9 & & & & 960.9 &  \\ 
$N_{dof}$ & & & 897 & & & & 912 &  \\ 
\hline 
$t_0$ (MJD) & 56020.61 & 56023.27 & 56022.24 & [-1.99, 2.59]  & 56019.46 & 56023.40 & 56020.22 & [-2.14, 2.29]  \\ 
$u_0$ & 0.23 & 0.32 & 0.29 & [-0.05, 0.06]  & 0.01 & 0.04 & 0.02 & [-0.04, 0.04]  \\ 
$t_E$ (days) & 138.2 & 151.8 & 139.9 & [-7.9, 10.3]  & 170.3 & 189.3 & 176.3 & [-10.0, 14.3]  \\ 
$\pi_{E,E}$ & 0.036 & 0.016 & 0.031 & [-0.018, 0.016]  & -0.001 & -0.023 & -0.006 & [-0.017, 0.016]  \\ 
$\pi_{E,N}$ & -0.021 & -0.138 & -0.090 & [-0.065, 0.058]  & -0.149 & -0.155 & -0.155 & [-0.027, 0.031]  \\ 
$b_{SFF,I}$ & 1.045 & 0.951 & 1.043 & [-0.101, 0.100]  & 0.616 & 0.489 & 0.574 & [-0.075, 0.059]  \\ 
$I_{base}$ (mag) & 19.375 & 19.378 & 19.374 & [-0.003, 0.003]  & 19.375 & 19.374 & 19.375 & [-0.004, 0.004]  \\ 
$\varepsilon_{a,I}$ (mmag) & 30.0 & 37.4 & 32.4 & [-5.7, 6.1]  & 33.9 & 34.7 & 32.0 & [-7.1, 7.8]  \\ 
$\varepsilon_{m,I}$ & 1.25 & 1.20 & 1.22 & [-0.05, 0.06]  & 1.20 & 1.20 & 1.23 & [-0.07, 0.07]  \\ 
$b_{SFF,Kp}$ & & & &  & 0.98 & 0.75 & 0.91 & [-0.12, 0.07]  \\ 
$Kp_{base}$ (mag) & & & &  & 17.93 & 17.91 & 17.92 & [-0.02, 0.02]  \\ 
$\theta_E$ (mas) & & & &  & 0.67 & 1.35 & 0.53 & [-0.25, 0.45]  \\ 
$\pi_S$ (mas) & & & &  & 0.126 & 0.085 & 0.113 & [-0.021, 0.021]  \\ 
$\mu_{S,\alpha*}$ (mas/yr) & & & &  & -3.58 & -3.54 & -3.53 & [-0.06, 0.06]  \\ 
$\mu_{S,\delta}$ (mas/yr) & & & &  & -7.73 & -8.30 & -7.77 & [-0.18, 0.11]  \\ 
$x_{S0,\alpha*}$ (mas) & & & &  & 23.19 & 22.91 & 23.08 & [-0.16, 0.16]  \\ 
$x_{S0,\delta}$ (mas) & & & &  & 31.97 & 32.01 & 31.82 & [-0.22, 0.23]  \\ 
\tableline
$M_L$ ($\msun$) & & & &  & 0.55 & 1.06 & 0.43 & [-0.21, 0.40]  \\ 
$\pi_L$ (mas) & & & &  & 0.225 & 0.296 & 0.197 & [-0.047, 0.071]  \\ 
$\pi_{rel}$ (mas) & & & &  & 0.099 & 0.211 & 0.081 & [-0.040, 0.071]  \\ 
$\mu_{L,\alpha*}$ (mas/yr) & & & &  & -3.57 & -3.15 & -3.49 & [-0.12, 0.12]  \\ 
$\mu_{L,\delta}$ (mas/yr) & & & &  & -6.30 & -5.73 & -6.70 & [-0.47, 0.83]  \\ 
$\mu_{rel,\alpha*}$ (mas/yr) & & & &  & -0.01 & -0.39 & -0.04 & [-0.13, 0.11]  \\ 
$\mu_{rel,\delta}$ (mas/yr) & & & &  & -1.43 & -2.57 & -1.09 & [-0.93, 0.52]  \\ 
$I_{src}$ (mag) & 19.328 & 19.432 & 19.328 & [-0.098, 0.111]  & 19.901 & 20.151 & 19.977 & [-0.106, 0.153]  \\ 
$Kp_{src}$ (mag) & & & &  & 17.95 & 18.23 & 18.03 & [-0.08, 0.15] 
\enddata
\tablecomments{The combined OGLE + Keck fit contains only one
  solution and no significant secondary mode. 
  Note, the point of
  maximum likelihood, MAP, and 1D median values are reported. The 68\% confidence interval values
  are presented as [$x_{16\%} - x_{Med}$, $x_{84\%} - x_{Med}$]
  for some parameter x. 
  Some parameters, such as $t_0$, $\pi_S$, $b_{SFF,Kp}$,
  and $Kp_{src}$, have MAP values that fall outside the
  68\% CI region. 
}
\end{deluxetable*}

\subsection{OB150029}

Based on the OGLE + MOA + Keck PSPL fits,
OB150029 is a long-duration event ($t_E = 147.1$ days; 145.8 - 148.4 days, 68\% CI); however,
the lack of significant astrometric microlensing signal indicates that
this is source is most likely a low-mass star.
The fitter produced 3 modes for both the photometry only and joint fits. One returned an un-physical $b_{SFF,I}=1.2$ and left clear residuals in the light curve. We dropped this mode as it is disfavored by a Bayes factor of $>10^{1000}$. The remaining modes are split by positive and negative $\pi_{E,N}$, much like OB140613. The $\pi_{E,N}>0$ result is strongly preferred by a Bayes factor of $10^{110}$, so we drop the $\pi_{E,N}<0$ mode as well. 

\begin{figure}
\centering
\includegraphics[width=\textwidth]{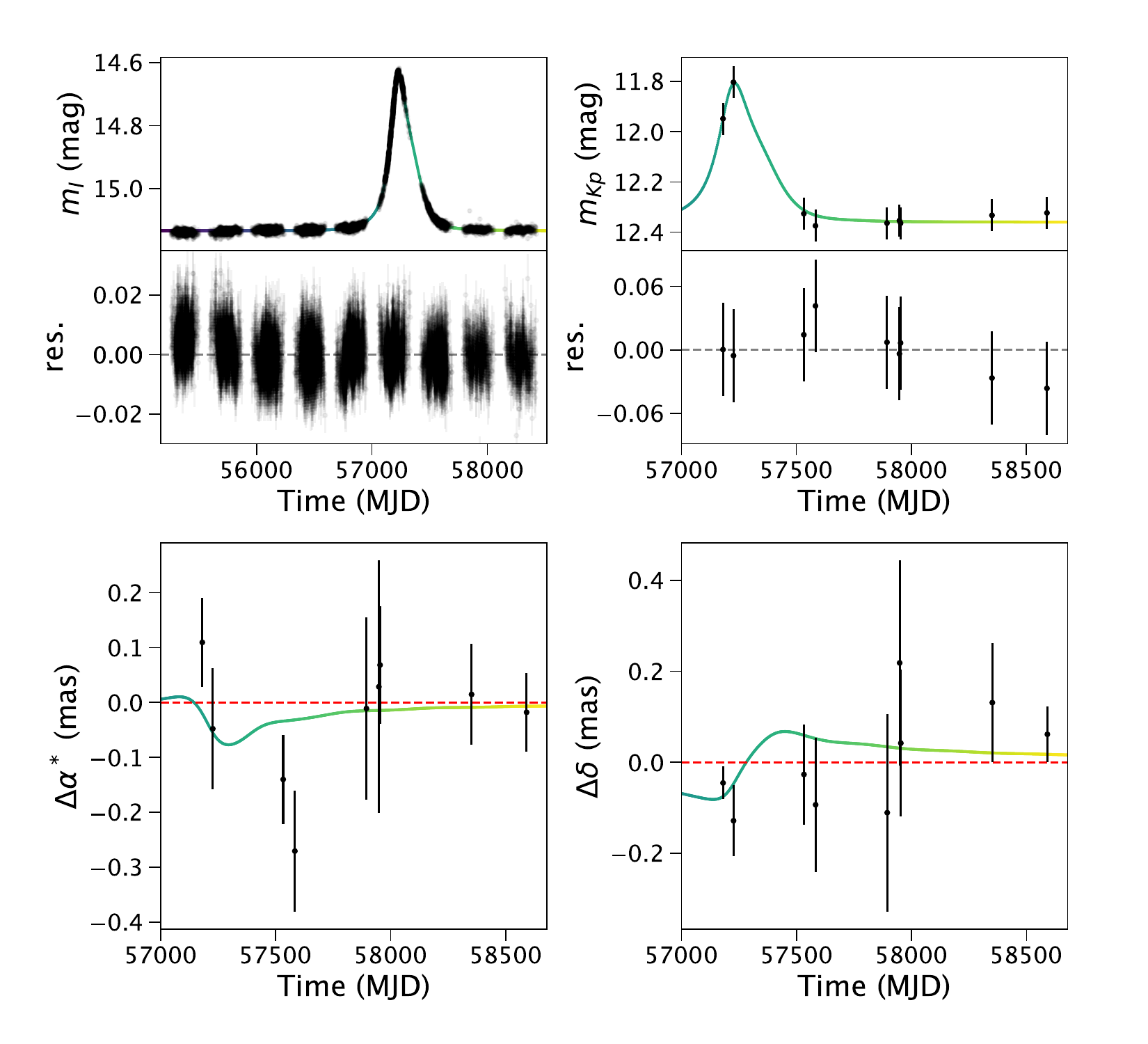}
\caption{The photometric + astrometric best-fit model for OB150029, following the conventions of Fig. \ref{fig:ob150211_results}.
\emph{Top left:} Keck image on July 19, 2017.
\emph{Top right:} the I-band magnitude of the target.
\emph{Bottom row:} the difference in RA \emph{(left)} and DEC \emph{(right)} from the top left's origin.}
\label{fig:ob150029_results}
\end{figure}

The best-fit model is shown in Figure \ref{fig:ob150029_results} and
the parameters and confidence intervals for seeing-limited photometry only and for joint photometry + astrometry fits with Keck are shown in Table \ref{tab:OB150029_pspl_fit}.
The moderate microlens parallax of $\pi_E = 0.171$ [0.169 - 0.173, 68\% CI] is also consistent with
a stellar lens. The upper mass limit for this target is $M_L < 0.53$
M$_\odot$ (99.73\% confidence) as shown in Figure \ref{fig:masses}.
The joint fit is of good quality, with a  $\chi^2_{dof} = 1.0$.

\begin{deluxetable*}{lrrrcrrrc}[htb!]
\tablecaption{OB150029 PSPL fits with seeing-limited photometry only and with added Keck photometry and astrometry.
  \label{tab:OB150029_pspl_fit}}
\tabletypesize{\scriptsize}
\tablehead{
  \colhead{} & 
  \multicolumn{4}{c}{OGLE + MOA Only} &
  \multicolumn{4}{c}{OGLE + MOA + Keck} \\ 
  \colhead{Parameter} & 
  \colhead{$\mathcal{L}_{max}$} & 
  \colhead{MAP} &
  \colhead{Median} &
  \colhead{68\% CI} &
  \colhead{$\mathcal{L}_{max}$} & 
  \colhead{MAP} &
  \colhead{Median} &
  \colhead{68\% CI}
}
\startdata
log$\mathcal{L}$ & 177236.82 & 177231.22 & 177237.42 & & 177540.34 & 177530.20 & 177540.79 &  \\ 
$\tilde{\chi}^2$ & 1.00 & 0.98 & 1.00 & & 1.00 & 1.01 & 1.00 &  \\ 
log$\mathcal{Z}$ & & & 177153.4 & & & & 177426.6 &  \\ 
$N_{dof}$ & 57680 & & & & 57698 & & &  \\ 
\hline 
$t_0$ (MJD) & 57260.88 & 57260.84 & 57260.82 & [-0.21, 0.20]  & 57260.69 & 57260.84 & 57260.69 & [-0.19, 0.20]  \\ 
$u_0$ & 0.42 & 0.43 & 0.42 & [-0.01, 0.01]  & 0.41 & 0.40 & 0.41 & [-0.01, 0.01]  \\ 
$t_E$ (days) & 135.3 & 135.4 & 135.6 & [-1.1, 1.0]  & 147.2 & 148.4 & 147.1 & [-1.3, 1.3]  \\ 
$\pi_{E,E}$ & 0.063 & 0.064 & 0.063 & [-0.002, 0.002]  & 0.057 & 0.054 & 0.057 & [-0.002, 0.002]  \\ 
$\pi_{E,N}$ & -0.221 & -0.219 & -0.221 & [-0.003, 0.003]  & 0.161 & 0.163 & 0.161 & [-0.002, 0.002]  \\ 
$b_{SFF,I}$ & 0.654 & 0.656 & 0.650 & [-0.012, 0.014]  & 0.542 & 0.529 & 0.543 & [-0.013, 0.014]  \\ 
$I_{base}$ (mag) & 15.133 & 15.134 & 15.134 & [-0.000, 0.000]  & 15.134 & 15.134 & 15.134 & [-0.000, 0.000]  \\ 
$\varepsilon_{a,I}$ (mmag) & 2.8 & 2.8 & 2.9 & [-0.2, 0.2]  & 2.8 & 3.0 & 2.8 & [-0.2, 0.2]  \\ 
$\varepsilon_{m,I}$ & 1.44 & 1.46 & 1.44 & [-0.04, 0.04]  & 1.44 & 1.38 & 1.44 & [-0.04, 0.04]  \\ 
$b_{SFF,R}$ & 0.55 & 0.55 & 0.55 & [-0.01, 0.01]  & 0.457 & 0.445 & 0.458 & [-0.011, 0.011]  \\ 
$R_{base}$ (mag) & -12.99 & -12.99 & -12.99 & [-0.00, 0.00]  & -12.985 & -12.985 & -12.985 & [-0.000, 0.000]  \\ 
$\varepsilon_{a,R}$ (mmag) & 2.7 & 2.9 & 2.7 & [-0.1, 0.1]  & 2.7 & 2.7 & 2.7 & [-0.1, 0.1]  \\ 
$\varepsilon_{m,R}$ & 1.76 & 1.76 & 1.76 & [-0.01, 0.01]  & 1.76 & 1.75 & 1.76 & [-0.01, 0.01]  \\ 
$b_{SFF,Kp}$ & & & &  & 0.61 & 0.57 & 0.61 & [-0.06, 0.06]  \\ 
$Kp_{base}$ (mag) & & & &  & 12.36 & 12.38 & 12.36 & [-0.02, 0.02]  \\ 
$\theta_E$ (mas) & & & &  & 0.28 & 0.32 & 0.28 & [-0.10, 0.13]  \\ 
$\pi_S$ (mas) & & & &  & 0.113 & 0.116 & 0.113 & [-0.022, 0.022]  \\ 
$\mu_{S,\alpha*}$ (mas/yr) & & & &  & -3.96 & -3.99 & -3.96 & [-0.04, 0.05]  \\ 
$\mu_{S,\delta}$ (mas/yr) & & & &  & -4.58 & -4.50 & -4.57 & [-0.09, 0.12]  \\ 
$x_{S0,\alpha*}$ (mas) & & & &  & -32.13 & -32.02 & -32.12 & [-0.05, 0.05]  \\ 
$x_{S0,\delta}$ (mas) & & & &  & 58.19 & 58.16 & 58.19 & [-0.03, 0.04]  \\ 
 & & & &  & & & &  \\ 
\tableline$M_L$ ($\msun$) & & & &  & 0.2 & 0.2 & 0.2 & [-0.1, 0.1]  \\ 
$\pi_L$ (mas) & & & &  & 0.161 & 0.171 & 0.162 & [-0.029, 0.030]  \\ 
$\pi_{rel}$ (mas) & & & &  & 0.048 & 0.055 & 0.048 & [-0.018, 0.023]  \\ 
$\mu_{L,\alpha*}$ (mas/yr) & & & &  & -4.19 & -4.24 & -4.19 & [-0.07, 0.05]  \\ 
$\mu_{L,\delta}$ (mas/yr) & & & &  & -5.23 & -5.24 & -5.22 & [-0.21, 0.16]  \\ 
$\mu_{rel,\alpha*}$ (mas/yr) & & & &  & 0.23 & 0.25 & 0.23 & [-0.09, 0.11]  \\ 
$\mu_{rel,\delta}$ (mas/yr) & & & &  & 0.65 & 0.75 & 0.65 & [-0.24, 0.31]  \\ 
$I_{src}$ (mag) & 15.594 & 15.592 & 15.601 & [-0.024, 0.021]  & 15.798 & 15.825 & 15.797 & [-0.027, 0.026]  \\ 
$R_{src}$ (mag) & -12.34 & -12.34 & -12.33 & [-0.02, 0.02]  & -12.135 & -12.106 & -12.136 & [-0.027, 0.026]  \\ 
$Kp_{src}$ (mag) & & & &  & 12.90 & 12.99 & 12.90 & [-0.09, 0.10]  \\ 

\enddata
\tablecomments{The points of
  maximum likelihood ($\mathcal{L}_{max}$), Maximum A Posteriori (MAP), and the median value are given. The 68\% confidence interval values
  are presented as [$x_{16\%} - x_{Median}$, $x_{84\%} - x_{Median}$]
  for some parameter x. 
  Some parameters have $\mathcal{L}_{max}$ and/or MAP values that fall outside the
  68\% CI region.
}
\end{deluxetable*}

The event shows significant blended flux in Keck, with $b_{SFF,Kp}=0.61$ [0.55 - 0.67, 68\% CI], which may be due to a luminous lens.
We apply the method described in \S\ref{sec:res_ob150211} to evaluate the probabilities of different lens types, which results in 91\% star probability and 9\% brown dwarf.
It may also be that OB150029 contains a binary lens or a binary source, and the blended flux comes from a companion, but no clear residuals suggest binarity in the microlensing signal.
While the lens system could still involve a black hole lens,
it is more likely that the system contains a binary stellar lens as
there is little indication of astrometric microlensing. 
Given the low relative proper motion of 0.7 mas yr$^{-1}$ [0.44 - 1.03 mas yr$^{-1}$, 68\% CI], the
source and lens pair will likely not be resolved in the coming decades.

\section{Fits \& Future Prospects Discussion}
\label{sec:discussion}


\subsection{Tension Between OGLE and Keck Fits}
\label{sec:disc_phot_vs_ast}

Modeling microlensing events can be tricky due to the
high-dimensional, multi-modal, and non-linear nature
of the parameter space.
While significant work has been done to optimize
parameter estimation for photometry-only data sets, even the addition
of Gaussian Process noise in the fitting
process is not yet well-generalized for use in multi-event analysis.
Thus, it is worth exploring how the fits change between OGLE
photometry-only and OGLE + Keck photometry and astrometry fits.

Two of the four targets in our sample show statistically significant
discrepancies between OGLE(+MOA)-only and OGLE(+MOA) + Keck.
OB140613 and OB150029 yield statistically consistent results for both fits,
which notably are the two targets with MOA data in addition to OGLE.
The other 2 targets (OB150211 and OB120169) show
larger changes in $t_E$, $\pi_E$, and $b_{SFF,I}$ when
the Keck photometry and astrometry are added to the fit.
One of the primary differences is that the Keck photometry is
restricted to have only physically allowed source-flux-fractions of
$b_{SFF,Kp} \leq 1$.

In the case of OB150211, the OGLE and Keck light-curves have very
similar shapes and source-flux-fractions, thus the addition of
high-resolution photometry brought the OGLE data into the valid $b_{SFF,I} \leq 1$ range. For OB120169, the OGLE light curve shows a significantly lower source flux fraction than the Keck data, and so requiring a physical $b_{SFF,Kp} \leq 1$ results in a greatly reduced $b_{SFF,I}$. 
In this case, the addition of high-resolution Keck data allowed certain $b_{SFF}-t_E-\pi_E$ parameter space to be ruled out, reducing uncertainties due to their degeneracy in the final fit results.

\subsection{Astrometric Reference Frame Uncertainties}

As discussed in \S\ref{sec:ast_analysis}, the AO imaging and FlyStar astrometric alignment process results in {\it relative} astrometry, not absolute. The quality of this relative astrometry is dependent on both the image quality achieved each night and the quality of the reference stars available, with stars near in position and similar in brightness being ideal. OB150211's relative astrometric data quality suffered some from poor observing conditions (weather and instrument issues). It was also in a sparser field than the other candidates and lacking in reference stars as bright as the target.

The difficulty of converting to an absolute astrometric reference frame contributed additional uncertainty. With only up to 2 reference stars per target with Gaia parallax measurements, the shift into the absolute reference frame was highly uncertain. Gaia DR3 astrometric data quality is known to worsen in crowded fields toward the Galactic bulge \citep{Luna2023}.

Fitting the astrometric data from these alignments required the addition of a new parameter into the astrometric model, the reference frame parallax. This parameter is strongly correlated with the other astrometric parameters, and particularly the source parallax, which contributed to additional uncertainty in the final best-fit parameters.

Recent and ongoing upgrades to the Keck AO imaging process can significantly improve performance in these areas for newer targets. The OSIRIS instrument's field of view is roughly 4$\times$ that of NIRC2 which will capture more reference stars. While the more recent Keck observations in this work included OSIRIS data, there were too few epochs to reliably align these additional stars beyond the original NIRC2 field. 

\subsection{Future Analysis of This Sample}

In addition to Gaia DR4's improved sample of reference stars with parallax and proper motion, full astrometric time series will be provided for all of their detected sources. OB140613, OB150029, and OB150211 were all present in Gaia DR3, so we can expect this additional astrometric data source for them. OB120169 is the dimmest of our targets and may not be present in Gaia DR4 as it was not in DR3. 
OB140613 has a Gaia G magnitude of 19.6, a reported parallax of $-2.59\pm0.72$ mas/yr, and a Renormalized Unit Weight Error (RUWE) of 1.24. The RUWE is near the $>1.25$ threshold for a poor fit in EDR3 \citep{Penoyre2022}, and the significant negative parallax may suggest odd astrometric behavior measured by Gaia. OB150029 is our most optically bright target, with a G magnitude of 16.33, a reported parallax of $-1.89\pm0.35$, and a RUWE of 6.98. Interestingly, this target showed very little astrometric residual in the Keck data, but the very high Gaia RUWE may indicate some astrometric variance, perhaps due to a binary system or crowding issues in Gaia. A binary system may also explain the non-negligible blend flux detected in the Keck aperture. OB150211 has a G magnitude of 18.63, a parallax of $-0.18\pm0.24$, and a RUWE of 1.05, which does not suggest any significant deviation from a parallax and proper motion trajectory. The DR3 data collection period spanned from July 2014 to May 2017, which includes the photometric peak for all of our events but not much coverage of the astrometric baseline afterward. DR4 will extend this observation period to January 2020, capturing the return to baseline for all of our events.

A joint fit of OGLE photometry, Keck photometry+astrometry, and Gaia photometry+astrometry may glean a firmer mass measurement for OB150211 than OGLE+Keck alone could provide, which may allow us to confirm or rule out the black hole scenario. The addition of Gaia DR4 may also refine our model results for OB140613 and OB150029 to perhaps confirm whether the lens is a star or a white dwarf, or perhaps a binary lens or source. \citet{El-Badry2024} showed that the per field-of-view transit astrometric precision along the scan direction of Gaia epoch astrometry should be $1/\sqrt{8}\times$ that of the per CCD transit precisions reported by \citet{Holl2023}.
Based on G magnitudes, the expected per-epoch astometric precision would be $\sim0.2$ mas for OB150029, $\sim0.5$ mas for OB150211, and $\sim1$ mas for OB140613. However, the performance may be worse for targets in the Galactic bulge due to crowding \citep{Luna2023}. \citep{Kaczmarek2026} found that bright targets (G$<17$) with high RUWE or excess noise provide the best chance of BH lens detection, a criteria that only OB150029 fits here. The Keck data for our targets may improve the constraining power of our joint multi-telescope data fits.

OB140613 lies within the Roman Galactic Bulge Time Domain Survey fields \citep{rotac}, and will thus have late-time multi-filter, multi-epoch imaging available within the next couple of years. This will help characterize the source's stellar type, parallax, and proper motion. Additionally, all 4 of these targets are contained within the intended Roman Galactic Plane Survey fields \citep{GalacticPlaneSurveyDefinitionCommittee2025} and will be observed in several filters in at least 1 epoch. 

\subsection{Future Experimental Design}
\label{sec:disc_future_experiments}

The next sample of Keck AO-monitored BH candidates (N. S. Abrams et al. in prep.) was monitored primarily with OSIRIS and will thus have more reference stars available for the cross-epoch alignment.
Additionally, with Keck All sky Precision Adaptive optics (KAPA) commissioning underway in 2026 to further improve OSIRIS image quality, currently-monitored events are expected to have higher astrometric precision. Gaia DR4 is set to be released in December 2026, which will provide improved parallax and proper motion measurements of more stars, improving prospects for the relative-absolute astrometric reference frame conversion in future analyses.

As black hole lenses have larger masses than a typical stellar lens,
they will have correspondingly smaller microlensing parallax $\pi_E$.
As seen in the case of OB150211, it is difficult to determine 
anything more than an upper limit for small values of $\pi_E$; however, this
produces significant uncertainty when estimating a lens mass.
The systematic errors in ground-based photometric surveys due to
crowding, variable weather and atmospheric transmission, changing
airmass over the year, and instrumental variability remains a major
limitation, even when astrometric microlensing signals are detected.
The upcoming Nancy Grace Roman Space Telescope will deliver much more
precise, accurate, and stable photometry as well as astrometric
measurements without the need for expensive, targetted follow-up.

\section{BH Population Discussion}
\label{sec:disc_res_sample}

The four events we studied with Keck were chosen to have long Einstein crossing
times ($t_E$) in order to maximize the probability of finding black
holes. It appears that the main contaminant in our sample is low mass
lenses with very low relative proper motions between the source and
lens.
OB120169, OB140613, and OB150029 all have small $\mu_{rel}$, which is
consistent with the expected clumping of proper motions along the
Galactic Plane.
OB150211 has a somewhat larger proper motion for both source and lens that sits
toward the edge of the observed distribution of proper motions. 
This field has lower stellar density overall, leaving fewer stars
with which to align the Keck and Gaia frames,
resulting in greater uncertainties in the overall astrometric reference frame.

\citet{Lam:2020} predicted a 40\% success rate for astrometric microlensing searches for black holes that select events with $t_E>120$ days via PopSyCLE simulations of single stars toward the Galactic bulge. \citet{Rose2022} showed that the choice of an initial-final mass relation (IFMR) impacts this success rate, noting that for all three tested IFMRs, about one third of $t_E>150$ day events should have BH lenses. \citet{Abrams2025} showed that the addition of binary systems in these simulations did not significantly impact this BH detection success rate. However, these simulations did not include binary evolution, and thus black holes were all contained in binary systems. A more realistic BH binary fraction may result in a decreased BH fraction for long-timescale events, as binarity tends to increase timescales.

The full sample of completed Keck AO astrometric microlensing searches for black holes include the 4 candidates explored here plus 1 non-BH confirmation OB110022 from \citet{Lu:2016}. An additional 5 candidates were observed with HST, including confirmed BH OB110462 \citep{Lam2022, Sahu2022, Mroz2022, Lam2023, Sahu2025} and four non-detections from OB110037, OB110310, MB09260, and MB10364. Of these past targets, we exclude OB110022, OB110037, OB110310, and MB10364 due to their relatively short timescales ($t_E<100$ days). Since the allowed BH solution has less than 50\% probability for OB150211, we assume it is a non-detection here, in addition to our other 3 confirmed non-BH lenses. 
Out of 6 long $t_E$ BH candidates, this leaves 1 confirmed BH, or a success rate of 17\%. This result is a bit low given the expected 40\% success rate, but this was likely overestimated. 

We simulate microlensing surveys via PopSyCLE in fields surrounding the 6 candidates to compare the estimated black hole event success rate to observed results. We ran several simulated microlensing surveys toward each target with different random seeds for the star generation process.
The simulations were generated following \citet{Abrams2025} with multiplicity on and the following model updates. For the neutron star kick velocity distribution, we use a log-normal distribution rather than a Maxwellian, with $\mu = 5.67$ and $\sigma = 0.59$ \citep{Disberg2025}. However, we retain the Maxwellian black hole kick velocity distribution, with a mean of 100 km s$^{-1}$. As a simple approach to accounting for binary evolution, we apply a uniform black hole binary cut in which 90\% of all black holes generated lose their companions, based on StarTrack simulations of binary system evolution \citep{Olejak2020}.

\begin{figure}
\centering
\includegraphics[width=0.49\textwidth]{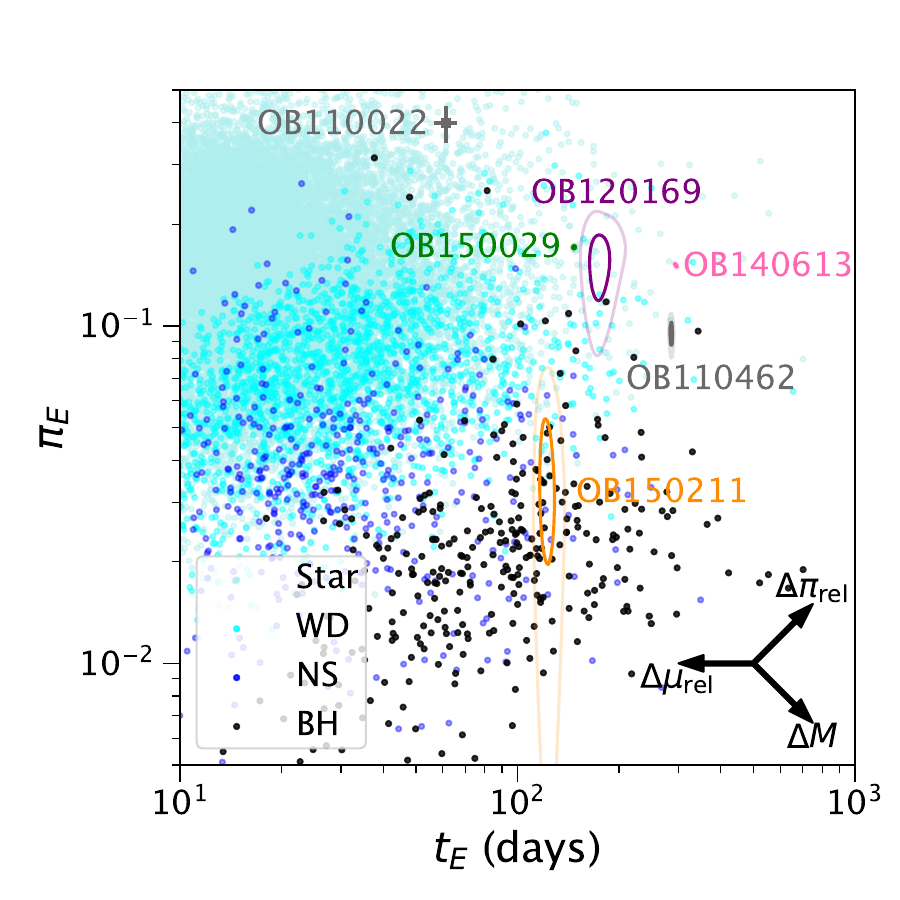}
\includegraphics[width=0.49\textwidth]{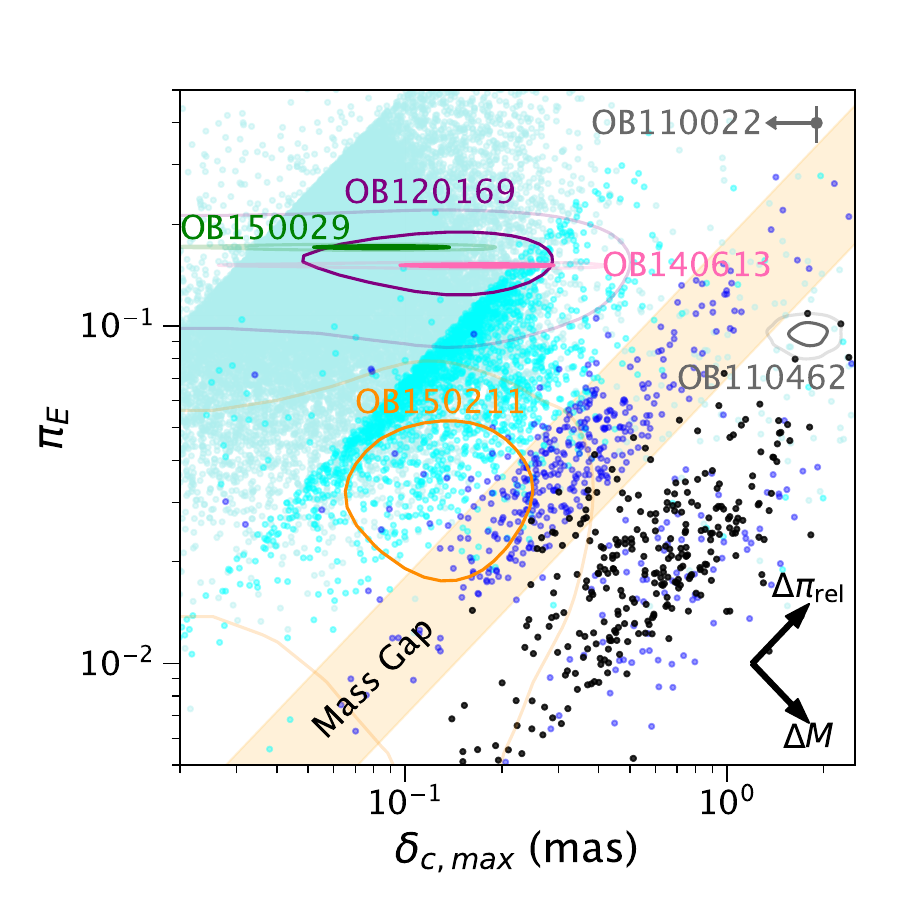}
    \caption{$\pi_E$-$t_E$ (left) and maximum astrometric shift ($\delta_{c,max}$) vs. $\pi_E$ (right) posteriors based on PSPL fits to OGLE photometry + Keck photometry and astrometry for each of the four targets.
    For each target, contours ares shown enclosing 39\% and 86\% probability, which corresponds to 1$\sigma$ and 2$\sigma$ for a 2D gaussian.
    We also show the only confirmed isolated stellar mass BH so far \citep[OB110462; fit posteriors from][]{Lam2023}, and the confirmed non-BH from our previous Keck search \citep[OB110022;][]{Lu:2016}. The compass arrows show how these observable parameters change with increasing proper motion $\mu_{rel}$, relative parallax $\pi_{rel}$, and lens mass $M$. Measuring astrometry, and thus $\delta_{c,max}$, eliminates $\mu_{rel}$ dependence and breaks the mass-distance degeneracy.
\label{fig:tE_piE}
}
\end{figure}

To approximate the OGLE survey candidates we drew from, we select single-peak microlensing events with baseline I-band magnitude $I_{\rm base}<21$, and magnification $\Delta I>0.1$. 
The observed vs. simulated distributions of
$\pi_E$ vs.~$t_E$ and
$\pi_E$ vs.~$\delta_{c,max}$ (maximum astrometric shift) shows
that OB150211 is the most probable black hole or neutron star in our
sample, while the others are likely lower in mass (Figure \ref{fig:tE_piE}).

\begin{figure*}
    \centering
    \includegraphics[width=0.5\textwidth]{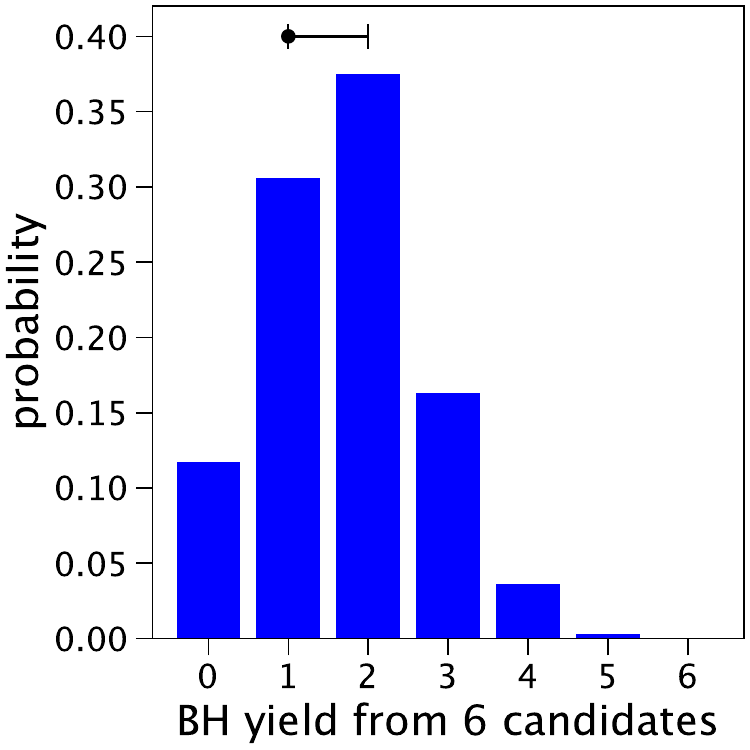}
    \caption{Black hole yields for PopSyCLE-simulated events mimicking our OGLE EWS event selection process for 100 random iterations of candidate selection ($t_E > 100$ days). The black point shows the confirmed 1 BH measurement with an error bar since OB150211 may be confirmed as a BH in the future with additional data.
    \label{fig:bh_yield}}
\end{figure*}

To determine whether our yield of 1 BH out of 6 candidates is consistent with our existing model, we mimic our candidate selection process on the mock OGLE events. For each target's simulation fields, we select 1 random event with $t_E>100$ days. The distribution of black hole yields from 100 iterations of this random selection is shown in Figure \ref{fig:bh_yield}. The mean and standard deviation of the BH yield is 1.7$\pm$1.0. Thus, a yield of 1 BH is consistent with expectations, and a larger sample size will be required to further refine the black hole population model. The simulation yield is also consistent with 2 BHs, which would be the true observational yield from the candidate set if the BH solution is proven for OB150211 in the future.

\begin{figure*}
    \centering
    \includegraphics[width=0.7\textwidth]{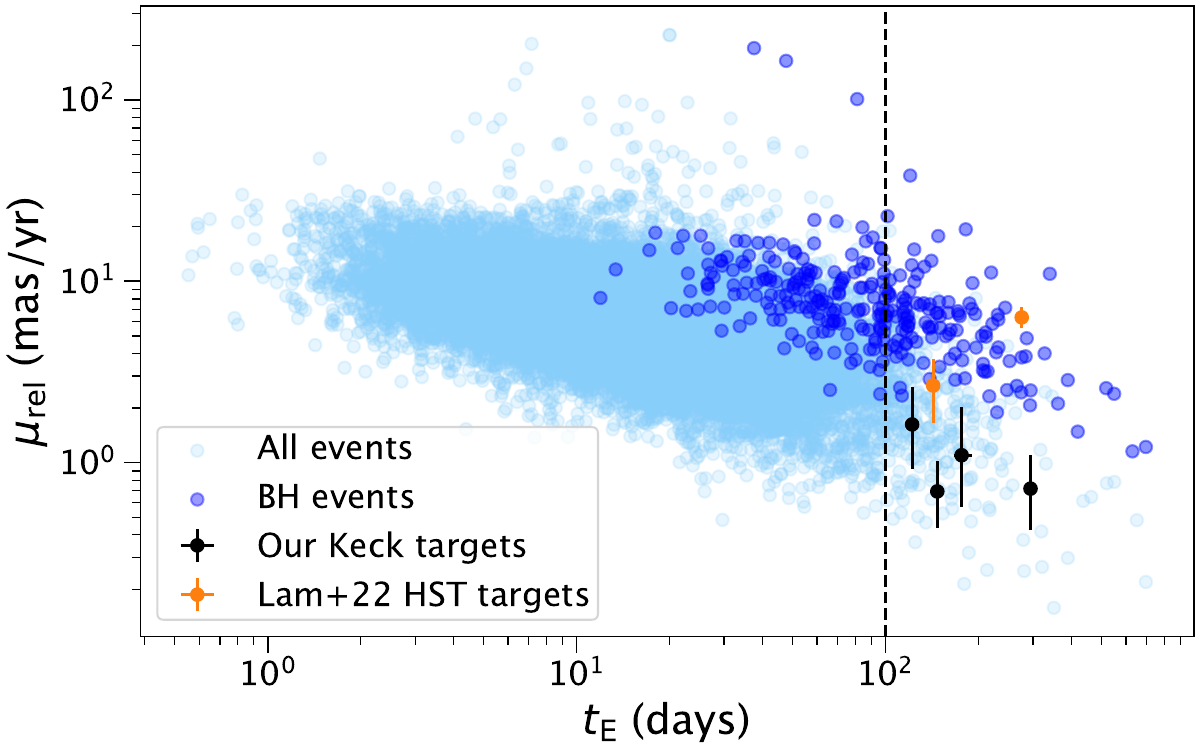}
    \caption{Timescale and relative proper motion distribution for all mock OGLE events (pale blue) and for BH lens mock events (dark blue). We show our Keck targets (black) and the $t_E>100$ day events from \citet{Lam2022b,Lam2023} (orange) with $1\sigma$ uncertainties.
    \label{fig:mu_rel_sim}}
\end{figure*}

Next, we evaluate whether the low relative proper motions of our non-BH contaminant events are consistent with simulated expectations. Figure \ref{fig:mu_rel_sim} shows the distribution of $t_E$ and $\mu_{\rm rel}$ across all of the mock OGLE events and the BH lens events. We confirm that simulated non-BH long timescale events are dominated by low relative proper motion stellar events ($\mu_{\rm rel} \lesssim 2$ mas/yr). The Keck event proper motions are consistent with non-BH simulated event proper motions for their given timescales. 

\section{Conclusions}
\label{sec:conclusions}

We have presented a search for astrometric microlensing signals in
four long-duration microlensing events monitored with Keck AO photometry
and astrometry from
2012-2021. Of the four, OB150211 remains a candidate for an
isolated stellar-mass black hole or neutron star. Its highly uncertain mass measurement
places it in allowable parameter space for stars, white dwarfs, neutron stars,
and black holes. 
Gaia DR4 may hold the answer to the lens' true nature. 
Additionally, new Keck data via OSIRIS with the KAPA AO upgrades may provide improved measurements of the target's baseline parallax and proper motion as well as the reference frame accuracy through nearby stars, though the observation quality during the event remains a limitation.
A black hole lens can be ruled out for the other three events: OB120169, OB140613, and OB150029.

All four of our events are likely stellar lenses with low relative proper motions, which are the primary contaminant for non-BH long $t_E$ ($>100$ day) events. When combined with the other two long-timescale BH candidates with prior completed astrometric analysis, the BH yield is 1$^{+1}_{-0}$ out of 6. Simulations provide an expected yield of 1.7$\pm$1.0, which is consistent with our result, regardless of whether OB150211 is ultimately confirmed as a BH. Thus, more candidates with complete astrometric measurements will be required to further refine our understanding of BH formation processes. 

\section*{Acknowledgements}

M.J.H., J.R.L., and N.S.A. acknowledge support from the Heising-Simons 
Foundation under grant No. 2022-3542.
J.R.L. and C.Y.L. acknowledge support by the National
Science Foundation under grant No. 1909641 and the National
Aeronautics and Space Administration (NASA) under contract
No. NNG16PJ26C issued through the WFIRST (now Roman)
Science Investigation Teams Program. 
N.S.A. acknowledges support from the H2H8 foundation.
Support for C.Y.L.'s work was provided by NASA through the NASA Hubble Fellowship grant HST-HF2-51609.001-A awarded by the Space Telescope Science Institute, which is operated by the Association of Universities for Research in Astronomy, Inc., for NASA, under contract NAS5-26555.
M.W.H. is supported by the Brinson Prize Fellowship.

Some of the data presented herein were obtained at the W.~M.~Keck
Observatory, which is operated as a scientific partnership among the
California Institute of Technology, the University of California and
NASA. The Observatory was
made possible by the generous financial support of the W.~M.~Keck
Foundation. The authors wish to recognize and acknowledge the very
significant cultural role and reverence that the summit of Maunakea
has always had within the indigenous Hawaiian community.  We are most
fortunate to have the opportunity to conduct observations from this
mountain.

This research used resources of the National Energy Research Scientific Computing Center (NERSC), a Department of Energy User Facility using NERSC awards HEP-ERCAP0030957 and HEP-ERCAP0036327.

The OGLE project has received funding from the Polish National Science
Centre grant OPUS-28 2024/55/B/ST9/00447 awarded to AU. 
The MOA project is supported by JSPS KAKENHI Grant Number 
JP16H06287, JP22H00153, JP23KK0060 and JP25H00668.

\bibliographystyle{aasjournalv7}
\bibliography{references}

@preamble{ " \newcommand{\noop}[1]{} " }

@ARTICLE{GalacticPlaneSurveyDefinitionCommittee2025,
       author = {{Galactic Plane Survey Definition Committee}, Roman},
        title = "{Roman Galactic Plane Survey Definition Committee Report}",
      journal = {arXiv e-prints},
         year = 2025,
        month = nov,
          eid = {arXiv:2511.07494},
        pages = {arXiv:2511.07494},
archivePrefix = {arXiv},
       eprint = {2511.07494},
 primaryClass = {astro-ph.GA},
       adsurl = {https://ui.adsabs.harvard.edu/abs/2025arXiv251107494G}
}

@ARTICLE{rotac,
       author = {{Roman Observations Time Allocation Committee} and {Core Community Survey Definition Committees}},
        title = "{Roman Observations Time Allocation Committee: Final Report and Recommendations}",
      journal = {arXiv e-prints},
         year = 2025,
        month = may,
          eid = {arXiv:2505.10574},
        pages = {arXiv:2505.10574},
          doi = {10.48550/arXiv.2505.10574},
archivePrefix = {arXiv},
       eprint = {2505.10574},
 primaryClass = {astro-ph.IM},
       adsurl = {https://ui.adsabs.harvard.edu/abs/2025arXiv250510574Z}
}

@ARTICLE{Lam2022b,
       author = {{Lam}, Casey Y. and {Lu}, Jessica R. and {Udalski}, Andrzej and {Bond}, Ian and {Bennett}, David P. and {Skowron}, Jan and {Mr{\'o}z}, Przemek and {Poleski}, Radek and {Sumi}, Takahiro and {Szyma{\'n}ski}, Micha{\l} K. and {Koz{\l}owski}, Szymon and {Pietrukowicz}, Pawe{\l} and {Soszy{\'n}ski}, Igor and {Ulaczyk}, Krzysztof and {Wyrzykowski}, {\L}ukasz and {Miyazaki}, Shota and {Suzuki}, Daisuke and {Koshimoto}, Naoki and {Rattenbury}, Nicholas J. and {Hosek}, Matthew W. and {Abe}, Fumio and {Barry}, Richard and {Bhattacharya}, Aparna and {Fukui}, Akihiko and {Fujii}, Hirosane and {Hirao}, Yuki and {Itow}, Yoshitaka and {Kirikawa}, Rintaro and {Kondo}, Iona and {Matsubara}, Yutaka and {Matsumoto}, Sho and {Muraki}, Yasushi and {Olmschenk}, Greg and {Ranc}, Cl{\'e}ment and {Okamura}, Arisa and {Satoh}, Yuki and {Silva}, Stela Ishitani and {Toda}, Taiga and {Tristram}, Paul J. and {Vandorou}, Aikaterini and {Yama}, Hibiki and {Abrams}, Natasha S. and {Agarwal}, Shrihan and {Rose}, Sam and {Terry}, Sean K.},
        title = "{Supplement: ``An Isolated Mass-gap Black Hole or Neutron Star Detected with Astrometric Microlensing'' (2022, ApJL, 933, L23)}",
      journal = {\apjs},
         year = 2022,
        month = jun,
       volume = {260},
       number = {2},
          eid = {55},
        pages = {55},
          doi = {10.3847/1538-4365/ac7441},
       adsurl = {https://ui.adsabs.harvard.edu/abs/2022ApJS..260...55L}
}

@ARTICLE{Kroupa2001,
       author = {{Kroupa}, Pavel},
        title = "{On the variation of the initial mass function}",
      journal = {\mnras},
         year = 2001,
        month = apr,
       volume = {322},
       number = {2},
        pages = {231-246},
          doi = {10.1046/j.1365-8711.2001.04022.x},
archivePrefix = {arXiv},
       eprint = {astro-ph/0009005},
 primaryClass = {astro-ph},
       adsurl = {https://ui.adsabs.harvard.edu/abs/2001MNRAS.322..231K}
}

@ARTICLE{Rybicki2024,
       author = {{Rybicki}, Krzysztof A. and {Shvartzvald}, Yossi and {Yee}, Jennifer C. and {Calchi Novati}, Sebastiano and {Ofek}, Eran O. and {Bond}, Ian A. and {Beichman}, Charles and {Bryden}, Geoff and {Carey}, Sean and {Henderson}, Calen and {Zhu}, Wei and {Fausnaugh}, Michael M. and {Wibking}, Benjamin and {Udalski}, Andrzej and {Poleski}, Radek and {Mr{\'o}z}, Przemek and {Szyma{\'n}ski}, Micha{\l} K. and {Soszy{\'n}ski}, Igor and {Pietrukowicz}, Pawe{\l} and {Koz{\l}owski}, Szymon and {Skowron}, Jan and {Ulaczyk}, Krzysztof and {Iwanek}, Patryk and {Wrona}, Marcin and {Ryu}, Yoon-Hyun and {Albrow}, Michael D. and {Chung}, Sun-Ju and {Gould}, Andrew and {Han}, Cheongho and {Hwang}, Kyu-Ha and {Jung}, Youn Kil and {Shin}, In-Gu and {Yang}, Hongjing and {Zang}, Weicheng and {Cha}, Sang-Mok and {Kim}, Dong-Jin and {Kim}, Hyoun-Woo and {Kim}, Seung-Lee and {Lee}, Chung-Uk and {Lee}, Dong-Joo and {Lee}, Yongseok and {Park}, Byeong-Gon and {Pogge}, Richard W. and {Abe}, Fumio and {Barry}, Richard and {Bennett}, David P. and {Bhattacharya}, Aparna and {Fukui}, Akihiko and {Hamada}, Ryusei and {Hamada}, Shunya and {Hamasaki}, Naoto and {Hirao}, Yuki and {Ishitani Silva}, Stela and {Itow}, Yoshitaka and {Kirikawa}, Rintaro and {Koshimoto}, Naoki and {Matsubara}, Yutaka and {Miyazaki}, Shota and {Muraki}, Yasushi and {Nagai}, Tutumi and {Nunota}, Kansuke and {Olmschenk}, Greg and {Ranc}, Clement and {Rattenbury}, Nicholas J. and {Satoh}, Yuki K. and {Sumi}, Takahiro and {Suzuki}, Daisuke and {Tristram}, Paul. J. and {Vandorou}, Aikaterini and {Yama}, Hibiki and {Wyrzykowski}, {\L}ukasz and {Howil}, Kornel and {Kruszy{\'n}ska}, Katarzyna},
        title = "{Analysis of the Full Spitzer Microlensing Sample. I. Dark Remnant Candidates and Gaia Predictions}",
      journal = {\apj},
         year = 2024,
        month = nov,
       volume = {975},
       number = {2},
          eid = {216},
        pages = {216},
          doi = {10.3847/1538-4357/ad7bb1},
archivePrefix = {arXiv},
       eprint = {2407.13740},
 primaryClass = {astro-ph.GA},
       adsurl = {https://ui.adsabs.harvard.edu/abs/2024ApJ...975..216R}
}

@ARTICLE{GaiaCollaboration2023,
       author = {{Gaia Collaboration} and {Vallenari}, A. and {Brown}, A.~G.~A. and {Prusti}, T. and {de Bruijne}, J.~H.~J. and {Arenou}, F. and {Babusiaux}, C. and {Biermann}, M. and {Creevey}, O.~L. and {Ducourant}, C. and {Evans}, D.~W. and {Eyer}, L. and {Guerra}, R. and {Hutton}, A. and {Jordi}, C. and {Klioner}, S.~A. and {Lammers}, U.~L. and {Lindegren}, L. and {Luri}, X. and {Mignard}, F. and {Panem}, C. and {Pourbaix}, D. and {Randich}, S. and {Sartoretti}, P. and {Soubiran}, C. and {Tanga}, P. and {Walton}, N.~A. and {Bailer-Jones}, C.~A.~L. and {Bastian}, U. and {Drimmel}, R. and {Jansen}, F. and {Katz}, D. and {Lattanzi}, M.~G. and {van Leeuwen}, F. and {Bakker}, J. and {Cacciari}, C. and {Casta{\~n}eda}, J. and {De Angeli}, F. and {Fabricius}, C. and {Fouesneau}, M. and {Fr{\'e}mat}, Y. and {Galluccio}, L. and {Guerrier}, A. and {Heiter}, U. and {Masana}, E. and {Messineo}, R. and {Mowlavi}, N. and {Nicolas}, C. and {Nienartowicz}, K. and {Pailler}, F. and {Panuzzo}, P. and {Riclet}, F. and {Roux}, W. and {Seabroke}, G.~M. and {Sordo}, R. and {Th{\'e}venin}, F. and {Gracia-Abril}, G. and {Portell}, J. and {Teyssier}, D. and {Altmann}, M. and {Andrae}, R. and {Audard}, M. and {Bellas-Velidis}, I. and {Benson}, K. and {Berthier}, J. and {Blomme}, R. and {Burgess}, P.~W. and {Busonero}, D. and {Busso}, G. and {C{\'a}novas}, H. and {Carry}, B. and {Cellino}, A. and {Cheek}, N. and {Clementini}, G. and {Damerdji}, Y. and {Davidson}, M. and {de Teodoro}, P. and {Nu{\~n}ez Campos}, M. and {Delchambre}, L. and {Dell'Oro}, A. and {Esquej}, P. and {Fern{\'a}ndez-Hern{\'a}ndez}, J. and {Fraile}, E. and {Garabato}, D. and {Garc{\'\i}a-Lario}, P. and {Gosset}, E. and {Haigron}, R. and {Halbwachs}, J.-L. and {Hambly}, N.~C. and {Harrison}, D.~L. and {Hern{\'a}ndez}, J. and {Hestroffer}, D. and {Hodgkin}, S.~T. and {Holl}, B. and {Jan{\ss}en}, K. and {Jevardat de Fombelle}, G. and {Jordan}, S. and {Krone-Martins}, A. and {Lanzafame}, A.~C. and {L{\"o}ffler}, W. and {Marchal}, O. and {Marrese}, P.~M. and {Moitinho}, A. and {Muinonen}, K. and {Osborne}, P. and {Pancino}, E. and {Pauwels}, T. and {Recio-Blanco}, A. and {Reyl{\'e}}, C. and {Riello}, M. and {Rimoldini}, L. and {Roegiers}, T. and {Rybizki}, J. and {Sarro}, L.~M. and {Siopis}, C. and {Smith}, M. and {Sozzetti}, A. and {Utrilla}, E. and {van Leeuwen}, M. and {Abbas}, U. and {{\'A}brah{\'a}m}, P. and {Abreu Aramburu}, A. and {Aerts}, C. and {Aguado}, J.~J. and {Ajaj}, M. and {Aldea-Montero}, F. and {Altavilla}, G. and {{\'A}lvarez}, M.~A. and {Alves}, J. and {Anders}, F. and {Anderson}, R.~I. and {Anglada Varela}, E. and {Antoja}, T. and {Baines}, D. and {Baker}, S.~G. and {Balaguer-N{\'u}{\~n}ez}, L. and {Balbinot}, E. and {Balog}, Z. and {Barache}, C. and {Barbato}, D. and {Barros}, M. and {Barstow}, M.~A. and {Bartolom{\'e}}, S. and {Bassilana}, J.-L. and {Bauchet}, N. and {Becciani}, U. and {Bellazzini}, M. and {Berihuete}, A. and {Bernet}, M. and {Bertone}, S. and {Bianchi}, L. and {Binnenfeld}, A. and {Blanco-Cuaresma}, S. and {Blazere}, A. and {Boch}, T. and {Bombrun}, A. and {Bossini}, D. and {Bouquillon}, S. and {Bragaglia}, A. and {Bramante}, L. and {Breedt}, E. and {Bressan}, A. and {Brouillet}, N. and {Brugaletta}, E. and {Bucciarelli}, B. and {Burlacu}, A. and {Butkevich}, A.~G. and {Buzzi}, R. and {Caffau}, E. and {Cancelliere}, R. and {Cantat-Gaudin}, T. and {Carballo}, R. and {Carlucci}, T. and {Carnerero}, M.~I. and {Carrasco}, J.~M. and {Casamiquela}, L. and {Castellani}, M. and {Castro-Ginard}, A. and {Chaoul}, L. and {Charlot}, P. and {Chemin}, L. and {Chiaramida}, V. and {Chiavassa}, A. and {Chornay}, N. and {Comoretto}, G. and {Contursi}, G. and {Cooper}, W.~J. and {Cornez}, T. and {Cowell}, S. and {Crifo}, F. and {Cropper}, M. and {Crosta}, M. and {Crowley}, C. and {Dafonte}, C. and {Dapergolas}, A. and {David}, M. and {David}, P. and {de Laverny}, P. and {De Luise}, F. and {De March}, R.},
        title = "{Gaia Data Release 3. Summary of the content and survey properties}",
      journal = {\aap},
         year = 2023,
        month = jun,
       volume = {674},
          eid = {A1},
        pages = {A1},
          doi = {10.1051/0004-6361/202243940},
archivePrefix = {arXiv},
       eprint = {2208.00211},
 primaryClass = {astro-ph.GA},
       adsurl = {https://ui.adsabs.harvard.edu/abs/2023A&A...674A...1G}
}

@ARTICLE{Freeman2023,
       author = {{Freeman}, Matthew S.~R. and {Lu}, Jessica R. and {Lyke}, Jim and {Gautam}, Abhimat and {Kupke}, Renate and {Ghez}, Andrea and {Sakai}, Shoko and {Anderson}, Jay and {Bellini}, Andrea},
        title = "{An Optical Distortion Solution for the Keck I OSIRIS Imager}",
      journal = {\aj},
         year = 2023,
        month = sep,
       volume = {166},
       number = {3},
          eid = {125},
        pages = {125},
          doi = {10.3847/1538-3881/aceaf7},
       adsurl = {https://ui.adsabs.harvard.edu/abs/2023AJ....166..125F}
}

@software{lu_2021_6677744,
  author       = {Lu, Jessica R. and Gautam, Abhimat K. and Chu, Devin and Terry, Sean K. and Do, Tuan},
  title        = {Keck-DataReductionPipelines/KAI: v1.0.0 Release of KAI},
  month        = mar,
  year         = 2021,
  publisher    = {Zenodo},
  version      = {v1.0.0},
  doi          = {10.5281/zenodo.6677744},
  url          = {https://doi.org/10.5281/zenodo.6677744},
}

@ARTICLE{Sahu2025,
       author = {{Sahu}, Kailash C. and {Anderson}, Jay and {Casertano}, Stefano and {Bond}, Howard E. and {Dominik}, Martin and {Calamida}, Annalisa and {Bellini}, Andrea and {Brown}, Thomas M. and {Ferguson}, Henry C. and {Rejkuba}, Marina},
        title = "{OGLE-2011-BLG-0462: An Isolated Stellar-mass Black Hole Confirmed Using New HST Astrometry and Updated Photometry}",
      journal = {\apj},
         year = 2025,
        month = apr,
       volume = {983},
       number = {2},
          eid = {104},
        pages = {104},
          doi = {10.3847/1538-4357/adbe6e},
archivePrefix = {arXiv},
       eprint = {2503.07820},
 primaryClass = {astro-ph.SR},
       adsurl = {https://ui.adsabs.harvard.edu/abs/2025ApJ...983..104S}
}

@ARTICLE{Lam2023,
       author = {{Lam}, Casey Y. and {Lu}, Jessica R.},
        title = "{A Reanalysis of the Isolated Black Hole Candidate OGLE-2011-BLG-0462/MOA-2011-BLG-191}",
      journal = {\apj},
         year = 2023,
        month = oct,
       volume = {955},
       number = {2},
          eid = {116},
        pages = {116},
          doi = {10.3847/1538-4357/aced4a},
archivePrefix = {arXiv},
       eprint = {2308.03302},
 primaryClass = {astro-ph.SR},
       adsurl = {https://ui.adsabs.harvard.edu/abs/2023ApJ...955..116L}
}

@ARTICLE{Surot2020,
       author = {{Surot}, F. and {Valenti}, E. and {Gonzalez}, O.~A. and {Zoccali}, M. and {S{\"o}kmen}, E. and {Hidalgo}, S.~L. and {Minniti}, D.},
        title = "{Mapping the stellar age of the Milky Way bulge with the VVV. III. High-resolution reddening map}",
      journal = {\aap},
         year = 2020,
        month = dec,
       volume = {644},
          eid = {A140},
        pages = {A140},
          doi = {10.1051/0004-6361/202038346},
archivePrefix = {arXiv},
       eprint = {2010.02723},
 primaryClass = {astro-ph.GA},
       adsurl = {https://ui.adsabs.harvard.edu/abs/2020A&A...644A.140S}
}

@ARTICLE{Choi2016,
       author = {{Choi}, Jieun and {Dotter}, Aaron and {Conroy}, Charlie and {Cantiello}, Matteo and {Paxton}, Bill and {Johnson}, Benjamin D.},
        title = "{Mesa Isochrones and Stellar Tracks (MIST). I. Solar-scaled Models}",
      journal = {\apj},
         year = 2016,
        month = jun,
       volume = {823},
       number = {2},
          eid = {102},
        pages = {102},
          doi = {10.3847/0004-637X/823/2/102},
archivePrefix = {arXiv},
       eprint = {1604.08592},
 primaryClass = {astro-ph.SR},
       adsurl = {https://ui.adsabs.harvard.edu/abs/2016ApJ...823..102C}
}

@ARTICLE{Dotter2016,
       author = {{Dotter}, Aaron},
        title = "{MESA Isochrones and Stellar Tracks (MIST) 0: Methods for the Construction of Stellar Isochrones}",
      journal = {\apjs},
         year = 2016,
        month = jan,
       volume = {222},
       number = {1},
          eid = {8},
        pages = {8},
          doi = {10.3847/0067-0049/222/1/8},
archivePrefix = {arXiv},
       eprint = {1601.05144},
 primaryClass = {astro-ph.SR},
       adsurl = {https://ui.adsabs.harvard.edu/abs/2016ApJS..222....8D}
}

@ARTICLE{Sumi2011,
       author = {{Sumi}, T. and {Kamiya}, K. and {Bennett}, D.~P. and {Bond}, I.~A. and {Abe}, F. and {Botzler}, C.~S. and {Fukui}, A. and {Furusawa}, K. and {Hearnshaw}, J.~B. and {Itow}, Y. and {Kilmartin}, P.~M. and {Korpela}, A. and {Lin}, W. and {Ling}, C.~H. and {Masuda}, K. and {Matsubara}, Y. and {Miyake}, N. and {Motomura}, M. and {Muraki}, Y. and {Nagaya}, M. and {Nakamura}, S. and {Ohnishi}, K. and {Okumura}, T. and {Perrott}, Y.~C. and {Rattenbury}, N. and {Saito}, To. and {Sako}, T. and {Sullivan}, D.~J. and {Sweatman}, W.~L. and {Tristram}, P.~J. and {Udalski}, A. and {Szyma{\'n}ski}, M.~K. and {Kubiak}, M. and {Pietrzy{\'n}ski}, G. and {Poleski}, R. and {Soszy{\'n}ski}, I. and {Wyrzykowski}, {\L}. and {Ulaczyk}, K. and {Microlensing Observations in Astrophysics (MOA) Collaboration}},
        title = "{Unbound or distant planetary mass population detected by gravitational microlensing}",
      journal = {\nat},
         year = 2011,
        month = may,
       volume = {473},
       number = {7347},
        pages = {349-352},
          doi = {10.1038/nature10092},
archivePrefix = {arXiv},
       eprint = {1105.3544},
 primaryClass = {astro-ph.EP},
       adsurl = {https://ui.adsabs.harvard.edu/abs/2011Natur.473..349S}
}

@ARTICLE{Mroz2022,
       author = {{Mr{\'o}z}, Przemek and {Udalski}, Andrzej and {Gould}, Andrew},
        title = "{Systematic Errors as a Source of Mass Discrepancy in Black Hole Microlensing Event OGLE-2011-BLG-0462}",
      journal = {\apjl},
         year = 2022,
        month = oct,
       volume = {937},
       number = {2},
          eid = {L24},
        pages = {L24},
          doi = {10.3847/2041-8213/ac90bb},
archivePrefix = {arXiv},
       eprint = {2207.10729},
 primaryClass = {astro-ph.SR},
       adsurl = {https://ui.adsabs.harvard.edu/abs/2022ApJ...937L..24M}
}

@ARTICLE{Lam2022,
       author = {{Lam}, Casey Y. and {Lu}, Jessica R. and {Udalski}, Andrzej and {Bond}, Ian and {Bennett}, David P. and {Skowron}, Jan and {Mr{\'o}z}, Przemek and {Poleski}, Radek and {Sumi}, Takahiro and {Szyma{\'n}ski}, Micha{\l} K. and {Koz{\l}owski}, Szymon and {Pietrukowicz}, Pawe{\l} and {Soszy{\'n}ski}, Igor and {Ulaczyk}, Krzysztof and {Wyrzykowski}, {\L}ukasz and {Miyazaki}, Shota and {Suzuki}, Daisuke and {Koshimoto}, Naoki and {Rattenbury}, Nicholas J. and {Hosek}, Matthew W. and {Abe}, Fumio and {Barry}, Richard and {Bhattacharya}, Aparna and {Fukui}, Akihiko and {Fujii}, Hirosane and {Hirao}, Yuki and {Itow}, Yoshitaka and {Kirikawa}, Rintaro and {Kondo}, Iona and {Matsubara}, Yutaka and {Matsumoto}, Sho and {Muraki}, Yasushi and {Olmschenk}, Greg and {Ranc}, Cl{\'e}ment and {Okamura}, Arisa and {Satoh}, Yuki and {Silva}, Stela Ishitani and {Toda}, Taiga and {Tristram}, Paul J. and {Vandorou}, Aikaterini and {Yama}, Hibiki and {Abrams}, Natasha S. and {Agarwal}, Shrihan and {Rose}, Sam and {Terry}, Sean K.},
        title = "{An Isolated Mass-gap Black Hole or Neutron Star Detected with Astrometric Microlensing}",
      journal = {\apjl},
         year = 2022,
        month = jul,
       volume = {933},
       number = {1},
          eid = {L23},
        pages = {L23},
          doi = {10.3847/2041-8213/ac7442},
archivePrefix = {arXiv},
       eprint = {2202.01903},
 primaryClass = {astro-ph.GA},
       adsurl = {https://ui.adsabs.harvard.edu/abs/2022ApJ...933L..23L}
}

@ARTICLE{Sahu2022,
       author = {{Sahu}, Kailash C. and {Anderson}, Jay and {Casertano}, Stefano and {Bond}, Howard E. and {Udalski}, Andrzej and {Dominik}, Martin and {Calamida}, Annalisa and {Bellini}, Andrea and {Brown}, Thomas M. and {Rejkuba}, Marina and {Bajaj}, Varun and {Kains}, No{\'e} and {Ferguson}, Henry C. and {Fryer}, Chris L. and {Yock}, Philip and {Mr{\'o}z}, Przemek and {Koz{\l}owski}, Szymon and {Pietrukowicz}, Pawe{\l} and {Poleski}, Radek and {Skowron}, Jan and {Soszy{\'n}ski}, Igor and {Szyma{\'n}ski}, Micha{\l} K. and {Ulaczyk}, Krzysztof and {Wyrzykowski}, {\L}ukasz and {Barry}, Richard K. and {Bennett}, David P. and {Bond}, Ian A. and {Hirao}, Yuki and {Silva}, Stela Ishitani and {Kondo}, Iona and {Koshimoto}, Naoki and {Ranc}, Cl{\'e}ment and {Rattenbury}, Nicholas J. and {Sumi}, Takahiro and {Suzuki}, Daisuke and {Tristram}, Paul J. and {Vandorou}, Aikaterini and {Beaulieu}, Jean-Philippe and {Marquette}, Jean-Baptiste and {Cole}, Andrew and {Fouqu{\'e}}, Pascal and {Hill}, Kym and {Dieters}, Stefan and {Coutures}, Christian and {Dominis-Prester}, Dijana and {Bennett}, Clara and {Bachelet}, Etienne and {Menzies}, John and {Albrow}, Michael and {Pollard}, Karen and {Gould}, Andrew and {Yee}, Jennifer C. and {Allen}, William and {Almeida}, Leonardo A. and {Christie}, Grant and {Drummond}, John and {Gal-Yam}, Avishay and {Gorbikov}, Evgeny and {Jablonski}, Francisco and {Lee}, Chung-Uk and {Maoz}, Dan and {Manulis}, Ilan and {McCormick}, Jennie and {Natusch}, Tim and {Pogge}, Richard W. and {Shvartzvald}, Yossi and {J{\o}rgensen}, Uffe G. and {Alsubai}, Khalid A. and {Andersen}, Michael I. and {Bozza}, Valerio and {Novati}, Sebastiano Calchi and {Burgdorf}, Martin and {Hinse}, Tobias C. and {Hundertmark}, Markus and {Husser}, Tim-Oliver and {Kerins}, Eamonn and {Longa-Pe{\~n}a}, Penelope and {Mancini}, Luigi and {Penny}, Matthew and {Rahvar}, Sohrab and {Ricci}, Davide and {Sajadian}, Sedighe and {Skottfelt}, Jesper and {Snodgrass}, Colin and {Southworth}, John and {Tregloan-Reed}, Jeremy and {Wambsganss}, Joachim and {Wertz}, Olivier and {Tsapras}, Yiannis and {Street}, Rachel A. and {Bramich}, D.~M. and {Horne}, Keith and {Steele}, Iain A. and {RoboNet Collaboration}},
        title = "{An Isolated Stellar-mass Black Hole Detected through Astrometric Microlensing}",
      journal = {\apj},
         year = 2022,
        month = jul,
       volume = {933},
       number = {1},
          eid = {83},
        pages = {83},
          doi = {10.3847/1538-4357/ac739e},
archivePrefix = {arXiv},
       eprint = {2201.13296},
 primaryClass = {astro-ph.SR},
       adsurl = {https://ui.adsabs.harvard.edu/abs/2022ApJ...933...83S}
}

@ARTICLE{GaiaCollaboration2024,
       author = {{Gaia Collaboration} and {Panuzzo}, P. and {Mazeh}, T. and {Arenou}, F. and {Holl}, B. and {Caffau}, E. and {Jorissen}, A. and {Babusiaux}, C. and {Gavras}, P. and {Sahlmann}, J. and {Bastian}, U. and {Wyrzykowski}, {\L}. and {Eyer}, L. and {Leclerc}, N. and {Bauchet}, N. and {Bombrun}, A. and {Mowlavi}, N. and {Seabroke}, G.~M. and {Teyssier}, D. and {Balbinot}, E. and {Helmi}, A. and {Brown}, A.~G.~A. and {Vallenari}, A. and {Prusti}, T. and {de Bruijne}, J.~H.~J. and {Barbier}, A. and {Biermann}, M. and {Creevey}, O.~L. and {Ducourant}, C. and {Evans}, D.~W. and {Guerra}, R. and {Hutton}, A. and {Jordi}, C. and {Klioner}, S.~A. and {Lammers}, U. and {Lindegren}, L. and {Luri}, X. and {Mignard}, F. and {Nicolas}, C. and {Randich}, S. and {Sartoretti}, P. and {Smiljanic}, R. and {Tanga}, P. and {Walton}, N.~A. and {Aerts}, C. and {Bailer-Jones}, C.~A.~L. and {Cropper}, M. and {Drimmel}, R. and {Jansen}, F. and {Katz}, D. and {Lattanzi}, M.~G. and {Soubiran}, C. and {Th{\'e}venin}, F. and {van Leeuwen}, F. and {Andrae}, R. and {Audard}, M. and {Bakker}, J. and {Blomme}, R. and {Casta{\~n}eda}, J. and {De Angeli}, F. and {Fabricius}, C. and {Fouesneau}, M. and {Fr{\'e}mat}, Y. and {Galluccio}, L. and {Guerrier}, A. and {Heiter}, U. and {Masana}, E. and {Messineo}, R. and {Nienartowicz}, K. and {Pailler}, F. and {Riclet}, F. and {Roux}, W. and {Sordo}, R. and {Gracia-Abril}, G. and {Portell}, J. and {Altmann}, M. and {Benson}, K. and {Berthier}, J. and {Burgess}, P.~W. and {Busonero}, D. and {Busso}, G. and {Cacciari}, C. and {C{\'a}novas}, H. and {Carrasco}, J.~M. and {Carry}, B. and {Cellino}, A. and {Cheek}, N. and {Clementini}, G. and {Damerdji}, Y. and {Davidson}, M. and {de Teodoro}, P. and {Delchambre}, L. and {Dell'Oro}, A. and {Fraile Garcia}, E. and {Garabato}, D. and {Garc{\'\i}a-Lario}, P. and {Haigron}, R. and {Hambly}, N.~C. and {Harrison}, D.~L. and {Hatzidimitriou}, D. and {Hern{\'a}ndez}, J. and {Hestroffer}, D. and {Hodgkin}, S.~T. and {Jamal}, S. and {Jevardat de Fombelle}, G. and {Jordan}, S. and {Krone-Martins}, A. and {Lanzafame}, A.~C. and {L{\"o}ffler}, W. and {Lorca}, A. and {Marchal}, O. and {Marrese}, P.~M. and {Moitinho}, A. and {Muinonen}, K. and {Nu{\~n}ez Campos}, M. and {Oreshina-Slezak}, I. and {Osborne}, P. and {Pancino}, E. and {Pauwels}, T. and {Recio-Blanco}, A. and {Riello}, M. and {Rimoldini}, L. and {Robin}, A.~C. and {Roegiers}, T. and {Sarro}, L.~M. and {Schultheis}, M. and {Smith}, M. and {Sozzetti}, A. and {Utrilla}, E. and {van Leeuwen}, M. and {Weingrill}, K. and {Abbas}, U. and {{\'A}brah{\'a}m}, P. and {Abreu Aramburu}, A. and {Ahmed}, S. and {Altavilla}, G. and {{\'A}lvarez}, M.~A. and {Anders}, F. and {Anderson}, R.~I. and {Anglada Varela}, E. and {Antoja}, T. and {Baig}, S. and {Baines}, D. and {Baker}, S.~G. and {Balaguer-N{\'u}{\~n}ez}, L. and {Balog}, Z. and {Barache}, C. and {Barros}, M. and {Barstow}, M.~A. and {Bartolom{\'e}}, S. and {Bashi}, D. and {Bassilana}, J.-L. and {Baudeau}, N. and {Becciani}, U. and {Bedin}, L.~R. and {Bellas-Velidis}, I. and {Bellazzini}, M. and {Beordo}, W. and {Bernet}, M. and {Bertolotto}, C. and {Bertone}, S. and {Bianchi}, L. and {Binnenfeld}, A. and {Blanco-Cuaresma}, S. and {Bland-Hawthorn}, J. and {Blazere}, A. and {Boch}, T. and {Bossini}, D. and {Bouquillon}, S. and {Bragaglia}, A. and {Braine}, J. and {Bratsolis}, E. and {Breedt}, E. and {Bressan}, A. and {Brouillet}, N. and {Brugaletta}, E. and {Bucciarelli}, B. and {Butkevich}, A.~G. and {Buzzi}, R. and {Camut}, A. and {Cancelliere}, R. and {Cantat-Gaudin}, T. and {Capilla Guilarte}, D. and {Carballo}, R. and {Carlucci}, T. and {Carnerero}, M.~I. and {Carretero}, J. and {Carton}, S. and {Casamiquela}, L. and {Casey}, A. and {Castellani}, M. and {Castro-Ginard}, A. and {Ceraj}, L. and {Cesare}, V. and {Charlot}, P. and {Chaudet}, C. and {Chemin}, L. and {Chiavassa}, A. and {Chornay}, N. and {Chosson}, D.},
        title = "{Discovery of a dormant 33 solar-mass black hole in pre-release Gaia astrometry}",
      journal = {\aap},
         year = 2024,
        month = jun,
       volume = {686},
          eid = {L2},
        pages = {L2},
          doi = {10.1051/0004-6361/202449763},
archivePrefix = {arXiv},
       eprint = {2404.10486},
 primaryClass = {astro-ph.GA},
       adsurl = {https://ui.adsabs.harvard.edu/abs/2024A&A...686L...2G}
}

@ARTICLE{Chakrabarti2023,
       author = {{Chakrabarti}, Sukanya and {Simon}, Joshua D. and {Craig}, Peter A. and {Reggiani}, Henrique and {Brandt}, Timothy D. and {Guhathakurta}, Puragra and {Dalba}, Paul A. and {Kirby}, Evan N. and {Chang}, Philip and {Hey}, Daniel R. and {Savino}, Alessandro and {Geha}, Marla and {Thompson}, Ian B.},
        title = "{A Noninteracting Galactic Black Hole Candidate in a Binary System with a Main-sequence Star}",
      journal = {\aj},
         year = 2023,
        month = jul,
       volume = {166},
       number = {1},
          eid = {6},
        pages = {6},
          doi = {10.3847/1538-3881/accf21},
archivePrefix = {arXiv},
       eprint = {2210.05003},
 primaryClass = {astro-ph.GA},
       adsurl = {https://ui.adsabs.harvard.edu/abs/2023AJ....166....6C}
}

@ARTICLE{El-Badry2023a,
       author = {{El-Badry}, Kareem and {Rix}, Hans-Walter and {Cendes}, Yvette and {Rodriguez}, Antonio C. and {Conroy}, Charlie and {Quataert}, Eliot and {Hawkins}, Keith and {Zari}, Eleonora and {Hobson}, Melissa and {Breivik}, Katelyn and {Rau}, Arne and {Berger}, Edo and {Shahaf}, Sahar and {Seeburger}, Rhys and {Burdge}, Kevin B. and {Latham}, David W. and {Buchhave}, Lars A. and {Bieryla}, Allyson and {Bashi}, Dolev and {Mazeh}, Tsevi and {Faigler}, Simchon},
        title = "{A red giant orbiting a black hole}",
      journal = {\mnras},
         year = 2023,
        month = may,
       volume = {521},
       number = {3},
        pages = {4323-4348},
          doi = {10.1093/mnras/stad799},
archivePrefix = {arXiv},
       eprint = {2302.07880},
 primaryClass = {astro-ph.SR},
       adsurl = {https://ui.adsabs.harvard.edu/abs/2023MNRAS.521.4323E}
}

@ARTICLE{El-Badry2023b,
       author = {{El-Badry}, Kareem and {Rix}, Hans-Walter and {Quataert}, Eliot and {Howard}, Andrew W. and {Isaacson}, Howard and {Fuller}, Jim and {Hawkins}, Keith and {Breivik}, Katelyn and {Wong}, Kaze W.~K. and {Rodriguez}, Antonio C. and {Conroy}, Charlie and {Shahaf}, Sahar and {Mazeh}, Tsevi and {Arenou}, Fr{\'e}d{\'e}ric and {Burdge}, Kevin B. and {Bashi}, Dolev and {Faigler}, Simchon and {Weisz}, Daniel R. and {Seeburger}, Rhys and {Almada Monter}, Silvia and {Wojno}, Jennifer},
        title = "{A Sun-like star orbiting a black hole}",
      journal = {\mnras},
         year = 2023,
        month = jan,
       volume = {518},
       number = {1},
        pages = {1057-1085},
          doi = {10.1093/mnras/stac3140},
archivePrefix = {arXiv},
       eprint = {2209.06833},
 primaryClass = {astro-ph.SR},
       adsurl = {https://ui.adsabs.harvard.edu/abs/2023MNRAS.518.1057E}
}

@ARTICLE{Tanikawa2023,
       author = {{Tanikawa}, Ataru and {Hattori}, Kohei and {Kawanaka}, Norita and {Kinugawa}, Tomoya and {Shikauchi}, Minori and {Tsuna}, Daichi},
        title = "{Search for a Black Hole Binary in Gaia DR3 Astrometric Binary Stars with Spectroscopic Data}",
      journal = {\apj},
         year = 2023,
        month = apr,
       volume = {946},
       number = {2},
          eid = {79},
        pages = {79},
          doi = {10.3847/1538-4357/acbf36},
archivePrefix = {arXiv},
       eprint = {2209.05632},
 primaryClass = {astro-ph.SR},
       adsurl = {https://ui.adsabs.harvard.edu/abs/2023ApJ...946...79T}
}

@ARTICLE{TheLIGOScientificCollaboration2025d,
       author = {{The LIGO Scientific Collaboration} and {the Virgo Collaboration} and {the KAGRA Collaboration} and {Abac}, A.~G. and {Abouelfettouh}, I. and {Acernese}, F. and {Ackley}, K. and {Adamcewicz}, C. and {Adhicary}, S. and {Adhikari}, D. and {Adhikari}, N. and {Adhikari}, R.~X. and {Adkins}, V.~K. and {Afroz}, S. and {Agarwal}, D. and {Agathos}, M. and {Aghaei Abchouyeh}, M. and {Aguiar}, O.~D. and {Ahmadzadeh}, S. and {Aiello}, L. and {Ain}, A. and {Ajith}, P. and {Akutsu}, T. and {Albanesi}, S. and {Alfaidi}, R.~A. and {Al-Jodah}, A. and {All{\'e}n{\'e}}, C. and {Allocca}, A. and {Al-Shammari}, S. and {Altin}, P.~A. and {Alvarez-Lopez}, S. and {Amarasinghe}, O. and {Amato}, A. and {Amra}, C. and {Ananyeva}, A. and {Anderson}, S.~B. and {Anderson}, W.~G. and {Andia}, M. and {Ando}, M. and {Andrade}, T. and {Andr{\'e}s-Carcasona}, M. and {Andri{\'c}}, T. and {Anglin}, J. and {Ansoldi}, S. and {Antelis}, J.~M. and {Antier}, S. and {Aoumi}, M. and {Appavuravther}, E.~Z. and {Appert}, S. and {Apple}, S.~K. and {Arai}, K. and {Araya}, A. and {Araya}, M.~C. and {Arca Sedda}, M. and {Areeda}, J.~S. and {Argianas}, L. and {Aritomi}, N. and {Armato}, F. and {Armstrong}, S. and {Arnaud}, N. and {Arogeti}, M. and {Aronson}, S.~M. and {Arun}, K.~G. and {Ashton}, G. and {Aso}, Y. and {Assiduo}, M. and {Assis de Souza Melo}, S. and {Aston}, S.~M. and {Astone}, P. and {Attadio}, F. and {Aubin}, F. and {AultONeal}, K. and {Avallone}, G. and {Babak}, S. and {Badaracco}, F. and {Badger}, C. and {Bae}, S. and {Bagnasco}, S. and {Bagui}, E. and {Baiotti}, L. and {Bajpai}, R. and {Baka}, T. and {Baker}, T. and {Ball}, M. and {Ballardin}, G. and {Ballmer}, S.~W. and {Banagiri}, S. and {Banerjee}, B. and {Bankar}, D. and {Baptiste}, T.~M. and {Baral}, P. and {Barayoga}, J.~C. and {Barish}, B.~C. and {Barker}, D. and {Barman}, N. and {Barneo}, P. and {Barone}, F. and {Barr}, B. and {Barsotti}, L. and {Barsuglia}, M. and {Barta}, D. and {Bartoletti}, A.~M. and {Barton}, M.~A. and {Bartos}, I. and {Basak}, S. and {Basalaev}, A. and {Bassiri}, R. and {Basti}, A. and {Bates}, D.~E. and {Bawaj}, M. and {Baxi}, P. and {Bayley}, J.~C. and {Baylor}, A.~C. and {Baynard}, II, P.~A. and {Bazzan}, M. and {Bedakihale}, V.~M. and {Beirnaert}, F. and {Bejger}, M. and {Belardinelli}, D. and {Bell}, A.~S. and {Bellie}, D.~S. and {Bellizzi}, L. and {Beltran-Martinez}, D. and {Benoit}, W. and {Bentara}, I. and {Bentley}, J.~D. and {Ben Yaala}, M. and {Bera}, S. and {Bergamin}, F. and {Berger}, B.~K. and {Bernuzzi}, S. and {Beroiz}, M. and {Berry}, C.~P.~L. and {Bersanetti}, D. and {Bertolini}, A. and {Betzwieser}, J. and {Beveridge}, D. and {Bevilacqua}, G. and {Bevins}, N. and {Bhandare}, R. and {Bhatt}, R. and {Bhattacharjee}, D. and {Bhaumik}, S. and {Bhowmick}, S. and {Biancalana}, V. and {Bianchi}, A. and {Bilenko}, I.~A. and {Billingsley}, G. and {Binetti}, A. and {Bini}, S. and {Binu}, C. and {Birnholtz}, O. and {Biscoveanu}, S. and {Bisht}, A. and {Bitossi}, M. and {Bizouard}, M. -A. and {Blaber}, S. and {Blackburn}, J.~K. and {Blagg}, L.~A. and {Blair}, C.~D. and {Blair}, D.~G. and {Bobba}, F. and {Bode}, N. and {Boileau}, G. and {Boldrini}, M. and {Bolingbroke}, G.~N. and {Bolliand}, A. and {Bonavena}, L.~D. and {Bondarescu}, R. and {Bondu}, F. and {Bonilla}, E. and {Bonilla}, M.~S. and {Bonino}, A. and {Bonnand}, R. and {Booker}, P. and {Borchers}, A. and {Borhanian}, S. and {Boschi}, V. and {Bose}, S. and {Bossilkov}, V. and {Boudon}, A. and {Bozzi}, A. and {Bradaschia}, C. and {Brady}, P.~R. and {Branch}, A. and {Branchesi}, M. and {Braun}, I. and {Briant}, T. and {Brillet}, A. and {Brinkmann}, M. and {Brockill}, P. and {Brockmueller}, E. and {Brooks}, A.~F. and {Brown}, B.~C. and {Brown}, D.~D. and {Brozzetti}, M.~L. and {Brunett}, S. and {Bruno}, G. and {Bruntz}, R. and {Bryant}, J.},
        title = "{GWTC-4.0: Population Properties of Merging Compact Binaries}",
      journal = {arXiv e-prints},
         year = 2025,
        month = aug,
          eid = {arXiv:2508.18083},
        pages = {arXiv:2508.18083},
          doi = {10.48550/arXiv.2508.18083},
archivePrefix = {arXiv},
       eprint = {2508.18083},
 primaryClass = {astro-ph.HE},
       adsurl = {https://ui.adsabs.harvard.edu/abs/2025arXiv250818083T}
}

@ARTICLE{Agol:2002,
   author = {{Agol}, E. and {Kamionkowski}, M.},
    title = "{X-rays from isolated black holes in the Milky Way}",
  journal = {\mnras},
   eprint = {astro-ph/0109539},
     year = 2002,
    month = aug,
   volume = 334,
    pages = {553-562},
      doi = {10.1046/j.1365-8711.2002.05523.x},
   adsurl = {http://adsabs.harvard.edu/abs/2002MNRAS.334..553A}
}

@ARTICLE{CalchiNovati:2015,
       author = {{Calchi Novati}, S. and {Gould}, A. and {Yee}, J.~C. and {Beichman}, C. and
         {Bryden}, G. and {Carey}, S. and {Fausnaugh}, M. and {Gaudi}, B.~S. and
         {Henderson}, C.~B. and {Pogge}, R.~W. and {Shvartzvald}, Y. and
         {Wibking}, B. and {Zhu}, W. and {Spitzer Team} and {Udalski}, A. and
         {Poleski}, R. and {Pawlak}, M. and {Szyma{\'n}ski}, M.~K. and
         {Skowron}, J. and {Mr{\'o}z}, P. and {Koz{\l}owski}, S. and
         {Wyrzykowski}, {\L}. and {Pietrukowicz}, P. and {Pietrzy{\'n}ski}, G. and
         {Soszy{\'n}ski}, I. and {Ulaczyk}, K. and {OGLE Group}},
        title = "{Spitzer IRAC Photometry for Time Series in Crowded Fields}",
      journal = {\apj},
         year = "2015",
        month = "Dec",
       volume = {814},
       number = {2},
          eid = {92},
        pages = {92},
          doi = {10.1088/0004-637X/814/2/92},
archivePrefix = {arXiv},
       eprint = {1509.00037},
 primaryClass = {astro-ph.EP},
       adsurl = {https://ui.adsabs.harvard.edu/abs/2015ApJ...814...92C}
}

@ARTICLE{Casares:2014,
   author = {{Casares}, J. and {Jonker}, P.~G.},
    title = "{Mass Measurements of Stellar and Intermediate-Mass Black Holes}",
  journal = {\ssr},
archivePrefix = "arXiv",
   eprint = {1311.5118},
 primaryClass = "astro-ph.HE",
     year = 2014,
    month = sep,
   volume = 183,
    pages = {223-252},
      doi = {10.1007/s11214-013-0030-6},
   adsurl = {http://adsabs.harvard.edu/abs/2014SSRv..183..223C}
}

@ARTICLE{Diolaiti:2000,
   author = {{Diolaiti}, E. and {Bendinelli}, O. and {Bonaccini}, D. and 
	{Close}, L. and {Currie}, D. and {Parmeggiani}, G.},
    title = "{StarFinder: a code to analyse isoplanatic high-resolution stellar fields.}",
  journal = {The Messenger},
     year = 2000,
    month = jun,
   volume = 100,
    pages = {23-27},
   adsurl = {http://adsabs.harvard.edu/abs/2000Msngr.100...23D}
}

@ARTICLE{DiStefano:1995,
   author = {{Di Stefano}, R. and {Esin}, A.~A.},
    title = "{Blending of Light in Gravitational Microlensing Events}",
  journal = {\apjl},
   eprint = {astro-ph/9506092},
     year = 1995,
    month = jul,
   volume = 448,
    pages = {L1},
      doi = {10.1086/309588},
   adsurl = {http://adsabs.harvard.edu/abs/1995ApJ...448L...1D}
}

@ARTICLE{Dominik:2000,
       author = {{Dominik}, Martin and {Sahu}, Kailash C.},
        title = "{Astrometric Microlensing of Stars}",
      journal = {\apj},
         year = "2000",
        month = "May",
       volume = {534},
       number = {1},
        pages = {213-226},
          doi = {10.1086/308716},
       adsurl = {https://ui.adsabs.harvard.edu/abs/2000ApJ...534..213D}
}

@ARTICLE{Feroz:2009,
       author = {{Feroz}, F. and {Hobson}, M.~P. and {Bridges}, M.},
        title = "{MULTINEST: an efficient and robust Bayesian inference tool for cosmology and particle physics}",
      journal = {\mnras},
         year = "2009",
        month = "Oct",
       volume = {398},
       number = {4},
        pages = {1601-1614},
          doi = {10.1111/j.1365-2966.2009.14548.x},
archivePrefix = {arXiv},
       eprint = {0809.3437},
 primaryClass = {astro-ph},
       adsurl = {https://ui.adsabs.harvard.edu/abs/2009MNRAS.398.1601F}
}

@ARTICLE{Feroz:2013,
       author = {{Feroz}, F. and {Hobson}, M.~P. and {Cameron}, E. and {Pettitt}, A.~N.},
        title = "{Importance Nested Sampling and the MultiNest Algorithm}",
      journal = {arXiv e-prints},
         year = "2013",
        month = "Jun",
          eid = {arXiv:1306.2144},
        pages = {arXiv:1306.2144},
archivePrefix = {arXiv},
       eprint = {1306.2144},
 primaryClass = {astro-ph.IM},
       adsurl = {https://ui.adsabs.harvard.edu/abs/2013arXiv1306.2144F}
}

@ARTICLE{Fruchter:2002,
   author = {{Fruchter}, A.~S. and {Hook}, R.~N.},
    title = "{Drizzle: A Method for the Linear Reconstruction of Undersampled Images}",
  journal = {\pasp},
   eprint = {astro-ph/9808087},
     year = 2002,
    month = feb,
   volume = 114,
    pages = {144-152},
      doi = {10.1086/338393},
   adsurl = {http://adsabs.harvard.edu/abs/2002PASP..114..144F}
}

@ARTICLE{Koshimoto:2019,
       author = {{Koshimoto}, Naoki and {Bennett}, David},
        title = "{Evidence of Systematic Errors in $Spitzer$ Microlens Parallax Measurements}",
      journal = {arXiv e-prints},
         year = 2019,
        month = may,
          eid = {arXiv:1905.05794},
        pages = {arXiv:1905.05794},
archivePrefix = {arXiv},
       eprint = {1905.05794},
 primaryClass = {astro-ph.EP},
       adsurl = {https://ui.adsabs.harvard.edu/abs/2019arXiv190505794K}
}

@ARTICLE{Lam:2020,
       author = {{Lam}, Casey Y. and {Lu}, Jessica R. and {Hosek}, Matthew W., Jr. and
         {Dawson}, William A. and {Golovich}, Nathan R.},
        title = "{PopSyCLE: A New Population Synthesis Code for Compact Object Microlensing Events}",
      journal = {\apj},
         year = "2020",
        month = "Jan",
       volume = {889},
       number = {1},
          eid = {31},
        pages = {31},
          doi = {10.3847/1538-4357/ab5fd3},
archivePrefix = {arXiv},
       eprint = {1912.04510},
 primaryClass = {astro-ph.SR},
       adsurl = {https://ui.adsabs.harvard.edu/abs/2020ApJ...889...31L}
}

@ARTICLE{Lu:2016,
   author = {{Lu}, J.~R. and {Sinukoff}, E. and {Ofek}, E.~O. and {Udalski}, A. and 
	{Kozlowski}, S.},
    title = "{A Search For Stellar-mass Black Holes Via Astrometric Microlensing}",
  journal = {\apj},
archivePrefix = "arXiv",
   eprint = {1607.08284},
 primaryClass = "astro-ph.SR",
     year = 2016,
    month = oct,
   volume = 830,
      eid = {41},
    pages = {41},
      doi = {10.3847/0004-637X/830/1/41},
   adsurl = {http://adsabs.harvard.edu/abs/2016ApJ...830...41L}
}

@ARTICLE{Nelson:2020,
       author = {{Nelson}, Benjamin E. and {Ford}, Eric B. and {Buchner}, Johannes and
         {Cloutier}, Ryan and {D{\'\i}az}, Rodrigo F. and {Faria}, Jo{\~a}o P. and
         {Hara}, Nathan C. and {Rajpaul}, Vinesh M. and {Rukdee}, Surangkhana},
        title = "{Quantifying the Bayesian Evidence for a Planet in Radial Velocity Data}",
      journal = {\aj},
         year = 2020,
        month = feb,
       volume = {159},
       number = {2},
          eid = {73},
        pages = {73},
          doi = {10.3847/1538-3881/ab5190},
archivePrefix = {arXiv},
       eprint = {1806.04683},
 primaryClass = {astro-ph.EP},
       adsurl = {https://ui.adsabs.harvard.edu/abs/2020AJ....159...73N}
}

@ARTICLE{Paczynski:1986a,
   author = {{Paczynski}, B.},
    title = "{Gravitational microlensing at large optical depth}",
  journal = {\apj},
     year = 1986,
    month = feb,
   volume = 301,
    pages = {503-516},
      doi = {10.1086/163919},
   adsurl = {http://adsabs.harvard.edu/abs/1986ApJ...301..503P}
}

@ARTICLE{Service:2016,
   author = {{Service}, M. and {Lu}, J.~R. and {Campbell}, R. and {Sitarski}, B.~N. and 
	{Ghez}, A.~M. and {Anderson}, J.},
    title = "{A New Distortion Solution for NIRC2 on the Keck II Telescope}",
  journal = {\pasp},
     year = 2016,
    month = sep,
   volume = 128,
   number = 9,
    pages = {095004},
      doi = {10.1088/1538-3873/128/967/095004},
   adsurl = {http://adsabs.harvard.edu/abs/2016PASP..128i5004S}
}

@ARTICLE{Udalski:2003,
   author = {{Udalski}, A.},
    title = "{The Optical Gravitational Lensing Experiment. Real Time Data Analysis Systems in the OGLE-III Survey}",
  journal = {Acta Astron.},
   eprint = {astro-ph/0401123},
     year = 2003,
    month = dec,
   volume = 53,
    pages = {291-305},
   adsurl = {http://adsabs.harvard.edu/abs/2003AcA....53..291U}
}

@ARTICLE{Udalski:2015b,
       author = {{Udalski}, A. and {Szyma{\'n}ski}, M.~K. and {Szyma{\'n}ski}, G.},
        title = "{OGLE-IV: Fourth Phase of the Optical Gravitational Lensing Experiment}",
      journal = {\actaa},
         year = 2015,
        month = mar,
       volume = {65},
       number = {1},
        pages = {1-38},
          doi = {10.48550/arXiv.1504.05966},
archivePrefix = {arXiv},
       eprint = {1504.05966},
 primaryClass = {astro-ph.SR},
       adsurl = {https://ui.adsabs.harvard.edu/abs/2015AcA....65....1U}
}

@ARTICLE{vanDam:2006,
   author = {{van Dam}, M.~A. and {Bouchez}, A.~H. and {Le Mignant}, D. and 
	{Johansson}, E.~M. and {Wizinowich}, P.~L. and {Campbell}, R.~D. and 
	{Chin}, J.~C.~Y. and {Hartman}, S.~K. and {Lafon}, R.~E. and 
	{Stomski}, Jr., P.~J. and {Summers}, D.~M.},
    title = "{The W. M. Keck Observatory Laser Guide Star Adaptive Optics System: Performance Characterization}",
  journal = {\pasp},
     year = 2006,
    month = feb,
   volume = 118,
    pages = {310-318},
      doi = {10.1086/499498},
   adsurl = {http://adsabs.harvard.edu/abs/2006PASP..118..310V}
}

@INPROCEEDINGS{Witzel:2016,
       author = {{Witzel}, Gunther and {Lu}, Jessica R. and {Ghez}, Andrea M. and
         {Martinez}, Gregory D. and {Fitzgerald}, Michael P. and
         {Britton}, Matthew and {Sitarski}, Breann N. and {Do}, Tuan and
         {Campbell}, Randall D. and {Service}, Maxwell},
        title = "{The AIROPA software package: milestones for testing general relativity in the strong gravity regime with AO}",
    booktitle = {Adaptive Optics Systems V},
         year = "2016",
       series = {Society of Photo-Optical Instrumentation Engineers (SPIE) Conference Series},
       volume = {9909},
        month = "Jul",
          eid = {99091O},
        pages = {99091O},
          doi = {10.1117/12.2233872},
       adsurl = {https://ui.adsabs.harvard.edu/abs/2016SPIE.9909E..1OW}
}

@ARTICLE{Wozniak:2000,
   author = {{Wozniak}, P.~R.},
    title = "{Difference Image Analysis of the OGLE-II Bulge Data. I. The Method}",
  journal = {Acta Astron.},
   eprint = {astro-ph/0012143},
     year = 2000,
    month = dec,
   volume = 50,
    pages = {421-450},
   adsurl = {http://adsabs.harvard.edu/abs/2000AcA....50..421W}
}

@ARTICLE{Yee:2015a,
       author = {{Yee}, Jennifer C. and {Gould}, Andrew and {Beichman}, Charles and
         {Calchi Novati}, Sebastiano and {Carey}, Sean and {Gaudi}, B. Scott and
         {Henderson}, Calen B. and {Nataf}, David and {Penny}, Matthew and
         {Shvartzvald}, Yossi and {Zhu}, Wei},
        title = "{Criteria for Sample Selection to Maximize Planet Sensitivity and Yield from Space-Based Microlens Parallax Surveys}",
      journal = {\apj},
         year = "2015",
        month = "Sep",
       volume = {810},
       number = {2},
          eid = {155},
        pages = {155},
          doi = {10.1088/0004-637X/810/2/155},
archivePrefix = {arXiv},
       eprint = {1505.00014},
 primaryClass = {astro-ph.EP},
       adsurl = {https://ui.adsabs.harvard.edu/abs/2015ApJ...810..155Y}
}

@ARTICLE{CorralSantana:2016,
   author = {{Corral-Santana}, J.~M. and {Casares}, J. and {Mu{\~n}oz-Darias}, T. and 
	{Bauer}, F.~E. and {Mart{\'{\i}}nez-Pais}, I.~G. and {Russell}, D.~M.
	},
    title = "{BlackCAT: A catalogue of stellar-mass black holes in X-ray transients}",
  journal = {\aap},
archivePrefix = "arXiv",
   eprint = {1510.08869},
 primaryClass = "astro-ph.HE",
     year = 2016,
    month = mar,
   volume = 587,
      eid = {A61},
    pages = {A61},
      doi = {10.1051/0004-6361/201527130},
   adsurl = {http://adsabs.harvard.edu/abs/2016A%26A...587A..61C}
}

@ARTICLE{Elbert:2018,
       author = {{Elbert}, Oliver D. and {Bullock}, James S. and {Kaplinghat}, Manoj},
        title = "{Counting black holes: The cosmic stellar remnant population and
        implications for LIGO}",
      journal = {\mnras},
         year = 2018,
        month = Jan,
       volume = {473},
        pages = {1186-1194},
          doi = {10.1093/mnras/stx1959},
 primaryClass = {Astrophysics - Astrophysics of Galaxies},
       adsurl = {https://ui.adsabs.harvard.edu/#abs/2018MNRAS.473.1186E}
}

@ARTICLE{Hog:1995,
   author = {{Hog}, E. and {Novikov}, I.~D. and {Polnarev}, A.~G.},
    title = "{MACHO photometry and astrometry.}",
  journal = {\aap},
     year = 1995,
    month = feb,
   volume = 294,
    pages = {287-294},
   adsurl = {http://adsabs.harvard.edu/abs/1995A%26A...294..287H}
}

@ARTICLE{Miyamoto:1995,
   author = {{Miyamoto}, M. and {Yoshii}, Y.},
    title = "{Astrometry for Determining the MACHO Mass and Trajectory}",
  journal = {\aj},
     year = 1995,
    month = sep,
   volume = 110,
    pages = {1427},
      doi = {10.1086/117616},
   adsurl = {http://adsabs.harvard.edu/abs/1995AJ....110.1427M}
}

@ARTICLE{Walker:1995,
   author = {{Walker}, M.~A.},
    title = "{Microlensed Image Motions}",
  journal = {\apj},
     year = 1995,
    month = nov,
   volume = 453,
    pages = {37},
      doi = {10.1086/176367},
   adsurl = {http://adsabs.harvard.edu/abs/1995ApJ...453...37W}
}

@ARTICLE{Wizinowich:2006,
  author =	 {{Wizinowich}, P.~L. and {Le Mignant}, D. and
                  {Bouchez}, A.~H. and {Campbell}, R.~D. and {Chin},
                  J.~C.~Y. and {Contos}, A.~R. and {van Dam},
                  M.~A. and {Hartman}, S.~K. and {Johansson},
                  E.~M. and {Lafon}, R.~E. and {Lewis}, H. and
                  {Stomski}, P.~J. and {Summers}, D.~M. and {Brown},
                  C.~G. and {Danforth}, P.~M. and {Max}, C.~E. and
                  {Pennington}, D.~M.  },
  title =	 "{The W. M. Keck Observatory Laser Guide Star
                  Adaptive Optics System: Overview}",
  journal =	 {\pasp},
  year =	 2006,
  month =	 feb,
  volume =	 118,
  pages =	 {297-309},
  doi =		 {10.1086/499290},
  adsurl =
                  {http://adsabs.harvard.edu/cgi-bin/nph-bib_query?bibcode=2006PASP..118..297W&db_key=AST}
}

@ARTICLE{Heger:2003,
       author = {{Heger}, A. and {Fryer}, C.~L. and {Woosley}, S.~E. and {Langer}, N. and
         {Hartmann}, D.~H.},
        title = "{How Massive Single Stars End Their Life}",
      journal = {\apj},
         year = "2003",
        month = "Jul",
       volume = {591},
       number = {1},
        pages = {288-300},
          doi = {10.1086/375341},
archivePrefix = {arXiv},
       eprint = {astro-ph/0212469},
 primaryClass = {astro-ph},
       adsurl = {https://ui.adsabs.harvard.edu/abs/2003ApJ...591..288H}
}

@ARTICLE{Sharma:2011,
   author = {{Sharma}, S. and {Bland-Hawthorn}, J. and {Johnston}, K.~V. and 
	{Binney}, J.},
    title = "{Galaxia: A Code to Generate a Synthetic Survey of the Milky Way}",
  journal = {\apj},
archivePrefix = "arXiv",
   eprint = {1101.3561},
     year = 2011,
    month = mar,
   volume = 730,
      eid = {3},
    pages = {3},
      doi = {10.1088/0004-637X/730/1/3},
   adsurl = {http://adsabs.harvard.edu/abs/2011ApJ...730....3S}
}

@ARTICLE{Wyrzykowski:2016,
   author = {{Wyrzykowski}, {\L}. and {Kostrzewa-Rutkowska}, Z. and {Skowron}, J. and 
	{Rybicki}, K.~A. and {Mr{\'o}z}, P. and {Koz{\l}owski}, S. and 
	{Udalski}, A. and {Szyma{\'n}ski}, M.~K. and {Pietrzy{\'n}ski}, G. and 
	{Soszy{\'n}ski}, I. and {Ulaczyk}, K. and {Pietrukowicz}, P. and 
	{Poleski}, R. and {Pawlak}, M. and {I{\l}kiewicz}, K. and {Rattenbury}, N.~J.
	},
    title = "{Black hole, neutron star and white dwarf candidates from microlensing with OGLE-III}",
  journal = {\mnras},
archivePrefix = "arXiv",
   eprint = {1509.04899},
 primaryClass = "astro-ph.SR",
     year = 2016,
    month = may,
   volume = 458,
    pages = {3012-3026},
      doi = {10.1093/mnras/stw426},
   adsurl = {http://adsabs.harvard.edu/abs/2016MNRAS.458.3012W}
}

@Article{Abbott:2018,
author="Abbott, B. P.
and Abbott, R.
and Abbott, T. D.
and Abernathy, M. R.
and Acernese, F.
and Ackley, K.
and Adams, C.
and Adams, T.
and Addesso, P.
and Adhikari, R. X.
and Adya, V. B.
and Affeldt, C.
and Agathos, M.
and Agatsuma, K.
and Aggarwal, N.
and Aguiar, O. D.
and Aiello, L.
and Ain, A.
and Ajith, P.
and Akutsu, T.
and Allen, B.
and Allocca, A.
and Altin, P. A.
and Ananyeva, A.
and Anderson, S. B.
and Anderson, W. G.
and Ando, M.
and Appert, S.
and Arai, K.
and Araya, A.
and Araya, M. C.
and Areeda, J. S.
and Arnaud, N.
and Arun, K. G.
and Asada, H.
and Ascenzi, S.
and Ashton, G.
and Aso, Y.
and Ast, M.
and Aston, S. M.
and Astone, P.
and Atsuta, S.
and Aufmuth, P.
and Aulbert, C.
and Avila-Alvarez, A.
and Awai, K.
and Babak, S.
and Bacon, P.
and Bader, M. K. M.
and Baiotti, L.
and Baker, P. T.
and Baldaccini, F.
and Ballardin, G.
and Ballmer, S. W.
and Barayoga, J. C.
and Barclay, S. E.
and Barish, B. C.
and Barker, D.
and Barone, F.
and Barr, B.
and Barsotti, L.
and Barsuglia, M.
and Barta, D.
and Bartlett, J.
and Barton, M. A.
and Bartos, I.
and Bassiri, R.
and Basti, A.
and Batch, J. C.
and Baune, C.
and Bavigadda, V.
and Bazzan, M.
and B{\'e}csy, B.
and Beer, C.
and Bejger, M.
and Belahcene, I.
and Belgin, M.
and Bell, A. S.
and Berger, B. K.
and Bergmann, G.
and Berry, C. P. L.
and Bersanetti, D.
and Bertolini, A.
and Betzwieser, J.
and Bhagwat, S.
and Bhandare, R.
and Bilenko, I. A.
and Billingsley, G.
and Billman, C. R.
and Birch, J.
and Birney, R.
and Birnholtz, O.
and Biscans, S.
and Bisht, A.
and Bitossi, M.
and Biwer, C.
and Bizouard, M. A.
and Blackburn, J. K.
and Blackman, J.
and Blair, C. D.
and Blair, D. G.
and Blair, R. M.
and Bloemen, S.
and Bock, O.
and Boer, M.
and Bogaert, G.
and Bohe, A.
and Bondu, F.
and Bonnand, R.
and Boom, B. A.
and Bork, R.
and Boschi, V.
and Bose, S.
and Bouffanais, Y.
and Bozzi, A.
and Bradaschia, C.
and Brady, P. R.
and Braginsky, V. B.
and Branchesi, M.
and Brau, J. E.
and Briant, T.
and Brillet, A.
and Brinkmann, M.
and Brisson, V.
and Brockill, P.
and Broida, J. E.
and Brooks, A. F.
and Brown, D. A.
and Brown, D. D.
and Brown, N. M.
and Brunett, S.
and Buchanan, C. C.
and Buikema, A.
and Bulik, T.
and Bulten, H. J.
and Buonanno, A.
and Buskulic, D.
and Buy, C.
and Byer, R. L.
and Cabero, M.
and Cadonati, L.
and Cagnoli, G.
and Cahillane, C.
and Calder{\'o}n Bustillo, J.
and Callister, T. A.
and Calloni, E.
and Camp, J. B.
and Cannon, K. C.
and Cao, H.
and Cao, J.
and Capano, C. D.
and Capocasa, E.
and Carbognani, F.
and Caride, S.
and Casanueva Diaz, J.
and Casentini, C.
and Caudill, S.
and Cavagli{\`a}, M.
and Cavalier, F.
and Cavalieri, R.
and Cella, G.
and Cepeda, C. B.
and Cerboni Baiardi, L.
and Cerretani, G.
and Cesarini, E.
and Chamberlin, S. J.
and Chan, M.
and Chao, S.
and Charlton, P.
and Chassande-Mottin, E.
and Cheeseboro, B. D.
and Chen, H. Y.
and Chen, Y.
and Cheng, H.-P.
and Chincarini, A.
and Chiummo, A.
and Chmiel, T.
and Cho, H. S.
and Cho, M.
and Chow, J. H.
and Christensen, N.
and Chu, Q.
and Chua, A. J. K.
and Chua, S.
and Chung, S.
and Ciani, G.
and Clara, F.
and Clark, J. A.
and Cleva, F.
and Cocchieri, C.
and Coccia, E.
and Cohadon, P.-F.
and Colla, A.
and Collette, C. G.
and Cominsky, L.
and Constancio, M.
and Conti, L.
and Cooper, S. J.
and Corbitt, T. R.
and Cornish, N.
and Corsi, A.
and Cortese, S.
and Costa, C. A.
and Coughlin, M. W.
and Coughlin, S. B.
and Coulon, J.-P.
and Countryman, S. T.
and Couvares, P.
and Covas, P. B.
and Cowan, E. E.
and Coward, D. M.
and Cowart, M. J.
and Coyne, D. C.
and Coyne, R.
and Creighton, J. D. E.
and Creighton, T. D.
and Cripe, J.
and Crowder, S. G.
and Cullen, T. J.
and Cumming, A.
and Cunningham, L.
and Cuoco, E.
and Canton, T. Dal
and Danilishin, S. L.
and D'Antonio, S.
and Danzmann, K.
and Dasgupta, A.
and Da Silva Costa, C. F.
and Dattilo, V.
and Dave, I.
and Davier, M.
and Davies, G. S.
and Davis, D.
and Daw, E. J.
and Day, B.
and Day, R.
and De, S.
and DeBra, D.
and Debreczeni, G.
and Degallaix, J.
and De Laurentis, M.
and Del{\'e}glise, S.
and Del Pozzo, W.
and Denker, T.
and Dent, T.
and Dergachev, V.
and De Rosa, R.
and DeRosa, R. T.
and DeSalvo, R.
and Devine, R. C.
and Dhurandhar, S.
and D{\'i}az, M. C.
and Fiore, L. Di
and Giovanni, M. Di
and Girolamo, T. Di
and Lieto, A. Di
and Pace, S. Di
and Palma, I. Di
and Virgilio, A. Di
and Doctor, Z.
and Doi, K.
and Dolique, V.
and Donovan, F.
and Dooley, K. L.
and Doravari, S.
and Dorrington, I.
and Douglas, R.
and Dovale {\'A}lvarez, M.
and Downes, T. P.
and Drago, M.
and Drever, R. W. P.
and Driggers, J. C.
and Du, Z.
and Ducrot, M.
and Dwyer, S. E.
and Eda, K.
and Edo, T. B.
and Edwards, M. C.
and Effler, A.
and Eggenstein, H.-B.
and Ehrens, P.
and Eichholz, J.
and Eikenberry, S. S.
and Eisenstein, R. A.
and Essick, R. C.
and Etienne, Z.
and Etzel, T.
and Evans, M.
and Evans, T. M.
and Everett, R.
and Factourovich, M.
and Fafone, V.
and Fair, H.
and Fairhurst, S.
and Fan, X.
and Farinon, S.
and Farr, B.
and Farr, W. M.
and Fauchon-Jones, E. J.
and Favata, M.",
title="Prospects for observing and localizing gravitational-wave transients with Advanced LIGO, Advanced Virgo and KAGRA",
journal="Living Reviews in Relativity",
year="2018",
month="Apr",
day="26",
volume="21",
number="1",
pages="3",
issn="1433-8351",
doi="10.1007/s41114-018-0012-9",
url="https://doi.org/10.1007/s41114-018-0012-9"
}

@ARTICLE{Gould:2004,
       author = {{Gould}, Andrew},
        title = "{Resolution of the MACHO-LMC-5 Puzzle: The Jerk-Parallax Microlens Degeneracy}",
      journal = {\apj},
         year = "2004",
        month = "May",
       volume = {606},
       number = {1},
        pages = {319-325},
          doi = {10.1086/382782},
archivePrefix = {arXiv},
       eprint = {astro-ph/0311548},
 primaryClass = {astro-ph},
       adsurl = {https://ui.adsabs.harvard.edu/abs/2004ApJ...606..319G}
}

@INPROCEEDINGS{Larkin:2006,
       author = {{Larkin}, James and {Barczys}, Matthew and {Krabbe}, Alfred and {Adkins}, Sean and {Aliado}, Ted and {Amico}, Paola and {Brims}, George and {Campbell}, Randy and {Canfield}, John and {Gasaway}, Thomas and {Honey}, Allan and {Iserlohe}, Christof and {Johnson}, Chris and {Kress}, Evan and {LaFreniere}, David and {Lyke}, James and {Magnone}, Ken and {Magnone}, Nick and {McElwain}, Michael and {Moon}, Juleen and {Quirrenbach}, Andreas and {Skulason}, Gunnar and {Song}, Inseok and {Spencer}, Michael and {Weiss}, Jason and {Wright}, Shelley},
        title = "{OSIRIS: a diffraction limited integral field spectrograph for Keck}",
    booktitle = {Society of Photo-Optical Instrumentation Engineers (SPIE) Conference Series},
         year = 2006,
       editor = {{McLean}, Ian S. and {Iye}, Masanori},
       series = {Society of Photo-Optical Instrumentation Engineers (SPIE) Conference Series},
       volume = {6269},
        month = jun,
          eid = {62691A},
        pages = {62691A},
          doi = {10.1117/12.672061},
       adsurl = {https://ui.adsabs.harvard.edu/abs/2006SPIE.6269E..1AL}
}

@INPROCEEDINGS{Arriaga:2018,
       author = {{Arriaga}, Pauline and {Fitzgerald}, Michael and {Johnson}, Chris and {Weiss}, Jason and {Lyke}, James E.},
        title = "{Upgrade and characterization of the OSIRIS imager detector}",
    booktitle = {Ground-based and Airborne Instrumentation for Astronomy VII},
         year = 2018,
       editor = {{Evans}, Christopher J. and {Simard}, Luc and {Takami}, Hideki},
       series = {Society of Photo-Optical Instrumentation Engineers (SPIE) Conference Series},
       volume = {10702},
        month = jul,
          eid = {107022U},
        pages = {107022U},
          doi = {10.1117/12.2313101},
       adsurl = {https://ui.adsabs.harvard.edu/abs/2018SPIE10702E..2UA}
}

@ARTICLE{Golovich:2022,
       author = {{Golovich}, Nathan and {Dawson}, William and {Bartoli{\'c}}, Fran and {Lam}, Casey Y. and {Lu}, Jessica R. and {Medford}, Michael S. and {Schneider}, Michael D. and {Chapline}, George and {Schlafly}, Edward F. and {Drlica-Wagner}, Alex and {Pruett}, Kerianne},
        title = "{A Reanalysis of Public Galactic Bulge Gravitational Microlensing Events from OGLE-III and -IV}",
      journal = {\apjs},
         year = 2022,
        month = may,
       volume = {260},
       number = {1},
          eid = {2},
        pages = {2},
          doi = {10.3847/1538-4365/ac5969},
archivePrefix = {arXiv},
       eprint = {2009.07927},
 primaryClass = {astro-ph.GA},
       adsurl = {https://ui.adsabs.harvard.edu/abs/2022ApJS..260....2G}
}

@ARTICLE{Li:2019,
       author = {{Li}, S. -S. and {Zang}, W. and {Udalski}, A. and {Shvartzvald}, Y. and {Huber}, D. and {Lee}, C. -U. and {Sumi}, T. and {Gould}, A. and {Mao}, S. and {Fouqu{\'e}}, P. and {Wang}, T. and {Dong}, S. and {J{\o}rgensen}, U.~G. and {Cole}, A. and {Mr{\'o}z}, P. and {Szyma{\'n}ski}, M.~K. and {Skowron}, J. and {Poleski}, R. and {Soszy{\'n}ski}, I. and {Pietrukowicz}, P. and {Koz{\l}owski}, S. and {Ulaczyk}, K. and {Rybicki}, K.~A. and {Iwanek}, P. and {Yee}, J.~C. and {Calchi Novati}, S. and {Beichman}, C.~A. and {Bryden}, G. and {Carey}, S. and {Gaudi}, B.~S. and {Henderson}, C.~B. and {Zhu}, W. and {Albrow}, M.~D. and {Chung}, S. -J. and {Han}, C. and {Hwang}, K. -H. and {Jung}, Y.~K. and {Ryu}, Y. -H. and {Shin}, I. -G. and {Cha}, S. -M. and {Kim}, D. -J. and {Kim}, H. -W. and {Kim}, S. -L. and {Lee}, D. -J. and {Lee}, Y. and {Park}, B. -G. and {Pogge}, R.~W. and {Bond}, I.~A. and {Abe}, F. and {Barry}, R. and {Bennett}, D.~P. and {Bhattacharya}, A. and {Donachie}, M. and {Fukui}, A. and {Hirao}, Y. and {Itow}, Y. and {Kondo}, I. and {Koshimoto}, N. and {Li}, M.~C.~A. and {Matsubara}, Y. and {Muraki}, Y. and {Miyazaki}, S. and {Nagakane}, M. and {Ranc}, C. and {Rattenbury}, N.~J. and {Suematsu}, H. and {Sullivan}, D.~J. and {Suzuki}, D. and {Tristram}, P.~J. and {Yonehara}, A. and {Christie}, G. and {Drummond}, J. and {Green}, J. and {Hennerley}, S. and {Natusch}, T. and {Porritt}, I. and {Bachelet}, E. and {Maoz}, D. and {Street}, R.~A. and {Tsapras}, Y. and {Bozza}, V. and {Dominik}, M. and {Hundertmark}, M. and {Peixinho}, N. and {Sajadian}, S. and {Burgdorf}, M.~J. and {Evans}, D.~F. and {Figuera Jaimes}, R. and {Fujii}, Y.~I. and {Haikala}, L.~K. and {Helling}, C. and {Henning}, T. and {Hinse}, T.~C. and {Mancini}, L. and {Longa-Pe{\~n}a}, P. and {Rahvar}, S. and {Rabus}, M. and {Skottfelt}, J. and {Snodgrass}, C. and {Southworth}, J. and {Unda-Sanzana}, E. and {von Essen}, C. and {Beaulieu}, J. -P. and {Blackman}, J. and {Hill}, K.},
        title = "{OGLE-2017-BLG-1186: first application of asteroseismology and Gaussian processes to microlensing}",
      journal = {\mnras},
         year = 2019,
        month = sep,
       volume = {488},
       number = {3},
        pages = {3308-3323},
          doi = {10.1093/mnras/stz1873},
archivePrefix = {arXiv},
       eprint = {1904.07718},
 primaryClass = {astro-ph.SR},
       adsurl = {https://ui.adsabs.harvard.edu/abs/2019MNRAS.488.3308L}
}

@ARTICLE{Hosek:2020,
       author = {{Hosek}, Matthew W., Jr. and {Lu}, Jessica R. and {Lam}, Casey Y. and {Gautam}, Abhimat K. and {Lockhart}, Kelly E. and {Kim}, Dongwon and {Jia}, Siyao},
        title = "{SPISEA: A Python-based Simple Stellar Population Synthesis Code for Star Clusters}",
      journal = {\aj},
         year = 2020,
        month = sep,
       volume = {160},
       number = {3},
          eid = {143},
        pages = {143},
          doi = {10.3847/1538-3881/aba533},
archivePrefix = {arXiv},
       eprint = {2006.06691},
 primaryClass = {astro-ph.SR},
       adsurl = {https://ui.adsabs.harvard.edu/abs/2020AJ....160..143H}
}

@ARTICLE{Bond:2001b,
       author = {{Bond}, I.~A. and {Abe}, F. and {Dodd}, R.~J. and {Hearnshaw}, J.~B. and {Honda}, M. and {Jugaku}, J. and {Kilmartin}, P.~M. and {Marles}, A. and {Masuda}, K. and {Matsubara}, Y. and {Muraki}, Y. and {Nakamura}, T. and {Nankivell}, G. and {Noda}, S. and {Noguchi}, C. and {Ohnishi}, K. and {Rattenbury}, N.~J. and {Reid}, M. and {Saito}, To. and {Sato}, H. and {Sekiguchi}, M. and {Skuljan}, J. and {Sullivan}, D.~J. and {Sumi}, T. and {Takeuti}, M. and {Watase}, Y. and {Wilkinson}, S. and {Yamada}, R. and {Yanagisawa}, T. and {Yock}, P.~C.~M.},
        title = "{Real-time difference imaging analysis of MOA Galactic bulge observations during 2000}",
      journal = {\mnras},
         year = 2001,
        month = nov,
       volume = {327},
       number = {3},
        pages = {868-880},
          doi = {10.1046/j.1365-8711.2001.04776.x},
archivePrefix = {arXiv},
       eprint = {astro-ph/0102181},
 primaryClass = {astro-ph},
       adsurl = {https://ui.adsabs.harvard.edu/abs/2001MNRAS.327..868B}
}

@ARTICLE{Sumi2003,
       author = {{Sumi}, T. and {Abe}, F. and {Bond}, I.~A. and {Dodd}, R.~J. and {Hearnshaw}, J.~B. and {Honda}, M. and {Honma}, M. and {Kan-ya}, Y. and {Kilmartin}, P.~M. and {Masuda}, K. and {Matsubara}, Y. and {Muraki}, Y. and {Nakamura}, T. and {Nishi}, R. and {Noda}, S. and {Ohnishi}, K. and {Petterson}, O.~K.~L. and {Rattenbury}, N.~J. and {Reid}, M. and {Saito}, To. and {Saito}, Y. and {Sato}, H. and {Sekiguchi}, M. and {Skuljan}, J. and {Sullivan}, D.~J. and {Takeuti}, M. and {Tristram}, P.~J. and {Wilkinson}, S. and {Yanagisawa}, T. and {Yock}, P.~C.~M.},
        title = "{Microlensing Optical Depth toward the Galactic Bulge from Microlensing Observations in Astrophysics Group Observations during 2000 with Difference Image Analysis}",
      journal = {\apj},
         year = 2003,
        month = jul,
       volume = {591},
       number = {1},
        pages = {204-227},
          doi = {10.1086/375212},
archivePrefix = {arXiv},
       eprint = {astro-ph/0207604},
 primaryClass = {astro-ph},
       adsurl = {https://ui.adsabs.harvard.edu/abs/2003ApJ...591..204S}
}

@ARTICLE{Terry2023,
       author = {{Terry}, Sean K. and {Lu}, Jessica R. and {Turri}, Paolo and {Ciurlo}, Anna and {Gautam}, Abhimat and {Do}, Tuan and {Fitzgerald}, Michael P. and {Ghez}, Andrea and {Hosek}, Matthew and {Witzel}, Gunther},
        title = "{AIROPA IV: Validating point spread function reconstruction on various science cases}",
      journal = {Journal of Astronomical Telescopes, Instruments, and Systems},
         year = 2023,
        month = jan,
       volume = {9},
          eid = {018003},
        pages = {018003},
          doi = {10.1117/1.JATIS.9.1.018003},
archivePrefix = {arXiv},
       eprint = {2209.05489},
 primaryClass = {astro-ph.IM},
       adsurl = {https://ui.adsabs.harvard.edu/abs/2023JATIS...9a8003T}
}

@ARTICLE{Turri2022,
       author = {{Turri}, Paolo and {Lu}, Jessica R. and {Witzel}, Gunther and {Ciurlo}, Anna and {Do}, Tuan and {Ghez}, Andrea M. and {Fitzgerald}, Michael P. and {Britton}, Matthew C. and {Ragland}, Sam and {Terry}, Sean K.},
        title = "{AIROPA III: testing simulated and on-sky data}",
      journal = {Journal of Astronomical Telescopes, Instruments, and Systems},
         year = 2022,
        month = jul,
       volume = {8},
          eid = {039002},
        pages = {039002},
          doi = {10.1117/1.JATIS.8.3.039002},
archivePrefix = {arXiv},
       eprint = {2207.00548},
 primaryClass = {astro-ph.IM},
       adsurl = {https://ui.adsabs.harvard.edu/abs/2022JATIS...8c9002T}
}

@ARTICLE{Ciurlo2022, 
       author = {{Ciurlo}, Anna and {Turri}, Paolo and {Witzel}, Gunther and {Lu}, Jessica R. and {Do}, Tuan and {Sitarski}, Breann N. and {Fitzgerald}, Michael P. and {Ghez}, Andrea M. and {Alvarez}, Carlos and {Terry}, Sean K. and {Doppmann}, Greg and {Lyke}, James E. and {Ragland}, Sam and {Campbell}, Randall and {Matthews}, Keith},
        title = "{AIROPA II: modeling instrumental aberrations for off-axis point spread functions in adaptive optics}",
      journal = {Journal of Astronomical Telescopes, Instruments, and Systems},
         year = 2022,
        month = jul,
       volume = {8},
          eid = {038007},
        pages = {038007},
          doi = {10.1117/1.JATIS.8.3.038007},
archivePrefix = {arXiv},
       eprint = {2210.10940},
 primaryClass = {astro-ph.IM},
       adsurl = {https://ui.adsabs.harvard.edu/abs/2022JATIS...8c8007C}
}

@ARTICLE{Abrams2025,
       author = {{Abrams}, Natasha S. and {Lu}, Jessica R. and {Lam}, Casey Y. and {Medford}, Michael S. and {Hosek}, Matthew W. and {Rose}, Sam},
        title = "{Assessing the Impact of Binary Systems on Microlensing Using SPISEA and PopSyCLE Population Simulations}",
      journal = {\apj},
         year = 2025,
        month = feb,
       volume = {980},
       number = {1},
          eid = {103},
        pages = {103},
          doi = {10.3847/1538-4357/ada5f9},
archivePrefix = {arXiv},
       eprint = {2501.03506},
 primaryClass = {astro-ph.SR},
       adsurl = {https://ui.adsabs.harvard.edu/abs/2025ApJ...980..103A}
}

@ARTICLE{Rose2022,
       author = {{Rose}, Sam and {Lam}, Casey Y. and {Lu}, Jessica R. and {Medford}, Michael and {Hosek}, Matthew W. and {Abrams}, Natasha S. and {Ramey}, Emily and {Vasylyev}, Sergiy S.},
        title = "{The Impact of Initial-Final Mass Relations on Black Hole Microlensing}",
      journal = {\apj},
         year = 2022,
        month = dec,
       volume = {941},
       number = {2},
          eid = {116},
        pages = {116},
          doi = {10.3847/1538-4357/aca09d},
archivePrefix = {arXiv},
       eprint = {2211.04471},
 primaryClass = {astro-ph.HE},
       adsurl = {https://ui.adsabs.harvard.edu/abs/2022ApJ...941..116R}
}

@ARTICLE{Huston2026,
       author = {{Huston}, Macy J. and {Crisp}, Alison L. and {Newman}, Marz and {Patlak}, Riley and {Penny}, Matthew T. and {Kluter}, Jonas and {Johnson}, Samson A. and {McGill}, Peter and {Smith}, Leigh C. and {Karkour}, Victor and {Abrams}, Natasha S. and {Fernandes}, Rachel B. and {Gaudi}, B. Scott and {Lam}, Casey Y. and {Lu}, Jessica R. and {McGee}, Carissma and {Calchi Novati}, Sebastiano and {Stassun}, Keivan G. and {Terry}, Sean K. and {Verma}, Himanshu and {Zohrabi}, Farzaneh},
        title = "{An Updated SynthPop Model for Microlensing Simulations I: Model Description \& Evaluation}",
      journal = {arXiv e-prints},
         year = 2026,
        month = mar,
          eid = {arXiv:2603.12219},
        pages = {arXiv:2603.12219},
          doi = {10.48550/arXiv.2603.12219},
archivePrefix = {arXiv},
       eprint = {2603.12219},
 primaryClass = {astro-ph.GA},
       adsurl = {https://ui.adsabs.harvard.edu/abs/2026arXiv260312219H}
}

@ARTICLE{synthpopI,
       author = {{Kl{\"u}ter}, Jonas and {Huston}, Macy J. and {Aronica}, Abigail and {Johnson}, Samson A. and {Penny}, Matthew T. and {Newman}, Marz and {Zohrabi}, Farzaneh and {Crisp}, Alison L. and {Chevis}, Allison},
        title = "{SYNTHPOP: A New Framework for Synthetic Milky Way Population Generation}",
      journal = {\aj},
         year = 2025,
        month = jun,
       volume = {169},
       number = {6},
          eid = {317},
        pages = {317},
          doi = {10.3847/1538-3881/adcd7a},
archivePrefix = {arXiv},
       eprint = {2411.18821},
 primaryClass = {astro-ph.IM},
       adsurl = {https://ui.adsabs.harvard.edu/abs/2025AJ....169..317K}
}

@ARTICLE{Wiktorowicz2019,
       author = {{Wiktorowicz}, Grzegorz and {Wyrzykowski}, {\L}ukasz and {Chruslinska}, Martyna and {Klencki}, Jakub and {Rybicki}, Krzysztof A. and {Belczynski}, Krzysztof},
        title = "{Populations of Stellar-mass Black Holes from Binary Systems}",
      journal = {\apj},
         year = 2019,
        month = nov,
       volume = {885},
       number = {1},
          eid = {1},
        pages = {1},
          doi = {10.3847/1538-4357/ab45e6},
archivePrefix = {arXiv},
       eprint = {1907.11431},
 primaryClass = {astro-ph.HE},
       adsurl = {https://ui.adsabs.harvard.edu/abs/2019ApJ...885....1W}
}

@ARTICLE{Giesers2019,
       author = {{Giesers}, Benjamin and {Kamann}, Sebastian and {Dreizler}, Stefan and {Husser}, Tim-Oliver and {Askar}, Abbas and {G{\"o}ttgens}, Fabian and {Brinchmann}, Jarle and {Latour}, Marilyn and {Weilbacher}, Peter M. and {Wendt}, Martin and {Roth}, Martin M.},
        title = "{A stellar census in globular clusters with MUSE: Binaries in NGC 3201}",
      journal = {\aap},
         year = 2019,
        month = dec,
       volume = {632},
          eid = {A3},
        pages = {A3},
          doi = {10.1051/0004-6361/201936203},
archivePrefix = {arXiv},
       eprint = {1909.04050},
 primaryClass = {astro-ph.SR},
       adsurl = {https://ui.adsabs.harvard.edu/abs/2019A&A...632A...3G}
}

@dataset{Minniti2023,
       author = {{Minniti}, D. and {Lucas}, P. and {Hempel}, M. and {The Vvv Science Team.}},
        title = "{VizieR Online Data Catalog: VISTA Variable in the Via Lactea Survey (VVV) DR4.2 (Minniti+, 2023)}",
 howpublished = {VizieR On-line Data Catalog: II/376.  Originally published in: 2012A\&A...537A.107S},
         year = 2023,
        month = jun,
          eid = {II/376},
       adsurl = {https://ui.adsabs.harvard.edu/abs/2023yCat.2376....0M}
}

@ARTICLE{Shan2019,
       author = {{Shan}, Yutong and {Yee}, Jennifer C. and {Udalski}, Andrzej and {Bond}, Ian A. and {Shvartzvald}, Yossi and {Shin}, In-Gu and {Jung}, Youn-Kil and {Calchi Novati}, Sebastiano and {Beichman}, Charles A. and {Carey}, Sean and {Gaudi}, B. Scott and {Gould}, Andrew and {Pogge}, Richard W. and {Spitzer Team} and {Poleski}, Rados{\l}aw and {Skowron}, Jan and {Koz{\l}owski}, Szymon and {Mr{\'o}z}, Przemys{\l}aw and {Pietrukowicz}, Pawe{\l} and {Szyma{\'n}ski}, Micha{\l} K. and {Soszy{\'n}ski}, Igor and {Ulaczyk}, Krzysztof and {Wyrzykowski}, {\L}ukasz and {OGLE Collaboration} and {Abe}, Fumio and {Barry}, Richard K. and {Bennett}, David P. and {Bhattacharya}, Aparna and {Donachie}, Martin and {Fukui}, Akihiko and {Hirao}, Yuki and {Itow}, Yoshitaka and {Kawasaki}, Kohei and {Kondo}, Iona and {Koshimoto}, Naoki and {Li}, Man Cheung Alex and {Matsubara}, Yutaka and {Muraki}, Yasushi and {Miyazaki}, Shota and {Nagakane}, Masayuki and {Ranc}, Cl{\'e}ment and {Rattenbury}, Nicholas J. and {Suematsu}, Haruno and {Sullivan}, Denis J. and {Sumi}, Takahiro and {Suzuki}, Daisuke and {Tristram}, Paul J. and {Yonehara}, Atsunori and {MOA Collaboration} and {Maoz}, Dan and {Kaspi}, Shai and {Friedmann}, Matan and {Wise Group}},
        title = "{OGLE-2014-BLG-0962 and a Comparison of Galactic Model Priors to Microlensing Data}",
      journal = {\apj},
         year = 2019,
        month = mar,
       volume = {873},
       number = {1},
          eid = {30},
        pages = {30},
          doi = {10.3847/1538-4357/ab0021},
       adsurl = {https://ui.adsabs.harvard.edu/abs/2019ApJ...873...30S}
}

@ARTICLE{Whitaker2026,
       author = {{Whitaker}, Matthew and {Kerr}, Evan and {Seth}, Anil and {H{\"a}berle}, Maximilian and {Strader}, Jay and {Anderson}, Jay and {Bellini}, Andrea and {Clontz}, Callie and {Freeman}, Zack and {Griggio}, Massimo and {Kamann}, Sebastian and {Libralato}, Mattia and {Neumayer}, Nadine and {Gonz{\'a}lez Prieto}, Elena and {Rodriguez}, Carl L. and {Saracino}, Sara and {Smith}, Peter and {van de Ven}, Glenn and {Wang}, Zixian},
        title = "{A Long Period Stellar-Mass Black Hole Binary in $ω$ Centauri}",
      journal = {arXiv e-prints},
         year = 2026,
        month = jun,
          eid = {arXiv:2606.18350},
        pages = {arXiv:2606.18350},
          doi = {10.48550/arXiv.2606.18350},
archivePrefix = {arXiv},
       eprint = {2606.18350},
 primaryClass = {astro-ph.GA},
       adsurl = {https://ui.adsabs.harvard.edu/abs/2026arXiv260618350W}
}

@ARTICLE{Giesers2018,
       author = {{Giesers}, Benjamin and {Dreizler}, Stefan and {Husser}, Tim-Oliver and {Kamann}, Sebastian and {Anglada Escud{\'e}}, Guillem and {Brinchmann}, Jarle and {Carollo}, C. Marcella and {Roth}, Martin M. and {Weilbacher}, Peter M. and {Wisotzki}, Lutz},
        title = "{A detached stellar-mass black hole candidate in the globular cluster NGC 3201}",
      journal = {\mnras},
         year = 2018,
        month = mar,
       volume = {475},
       number = {1},
        pages = {L15-L19},
          doi = {10.1093/mnrasl/slx203},
archivePrefix = {arXiv},
       eprint = {1801.05642},
 primaryClass = {astro-ph.SR},
       adsurl = {https://ui.adsabs.harvard.edu/abs/2018MNRAS.475L..15G}
}

@article{Disberg2025,
  author = {{Disberg}, Paul and {Mandel}, Ilya},
  title = {{The Kick Velocity Distribution of Isolated Neutron Stars}},
  journal = {\apjl},
  year = {2025},
  month = aug,
  volume = {989},
  number = {1},
  eid = {L8},
  pages = {L8},
  doi = {10.3847/2041-8213/adf286},
  archiveprefix = {arXiv},
  eprint = {2505.22102},
  primaryclass = {astro-ph.HE},
  adsurl = {https://ui.adsabs.harvard.edu/abs/2025ApJ...989L...8D},
}

@article{Luna2023,
  author = {{Luna}, Alonso and {Marchetti}, Tommaso and {Rejkuba}, Marina and {Minniti}, Dante},
  title = {{Astrometry in crowded fields towards the Galactic bulge}},
  journal = {\aap},
  year = {2023},
  month = sep,
  volume = {677},
  eid = {A185},
  pages = {A185},
  doi = {10.1051/0004-6361/202346257},
  archiveprefix = {arXiv},
  eprint = {2307.13719},
  primaryclass = {astro-ph.GA},
  adsurl = {https://ui.adsabs.harvard.edu/abs/2023A&A...677A.185L},
}

@article{Buchner2014,
  author = {{Buchner}, J. and {Georgakakis}, A. and {Nandra}, K. and {Hsu}, L. and {Rangel}, C. and {Brightman}, M. and {Merloni}, A. and {Salvato}, M. and {Donley}, J. and {Kocevski}, D.},
  title = {{X-ray spectral modelling of the AGN obscuring region in the CDFS: Bayesian model selection and catalogue}},
  journal = {\aap},
  year = {2014},
  month = apr,
  volume = {564},
  eid = {A125},
  pages = {A125},
  doi = {10.1051/0004-6361/201322971},
  archiveprefix = {arXiv},
  eprint = {1402.0004},
  primaryclass = {astro-ph.HE},
  adsurl = {https://ui.adsabs.harvard.edu/abs/2014A&A...564A.125B},
}

@article{Lu2026,
  author = {{Lu}, J.~R. and {Medford}, Michael and {Lam}, Casey Y. and {Bhadra}, T. Dex and {Huston}, Macy J. and {Abrams}, Natasha S. and {Broadberry}, Edward and {Chen}, Jeff and {Terry}, Sean K. and {Arredondo}, Nijaid and {Scharf}, Andrew and {Oliveira}, Raphael A.~P.},
  title = {{The BAGLE Python Package for Bayesian Analysis of Gravitational Lensing Events}},
  journal = {\aj},
  year = {2026},
  month = jul,
  volume = {172},
  number = {1},
  eid = {30},
  pages = {30},
  doi = {10.3847/1538-3881/ae665b},
  archiveprefix = {arXiv},
  eprint = {2512.03364},
  primaryclass = {astro-ph.SR},
  adsurl = {https://ui.adsabs.harvard.edu/abs/2026AJ....172...30L},
}

@article{Bhadra2026,
  author = {{Bhadra}, T. Dex and {Lu}, J.~R. and {Abrams}, Natasha S. and {Scharf}, Andrew and {Broadberry}, Edward and {Lam}, Casey and {Huston}, Macy J.},
  title = {{Modeling Binary Lenses and Sources with the BAGLE Python Package}},
  journal = {\apj},
  year = {2026},
  month = may,
  volume = {1002},
  number = {1},
  eid = {60},
  pages = {60},
  doi = {10.3847/1538-4357/ae5701},
  archiveprefix = {arXiv},
  eprint = {2512.03392},
  primaryclass = {astro-ph.SR},
  adsurl = {https://ui.adsabs.harvard.edu/abs/2026ApJ..1002...60B},
}

@article{Olejak2020,
  author = {{Olejak}, A. and {Belczynski}, K. and {Bulik}, T. and {Sobolewska}, M.},
  title = {{Synthetic catalog of black holes in the Milky Way}},
  journal = {\aap},
  year = {2020},
  month = jun,
  volume = {638},
  eid = {A94},
  pages = {A94},
  doi = {10.1051/0004-6361/201936557},
  archiveprefix = {arXiv},
  eprint = {1908.08775},
  primaryclass = {astro-ph.SR},
  adsurl = {https://ui.adsabs.harvard.edu/abs/2020A&A...638A..94O},
}

@article{Penoyre2022,
  author = {{Penoyre}, Zephyr and {Belokurov}, Vasily and {Evans}, N. Wyn},
  title = {{Astrometric identification of nearby binary stars - I. Predicted astrometric signals}},
  journal = {\mnras},
  year = {2022},
  month = jun,
  volume = {513},
  number = {2},
  pages = {2437-2456},
  doi = {10.1093/mnras/stac959},
  archiveprefix = {arXiv},
  eprint = {2111.10380},
  primaryclass = {astro-ph.SR},
  adsurl = {https://ui.adsabs.harvard.edu/abs/2022MNRAS.513.2437P},
}

@article{Holl2023,
  author = {{Holl}, B. and {Sozzetti}, A. and {Sahlmann}, J. and {Giacobbe}, P. and {S{\'e}gransan}, D. and {Unger}, N. and {Delisle}, J.-B. and {Barbato}, D. and {Lattanzi}, M.~G. and {Morbidelli}, R. and {Sosnowska}, D.},
  title = {{Gaia Data Release 3. Astrometric orbit determination with Markov chain Monte Carlo and genetic algorithms: Systems with stellar, sub-stellar, and planetary mass companions}},
  journal = {\aap},
  year = {2023},
  month = jun,
  volume = {674},
  eid = {A10},
  pages = {A10},
  doi = {10.1051/0004-6361/202244161},
  archiveprefix = {arXiv},
  eprint = {2206.05439},
  primaryclass = {astro-ph.EP},
  adsurl = {https://ui.adsabs.harvard.edu/abs/2023A&A...674A..10H},
}

@article{El-Badry2024,
  author = {{El-Badry}, Kareem and {Lam}, Casey and {Holl}, Berry and {Halbwachs}, Jean-Louis and {Rix}, Hans-Walter and {Mazeh}, Tsevi and {Shahaf}, Sahar},
  title = {{A generative model for Gaia astrometric orbit catalogs: selection functions for binary stars, giant planets, and compact object companions}},
  journal = {The Open Journal of Astrophysics},
  year = {2024},
  month = nov,
  volume = {7},
  eid = {100},
  pages = {100},
  doi = {10.33232/001c.125461},
  archiveprefix = {arXiv},
  eprint = {2411.00088},
  primaryclass = {astro-ph.SR},
  adsurl = {https://ui.adsabs.harvard.edu/abs/2024OJAp....7E.100E},
}

@article{Kaczmarek2026,
  author = {{Kaczmarek}, Zofia and {Sweeney}, David and {Wyrzykowski}, {\L}ukasz and {Bastian}, Ulrich},
  title = {{Predictions for astrometric microlensing in Gaia}},
  journal = {arXiv e-prints},
  year = {2026},
  month = sep,
  eid = {arXiv:2609.10705},
  pages = {arXiv:2609.10705},
  doi = {10.48550/arXiv.2609.10705},
  archiveprefix = {arXiv},
  eprint = {2609.10705},
  primaryclass = {astro-ph.GA},
  adsurl = {https://ui.adsabs.harvard.edu/abs/2026arXiv260910705K},
}

@article{Gould1994,
  author = {{Gould}, Andrew},
  title = {{Proper Motions of MACHOs}},
  journal = {\apjl},
  year = {1994},
  month = feb,
  volume = {421},
  pages = {L71},
  doi = {10.1086/187190},
  adsurl = {https://ui.adsabs.harvard.edu/abs/1994ApJ...421L..71G},
}

\appendix

\section{Microlensing Fit Priors}
\label{app:priors}

We perform a Bayesian analysis for model selection and parameter estimation as described in \S\ref{sec:fit_method}-\ref{sec:phot_fits}.
The priors used in the seeing-limited photometry only fits are listed in in the first 3 sections of Table \ref{tab:priors}, with the additional Keck photometry + astrometry parameters for the join fit in the final section. 

\begin{deluxetable*}{lcccc}
\tablecaption{Photometric and joint photometric-astrometric fit priors \label{tab:priors}}
\tablehead{
  \colhead{} & 
  \colhead{OB120169} & 
  \colhead{OB140613} & 
  \colhead{OB150029} & 
  \colhead{OB150211} 
}
\startdata
\multicolumn{5}{c}{\underline{Base Photometry Parameters}} \\
$t_0$ (MJD) & $U(55920, 56120)$ & $U(57050, 57250)$ & $U(57130, 57330)$ & $U(57125, 57325)$ \\
$u_0$ & $U(-1.5, 1.5)$ & $U(-1.5, 1.5)$ & $U(-1.5, 1.5)$ & $U(-1.5, 1.5)$ \\
$t_E$ (days) & $U(10, 1000)$ & $U(10, 1000)$ & $U(10, 1000)$ & $U(10, 1000)$ \\
$\pi_{E,E}$ & $U(-1, 1)$ & $U(-1, 1)$ & $U(-1, 1)$ & $U(-1, 1)$ \\
$\pi_{E,N}$ & $U(-1, 1)$ & $U(-1, 1)$ & $U(-1, 1)$ & $U(-1, 1)$ \\
$b_{SFF,I}$ & $U(0, 1.2)$ & $U(0, 1.2)$ & $U(0, 1.2)$ & $U(0, 1.2)$ \\
$M_{base,I}$ (mag) & $ N(19.4, 0.5) $ & $ N(18.2, 0.5) $ & $ N(15.1, 0.5) $ & $ N(17.3, 0.5) $ \\ \hline 
\multicolumn{5}{c}{\underline{MOA Photometry Parameters}} \\
$b_{SFF,R}$ & - & $U(0, 1.2)$ & $U(0, 1.2)$ & - \\
$M_{base,R}$ (inst. mag) & - & $ N(-10.2, 0.5) $ & $ N(-13.0, 0.5) $ & - \\ 
\hline
\multicolumn{5}{c}{\underline{Optional Error Modification Parameters}} \\
$\varepsilon_{a,I}$ (mag) & $U(0, 0.3)$ & $U(0, 0.3)$ & $U(0, 0.3)$ & $U(0, 0.3)$ \\
$\varepsilon_{m,I}$ & $U(1, 3)$ & $U(1, 3)$ & $U(1, 3)$ & - \\
$\varepsilon_{a,R}$ (mag) & - & $U(0, 0.3)$ & $U(0, 0.3)$ & - \\
$\varepsilon_{m,R}$ & - & $U(1, 3)$ & $U(1, 3)$ & - \\
\hline
\multicolumn{5}{c}{\underline{Additional Keck Photometry + Astrometry Parameters}} \\
$b_{SFF,K}$ & $U(0, 1)$ & $U(0, 1)$ & $U(0, 1)$ & $U(0.4, 1)$ \\
$M_{base,K}$ & $N(18.0, 0.5) $ & $N(14.3, 0.5)$ & $N(12.35, 0.5)$ & $N(11.25, 0.5)$ \\
$\log_{10}\theta_E$ (mas) & $N_T(-0.2, 0.3, -4, 4)$ & $N_T(-0.2, 0.3, -4, 4)$ & $N_T(-0.2, 0.3, -4, 4)$ & $N_T(-0.2, 0.3, -4, 4)$ \\
$\pi_S$ (mas) & $N_T(0.1126, 0.0213, -2.9, 90)$ & $N_T(0.1126, 0.0213, -2.9, 90)$ & $N_T(0.1126, 0.0213, -2.9, 90)$ & $N_T(0.1126, 0.0213, -2.9, 90)$ \\
$x_{S0,E}$ (arcsec) & $U(-0.020,0.050)$ & $U(-0.047,0.010)$ & $U(-0.072,-0.006)$ & $U(-0.017,0.054)$ \\
$x_{S0,N}$ (arcsec) & $U(-0.061,0.090)$ & $U(-0.057,0.047)$ & $U(0.011,0.089)$ & $U(-0.197,-0.050)$ \\
$\mu_{S,E}$ (mas/yr) & $U(-8.726,1.680)$ & $U(-5.923,-0.523)$ & $U(-9.788,1.692)$ & $U(-10.334,3.330)$ \\
$\mu_{S,N}$ (mas/yr) & $U(-18.040,2.705)$ & $U(-8.144,-3.725)$ & $U(-7.421,-2.124)$ & $U(-11.039,-3.382)$ 
\enddata
\tablecomments{Here we denote $U(a, b)$ to be a uniform PDF, equal to $\frac{1}{b-a}$ over $[a,b]$ and zero elsewhere.
$N(\mu, \sigma)$ is used to denote a normal PDF with mean $\mu$ and standard deviation $\sigma$. 
$N_T(\mu, \sigma, \alpha, \beta)$ is used to denote a truncated normal distribution with mean $\mu$, standard deviation $\sigma$, and upper and lower limits $\alpha$ and $\beta$ defined with respect to a standard normal distribution.}
\end{deluxetable*}

\end{document}